%% file: arxiv.tex
\documentclass[
aps, %
prl,
preprintnumbers,
twocolumn,
superscriptaddress,
nofootinbib,
floatfix,
10pt
]{revtex4-1}

\usepackage[abs]{overpic}
\usepackage{graphicx}
\usepackage{bm}
\usepackage{amssymb,amsmath}
\usepackage{mathrsfs}
\usepackage{latexsym}
\usepackage{float}
\usepackage{color}
\usepackage[normalem]{ulem} 
\usepackage{dcolumn}
\usepackage[colorlinks=true,citecolor=blue,urlcolor=blue]{hyperref}
\usepackage[usenames,dvipsnames]{xcolor}
\usepackage{slashed}
\usepackage{cancel}
\usepackage[dvipsnames]{xcolor}

\usepackage{amsfonts,amssymb,stmaryrd,latexsym,amsmath,braket}
\usepackage{graphicx,subfigure}
\usepackage{times}
\usepackage{slashed}
\usepackage{braket}
\usepackage{verbatim}
\usepackage[utf8]{inputenc}
\usepackage{color}
\usepackage[dvipsnames]{xcolor}

\begin{document}
\preprint{N3AS-24-028}
\title{Characterizing the nuclear models informed by PREX and CREX: a view from Bayesian inference} 

\author{Tianqi Zhao}
\email{tianqi.zhao@berkeley.edu}
\affiliation{Department of Physics, University of California Berkeley, Berkeley, California 94720, USA}
\affiliation{Department of Physics and Astronomy, Ohio University,
Athens, Ohio 45701, USA}

\author{Zidu Lin}
\email{zlin23@utk.edu}
\affiliation{University of Tennessee, Knoxville, Tennessee 37996, USA}

\author{Bharat Kumar}
\email{kumarbh@nitrkl.ac.in }
\affiliation{Department of Physics \& Astronomy, National Institute of Technology, Rourkela 769008, India}

\author{Andrew W. Steiner}
\email{awsteiner@utk.edu}
\affiliation{University of Tennessee, Knoxville, Tennessee 37996, USA}
\affiliation{Physics Division, Oak Ridge National Laboratory}

\author{Madappa Prakash}
\email{prakash@ohio.edu}
\affiliation{Department of Physics and Astronomy, Ohio University,
Athens, Ohio 45701, USA}

\date{\today}

\begin{abstract}
New measurements of the weak charge density distributions of $^{48}$Ca and $^{208}$Pb challenge existing nuclear models. In the post-PREX-CREX era, it is unclear if current models can simultaneously describe weak charge distributions along with accurate measurements of binding energy and charge radii. In this letter, we explore the parameter space of relativistic and non-relativistic models to study the differences between the electric and weak form factors, $\Delta F=F_{ch}-F_{W}$, in $^{48}$Ca and $^{208}$Pb. We show, for the first time, which aspects of mean-field models are the most important in determining the relative magnitude of the neutron skin in lead and calcium nuclei. We carefully disentangle the tension between the PREX-2/CREX constraints and the ability of the RMF and Skyrme models to accurately describe binding energies and charge radii. We find that the nuclear symmetry energy coefficient $S_\mathrm{V}$ and the isovector spin-orbit coefficient $b'_4$ play different roles in determining $\Delta F$ of $^{48}$Ca and $^{208}$Pb. Consequently, adjusting $S_\mathrm{V}$ or $b'_4$ shifts predicted $\Delta F$ values toward or away from PREX-2/CREX measurements. Additionally, $S_\mathrm{V}$ and the slope L are marginally correlated given the prior constraints of our Bayesian inference, allowing us to infer them separately from PREX-2/CREX data.

\end{abstract}

\maketitle

\emph{Nuclear models describing finite nuclei and nuclear matter --} 
One of the goals in nuclear physics is the construction of a unified theory that describes the basic properties of nuclei and neutron star (NS) observables. Historically, nuclear Hamiltonians have been constrained by binding energies, charge radii, and the energies of giant resonances. The requirement of simultaneously describing the charge distributions of a variety of isotopes has spurred the improvement of nuclear structure theories, including those based on \textit{ab initio} calculations, Skyrme effective interactions and relativistic mean field (RMF) calculations\cite{Lalazissis:1996rd,Konig:2023bzp,Sommer:2022sok,Malbrunot-Ettenauer:2021fnr,Reinhard:2022jby}. The Skyrme force is an effective interaction used in nuclear Hartree-Fock approximations, characterized by a zero range interaction with density-dependent terms. On the other hand, the RMF theory is a relativistic framework describing interacting nucleonic Dirac fields and mesonic mean fields. In RMF theory, the basic fields typically included are the scalar ($\sigma$), vector ($\omega$) and vector-isovector fields. The RMF can be directly related to Skyrme models by considering a non-relativistic expansion \cite{koepf1991spin}, which will be discussed in more details in this letter. Neutron distributions in finite nuclei are less sensitive to binding energies and charge radii \cite{Horowitz_2001}, leading to less-constrained isovector interactions in traditional nuclear models. Nonetheless, isospin-dependent interactions play a crucial role in describing NS properties. Recently, information about differences between the electric and weak form factors $\Delta F=F_{ch}-F_{W}$ of two neutron-rich magic nuclei, $^{208}$Pb and $^{48}$Ca, has been model-independently measured via parity-violating electron scattering experiments \cite{PREX:2021umo,CREX:2022kgg} and many theoretical models \cite{reed2024density,Reed2021,reinhard2022combined,Hagen:2015yea,Hu2022,mahzoon2017neutron,atkinson2020dispersive}  have been proposed to interpret the measurements from PREX-2 and CREX. However, it is important to note that these models generally succeed in describing either PREX-2 or CREX measurements, but face difficulties in simultaneously explaining both sets of data.

Can existing forms of relativistic RMF models and non-relativistic Skyrme models simultaneously describe the weak charge distributions inferred by PREX-2/CREX, while maintaining consistency with the basic ground state properties of finite nuclei? If so, what would be the implications for the isospin-dependent interactions of such models? In this letter,  we investigate this issue by extensively surveying the parameter space of both RMF and Skyrme models to study their dependence on \(S_\mathrm{V}\), \(L\) and isovector spin-orbit interactions. Using Bayesian inference, we analyze the information provided by PREX-2 and CREX measurements, detailed further in the supplemental material and data repository \cite{Zhao_CPREX}. Our approach also incorporates the fundamental ground state properties of \(^{48}\text{Ca}\), \(^{90}\text{Zr}\), and \(^{208}\text{Pb}\), in line with previous studies \cite{reinhard2022combined, Zhang2022, salinas2023bayesian}. Notably, we focus on the physical properties of the Skyrme and RMF models that are consistent with PREX-2/CREX experiments and quantify the extent to which these models can jointly reconcile PREX-2/CREX data with other existing nuclear measurements.

\emph{The picture of $S_\mathrm{v}-L$ inferred by joint PREX-2/CREX measurements --}
For densities of relevance to nuclei, the energy of asymmetric nuclear matter can be expressed as the sum of symmetric nuclear matter (SNM) and symmetry energies as
\begin{equation}
E(n_p,n_n)=E(n_p=n_n=n/2)+S(n_p,n_n)~,
\end{equation}
where $n_n$ and $n_p$ are the neutron and proton densities, and $n$ is their sum. The SNM energy $E$ can be expanded around the saturation density $n_0$ {, at which the SNM energy reaches its minimum}:
\begin{equation}
    E(n_p=n_n=n/2)=-B+\frac{K_0}{18n_0^2}(n-n_0)^2+... ~.
\end{equation}
Above, $B$ is the binding energy, and $K_0$ is the SNM incompressibility. To good approximation, the symmetry energy $S$ is given by the quadratic term in the expansion (asymmetry) parameter $(n_n - n_p)/n$:
\begin{equation}
S(n_p,n_n)=S(n) \left(\frac{n_n-n_p}{n}\right)^2+... ~,
\end{equation}
where $S(n)$ is the density-dependent symmetry energy. An alternative definition based on pure neutron matter (PNM), $S(n) = S(n_n = n, n_p = 0)$, has also been used \cite{gandolfi2014equation}. The deviation of $S(n)$ between these two definitions represents the breaking of the quadratic approximation, which is usually small for most Skyrme models \cite{lee1998nuclear, zuo1999asymmetric, Steiner06hs,drischler2014microscopic}. However, such a deviation can be larger in some other models \cite{cai2019relativistic}. In RMF models, the inclusion of the $\delta$-meson with large Yukawa coupling breaks the quadratic approximation significantly as shown in \cite{salinas2023impact}, and Table II in the supplemental material. Ignoring corrections that are quartic in the isospin asymmetry, $S(n)$ can be expanded around the saturation density as 
\begin{eqnarray}
    S(n)&=&S_\mathrm{V}+\frac{L}{3n_0}(n-n_0)+\frac{K_{sym}}{18n_0^2}(n-n_0)^2+... , 
\end{eqnarray}
where $S_\mathrm{V}$ is the bulk symmetry energy, $L$ refers to the slope of the symmetry energy slope, and $K_{sym}$ to the curvature of the symmetry energy. 

Neutron skin thicknesses of neutron-rich nuclei are primarily determined by $S(n)$~\cite{Steiner05ia}. The strong correlation between $L$ and the neutron skin thickness has been extensively studied using Skyrme and RMF models of nuclei \cite{horowitz2001parity, Reed2021}. The neutron skin can also be understood using a simple droplet model
in terms of the competition between the bulk symmetry term $S_\mathrm{V} \cdot I$ and the surface term $\Delta R_{np} \cdot Q$~ \cite{Myers:1980iht, Centelles:2008vu, Warda2009}.  Here, 
$I = (N - Z)/A$ with $N$ and $Z$ being the neutron and proton numbers in a nucleus and $A$ their sum,   
$\Delta R_{np}$ is the neutron-proton radius difference, and
$Q$ is the surface stiffness parameter. The first of these competing terms is closely related to the properties of bulk matter in the central regions of nuclei and the second relates also to the spin-orbit interaction.
 \begin{figure}
    \centering
    \includegraphics[width=\linewidth]{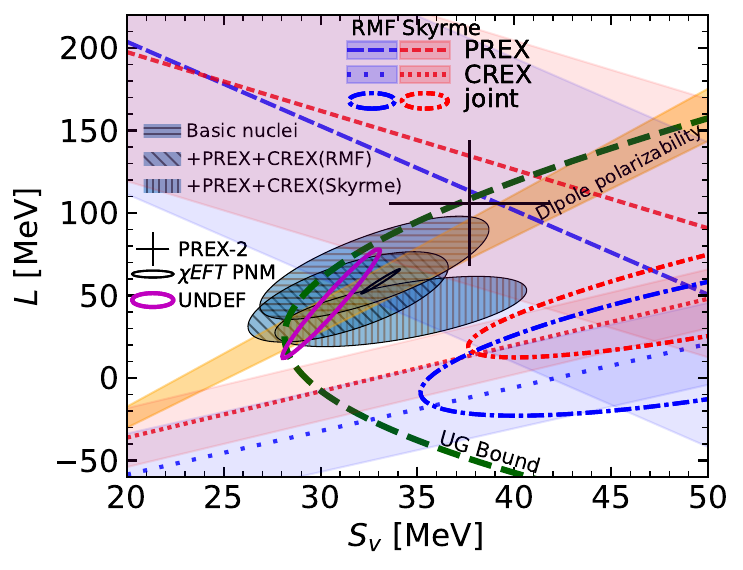}
    \caption{Symmetry energy parameters $S_\mathrm{V}$ and $L$ constrained by parity violating asymmetry of $^{48}$Ca (CREX) and $^{208}$Pb (PREX-2). The dashed (dotted) line and the corresponding band represent the mean and 1-$\sigma$ band of PREX-2 (CREX) alone based on the correlations in Eq. ~(\ref{eq:fit_Sv}) and Eq. ~(\ref{eq:fit_L}). Dash-dotted ellipse represents the 1-$\sigma$ contour of joint PREX-2/CREX constraints. The PREX-2 analysis~\cite{Reed2021} as well as constraints from the unitary gas conjecture~\cite{tews2017symmetry} and dipole polarizability~\cite{Piekarewicz2012} are shown for comparison. The horizontally shaded ellipse shows the 1-$\sigma$ contour of the posterior with basic nuclei constraints, corresponding to the yellow distribution shown in Fig.~\ref{fig:svLposterior}, whereas the other two shaded ellipses show the posterior with +PREX+CREX for Skyrme (vertical shade) and RMF (diagonal shade) models. The magenta ellipse shows the UNEDF correlation considering properties of more nuclei~\cite{kortelainen2010nuclear}. The small black ellipse shows the correlation from chiral EFT calculation of PNM~\cite{drischler2024bayesian}.}
    \label{fig:svLLinear}
\end{figure}
In Fig. \ref{fig:svLLinear}, we present the $S_\mathrm{V}$ and $L$ constrained by $\Delta F^{\mathrm{Ca}48}$ and $\Delta F^{\mathrm{Pb}208}$, without assuming the correlation between them. Given $\Delta F^{\mathrm{Ca}48}$, $\Delta F^{\mathrm{Pb}208}$, $S_\mathrm{V}$ and  L sampled from our Bayesian posterior, we find that $S_\mathrm{V}$ and $L$ can be well approximated by a linear combination of $\Delta F^{\mathrm{Ca}48}$ and $\Delta F^{\mathrm{Pb}208}$ as 
 \begin{eqnarray}
    S_\mathrm{V}&=&a\Delta F^{\mathrm{Ca}48}+b \Delta F^{\mathrm{Pb}208} +c \label{eq:fit_Sv}\\
    L  &=&a'\Delta F^{\mathrm{Ca}48}+b' \Delta F^{\mathrm{Pb}208} +c' \label{eq:fit_L}~.
\end{eqnarray}
\begin{table*}[]
    \centering
    \begin{tabular}{cccccccccc}
    \hline\hline
    &$a$&$b$&$c$&$a'$&$b'$&$c'$&$a''$&$b''$&$c''$\\
    \hline
RMF & $-575.2 \pm 5.1$ & $916.3 \pm 4.6$ & $32.2 \pm 3.7$ & $2938.7 \pm 43.5$ & $2420.6 \pm 33.9$ & $-149.8 \pm 25.6$ & $-31.8 \pm 0.2$ & $23.5 \pm 0.1$ & $0.94 \pm 0.11$\\
Skyrme & $-503.2 \pm 7.8$ & $945.2 \pm 5.5$ & $31.9 \pm 2.9$ & $1791.2 \pm 27.2$ & $2652.0 \pm 19.0$ & $-91.5 \pm 10.1$ & $-52.0 \pm 0.4$ & $27.8 \pm 0.3$ & $1.67 \pm 0.15$\\
    \hline\hline
    \end{tabular}
    \caption{Parameters for the fitting formulae of $S_\mathrm{V}$ (Eq. \ref{eq:fit_Sv}) and $L$ (Eq. \ref{eq:fit_L}) are in units of MeV, and for $b'_4$ (Eq. \ref{eq:fit_b'4}), they are in units of fm$^{4}$.}
    \label{tab:Sv_L_FF_fit}
\end{table*}
The slope coefficients $a$, $b$, $a'$, and $b'$, as well as the intercept parameters $c$ and $c'$, are determined by fitting them to the posterior distribution of $(\Delta F^{\mathrm{Ca}48}, \Delta F^{\mathrm{Pb}208}, S_\mathrm{V})$ and $(\Delta F^{\mathrm{Ca}48}, \Delta F^{\mathrm{Pb}208}, L)$, respectively. These parameters may reflect the inherent structure of RMF and Skyrme models and could be insensitive to the Bayesian constraints. All the sampled points from the posterior are presented in the supplemental material. The fitted slope and intercept parameters are summarized in Table \ref{tab:Sv_L_FF_fit}. The precision of the fitting formula in Eq. (\ref{eq:fit_Sv}) remains within 10\% (1-$\sigma$) for both Skyrme and RMF models. 
Given the mean and standard deviation of $\Delta F^{\mathrm{Pb}208}$ and $\Delta F^{\mathrm{Ca}48}$ measured by PREX-2/CREX and the linear relationship in Eq. (\ref{eq:fit_Sv}) and Eq. (\ref{eq:fit_L}), we obtain the mean and covariance of $S_\mathrm{V}$ and $L$ to be
\begin{eqnarray}
   (\bar S_\mathrm{V}, \bar L)&=& (56.7,~~66.8)_{Skyrme},~~~(53.8,~~30.9)_{RMF} \label{eq:SvL_cov} \\
   \sqrt{\mathrm{cov}} &=&
   \begin{pmatrix}
   19.6& 31.2\\
   31.2& 56.5
   \end{pmatrix}_{Skyrme},~~
   \begin{pmatrix}
   19.5& 31.0 \\
   31.0& 66.3
   \end{pmatrix}_{RMF} \nonumber
\end{eqnarray}
in units of MeV for Skyrme and RMF models, respectively. See the red and blue dash-dotted ellipses in Fig. \ref{fig:svLLinear}. The covariance is dominated by the experimental error of $\Delta F^{\mathrm{Ca}48}$ or $\Delta F^{\mathrm{Pb}208}$. The precision of the fitting formulas in Eq. (\ref{eq:fit_Sv}) and Eq. (\ref{eq:fit_L}) serves as a minor modification.

Comparing to previous PREX-2 analysis~\cite{Reed2021}, our constraints on $S_V$ and L are less affected by the basic finite nuclei ground state properties. The overlapping region of these two bands represents the range of ${S_\mathrm{V},L}$ inferred by the 1-$\sigma$ experimental measurements from  PREX-2/CREX, which is quantitatively described by Eq. (\ref{eq:SvL_cov}).
\begin{figure}
    \centering
    \includegraphics[width=\linewidth]{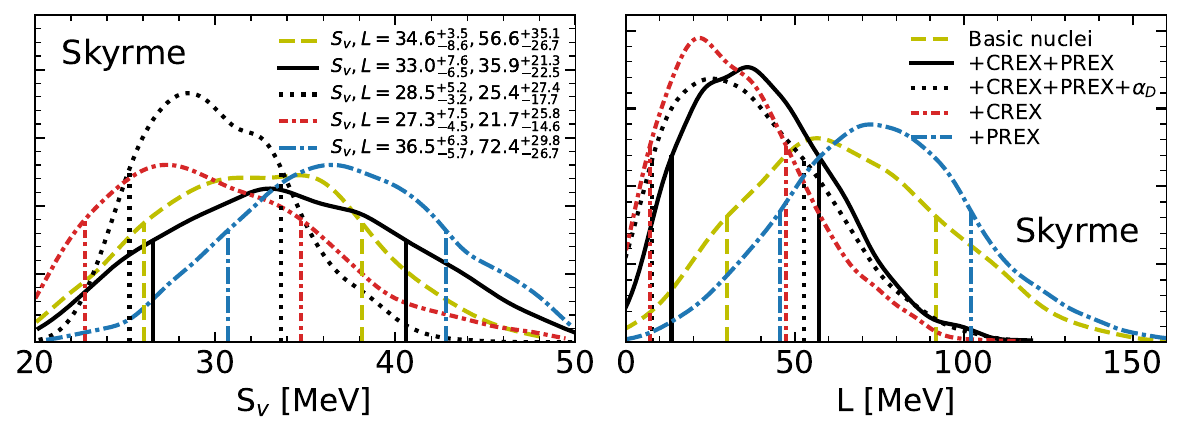}
    \includegraphics[width=\linewidth]{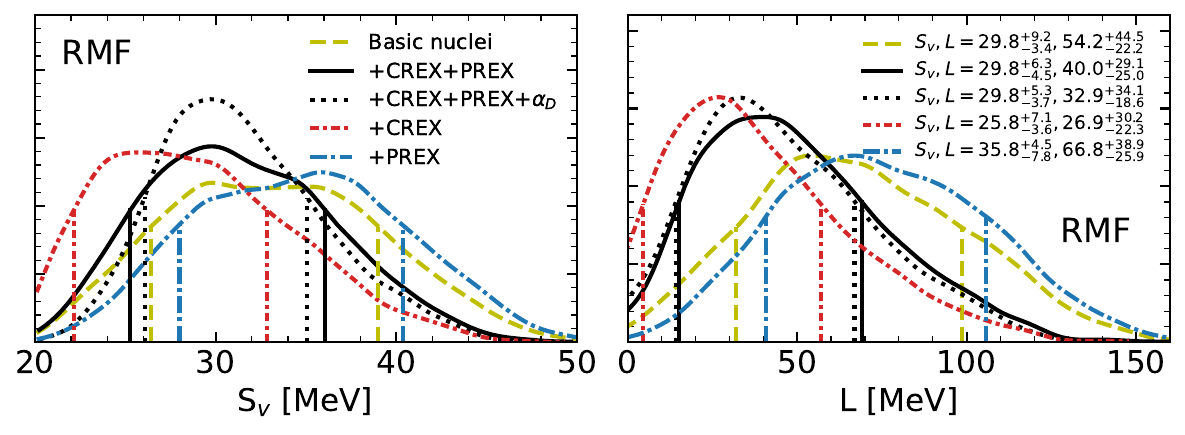}
    \caption{The posterior distribution of bulk symmetry energy $S_\mathrm{V}$ and symmetry energy slope $L$ at saturation density constrained by basic nuclei constraints (dashed yellow), +PREX experiment (dash-dotted blue), +CREX experiment (dash-dotted red), +PREX+CREX experiment (solid black). The dotted black curve includes additional constraint from dipole Polarizability \cite{Piekarewicz2012}.  {The vertical lines correspond to $1-\sigma$ asymmetric uncertainty.} Details of prior and likelihood are discussed in Supplemental Material.}
    \label{fig:svLposterior}
\end{figure}
In Fig. \ref{fig:svLposterior}, we analyze the evolution of the posterior distributions for $S_\mathrm{V}$ and $L$ under five different constraint sets: basic nuclear constraints as the "minimum setting", corresponding to horizontally shaded
ellipse in Fig. \ref{fig:svLLinear}; inclusion of either PREX-2 or CREX experimental data; inclusion of both PREX-2/CREX experimental data, corresponding to vertically and diagonally shaded
ellipse in Fig. \ref{fig:svLLinear}; and inclusion of additional dipole polarizability constraint. In the upper two panels for the Skyrme model, we observe that the range of $S_\mathrm{V}$ is approximately [20,50] MeV, whereas the range of $L$ spans from [0,150] MeV. In the upper right (Skyrme) and lower right (RMF) panels, the posteriors of $L$ exhibit similar trends. The posteriors of $L$ constrained by CREX (PREX-2) measurements has a mean smaller (larger) than that constrained by only basic properties of finite nuclei. Moreover, the posterior of $L$ constrained by both PREX-2 and CREX measurements have a mean lower than that without any neutron skin constraints, but slightly higher than the one that includes only the CREX constraint. This trend exists in both RMF and Skyrme models and aligns with previous studies using Skyrme\cite{Reinhard2021,Reinhard:2022jby,Zhang2022} and RMF models\cite{Reed2021,salinas2023bayesian,yuksel2023implications}. The confidence level integral suggests a probability of inconsistency of 93\% (1.8-$\sigma$) for Skyrme models and 76\% (1.2-$\sigma$) for RMF models, significantly lower than other analyses showing over 2-$\sigma$ inconsistency \cite{Zhang2022,salinas2023bayesian,yuksel2023implications}. Further details of our Bayesian inference and Monte Carlo sampling are provided in the supplemental material.

\emph{Impact of spin-orbit interactions --} 
Unlike the Hamiltonian of infinite nuclear matter, the Hamiltonian of finite nuclei incorporates terms proportional to the derivative of isoscalar (isovector) densities, primarily arising from spin-orbit interactions. The impact of spin-orbit interactions on single-particle spectra and the charge distributions of various neutron-rich isotopes has been acknowledged and discussed in prior studies \cite{Reinhard:1995zz,Horowitz:2012we}. However, the isospin-dependent spin-orbit interactions are not well constrained, in comparison with other interactions constrained by isobaric analog states (IAS) \cite{roca2018nuclear} and mirror nuclei \cite{sagawa2024qcd}. Since the weak charge form factor $F_W$ and the charge form factor $F_{ch}$ for $^{208}\mathrm{Pb}$ and $^{48}\mathrm{Ca}$ are very close in value, the resulting difference $\Delta F = F_W - F_{ch}$ is relatively small—roughly one order of magnitude smaller than $F_\mathrm{W}$ and $F_\mathrm{ch}$. As a result, the influence of spin-orbit interactions becomes crucial in accurately determining the distribution of $\{\Delta F_{^{48}\mathrm{Ca}}, \Delta F_{^{208}\mathrm{Pb}}\}$ in our study. For the first time (as far as we are aware), we investigate the influence of spin-orbit interactions on both $\Delta F^{\mathrm{Ca}48}$ and $\Delta F^{\mathrm{Pb}208}$ across a very large parameter space encompassing both RMF and Skyrme parameterizations. This investigation is based on the posterior distribution incorporating constraints from  PREX-2/CREX measurements.
\begin{figure*}
    \centering
    \includegraphics[width=0.4\linewidth]{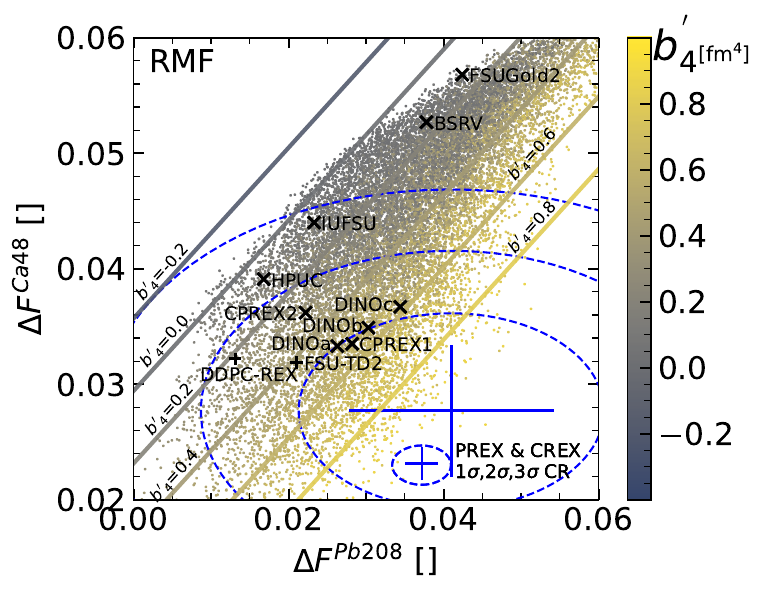}
    \includegraphics[width=0.4\linewidth]{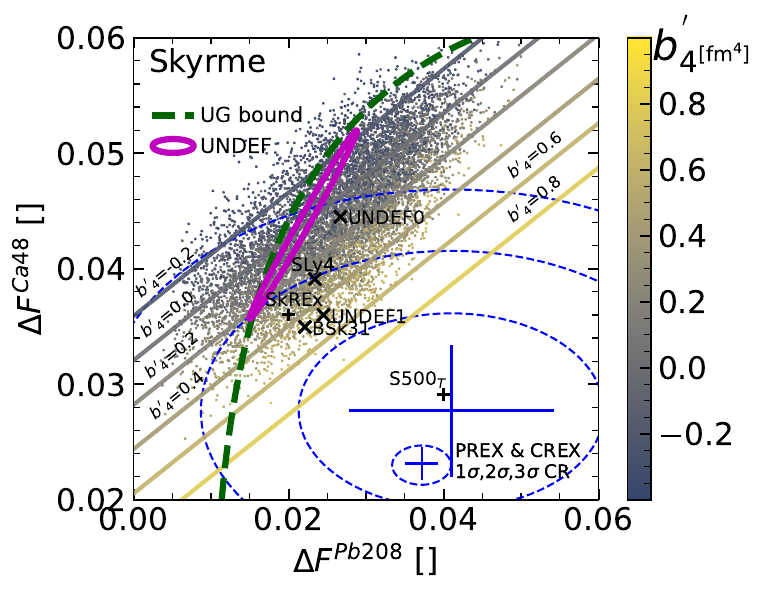}
    \caption{Charge and weak from factor difference of $^{48}$Ca and $^{208}$Pb for samples of RMF (left) and Skyrme (right) model colored by the isovector spin-orbit parameter $b'_4$ in unit fm$^4$. Colorful lines corresponds to the minimum $\chi^2$ fitting of linear function same as Eq.(\ref{eq:fit_Sv}) and Eq.(\ref{eq:fit_L}) with $b'_4=[-0.2,0,0.2,0.4,0.6,0.8]$. Constraints from unitary gas conjecture\cite{tews2017symmetry} and nuclei properties\cite{kortelainen2010nuclear} calculated from corresponding curves in Fig.\ref{fig:svLLinear} based on the correlation Eq.(\ref{eq:fit_Sv}) and Eq.(\ref{eq:fit_L}). Blue dashed line shows 1-,2- and 3-$\sigma$ confidence region of joint PREX-2/CREX results. `x' marker on the left panels are RMF models listed in Supplementary material. Reference and $b'_4$ for Skyrme models with marker `x' are discussed in the text. `+' marker represent the recently calibrated Skyrme model\cite{Zhang2022,Yue:2024srj}, the relativistic Hartree-Bogoliubov model\cite{yuksel2023implications} and RMF model with tensor interaction\cite{salinas2023impact} which take different form of interactions. For readability only part of RMF and Skyrme models are shown. A more comprehensive analysis of the $b'_4$ distribution of existing models is presented in the supplemental material.}
    \label{fig:b4posterior}
\end{figure*}
The spin-orbit interactions in Skyrme models are written as,
\begin{equation}
\begin{split}
    H_{\mathrm{SO}}=b_4 \textbf{J}\cdot{\nabla} n+b'_4(\textbf{J}_n\cdot\nabla n_n+\textbf{J}_p\cdot\nabla n_p),
\end{split}
\end{equation}
where the $b_4$ and $b'_4$ coefficients correspond to the isoscalar and isovector contributions of the spin-orbit potentials. To compare spin-orbit interactions in both relativistic and non-relativistic models in a parallel manner, we recast the Dirac equation in Schrödinger form using the Foldy-Wouthuysen transformation and the RMF spin-orbit contributions models can be approximated as:
\begin{eqnarray}
    V^{p/n}_{SO}&=&-\frac{2b_4}{r} \frac{d\rho_B}{dr}- \frac{b'_4}{r}\frac{d}{dr} \left[\rho_B\pm(\rho_p-\rho_n)\right]~,
\end{eqnarray}
with 
\begin{eqnarray}
    b_4&\approx&\frac{1}{4m^2}\left (\frac{g_\sigma^2}{m^2_\sigma}+\frac{g_\omega^2}{m^2_\omega}-\frac{g_\delta^2}{4m^2_\delta}-\frac{g_\rho^2}{4m^2_\rho}\right ),\\
    b'_4&\approx&\frac{1}{8m^2}\left (\frac{g_\delta^2}{m^2_\delta}+\frac{g_\rho^2}{m^2_\rho}\right ). \label{eq:b4'}
\end{eqnarray}
The detailed derivation of spin-orbit contributions in RMF models is provided in the supplemental material.

In Fig. \ref{fig:b4posterior}, we present the distributions of $\{\Delta F^{\mathrm{Ca}48}, \Delta F^{\mathrm{Pb}208}\}$ colored by $b'_4$  in  RMF and Skyrme models. As $b'_4$ increases, $\Delta F^{\mathrm{Pb}208}$ rises while $\Delta F^{\mathrm{Ca}48}$ drops, moving predictions closer to the PREX-2/CREX mean. Neutron skin thickness $\Delta R_{np}$ follows the same trend. The detailed proton and neutron density profiles are presented in the supplemental material. We can fit the parameter $b'_4$ as:
\begin{eqnarray}
    b'_4=a''\Delta F^{\mathrm{Ca}48}+b'' \Delta F^{\mathrm{Pb}208} +c'', \label{eq:fit_b'4}
\end{eqnarray}
with the fit parameters in Table \ref{tab:Sv_L_FF_fit}. The fit yields $c'' = 1.69 \pm 0.15$ (Skyrme) and $0.93 \pm 0.11$ (RMF), which is well within PREX-2/CREX errors. This correlation with $\Delta F$ in PREX-2/CREX gives $b'_4 = 1.37 \pm 0.49$ fm$^4$ in Skyrme model, corresponding to a 90\% lower bound of $b'_4 > 0.744$ fm$^4$. The widely used SLy4 model has $b_4 = b'_4 = 0.312$ fm$^{4}$ \cite{Chabanat:1997un}, which lies on the 2-$\sigma$ contour of the PREX-2/CREX measurement. Many successful Skyrme models \cite{reinhard1995nuclear,chabanat1997skyrme,erler2013energy,klupfel2009variations,kortelainen2012nuclear,goriely2016further} with $b_4 \neq b'_4$ have been developed, generally showing that larger $b'_4$ values correlate with better agreement with PREX-2/CREX. Most recently, an extended Skyrme model S500$_T$\cite{Yue:2024srj} with $b'_4 = 1.27$—about three times that of BSk31\cite{goriely2016further} and UNEDF1\cite{kortelainen2012nuclear}—was proposed to match the center of the PREX-2/CREX constraint, in agreement with the prediction of Eq. (\ref{eq:fit_b'4}). For RMF models, Eq.\ref{eq:fit_b'4} gives $b'_4 = 1.02 \pm 0.37$, corresponding to a 90\% lower bound of $b'_4 > 0.544$ fm$^4$ or $g_\delta^2 + 1.65 g_\rho^2 > 2433$ with Eq.\ref{eq:b4'}. Common RMF models without delta mesons (e.g., FSUGold2\cite{Chen_2014}, IUFSU\cite{fattoyev2010relativistic}) yield $b'_4 < 0.1$, outside the 2$\sigma$ region; see, e.g., recent calibrated HPUA-C~\cite{sharma2023new}. Even newer models like BSRV with weak isovector couplings remain insufficient \cite{kumar2023crex}. However, models like DINOa-c, with large isovector Yukawa couplings ($g_\delta^2 > 1100$ and $g_\rho^2 > 800$), all surpass the 90\% lower bound of $b'_4 > 0.544$ \cite{reed2024density}, and lie near the 1$\sigma$ PREX-2/CREX region as in Fig \ref{fig:b4posterior}. Exploring the parameter space with large $g_\delta$ and $g_\rho$, we identify a new parameter set, CPREX1 ($b'_4=0.652$), which resides on the 1-$\sigma$ upper (lower) edge of $\Delta F^{\mathrm{Ca}48}$ ($\Delta F^{\mathrm{Pb}208}$), and well within the 1-$\sigma$ joint contour. CPREX1 also yields reasonable nuclear properties for $^{48}\mathrm{Ca}$, $^{90}\mathrm{Zr}$, and $^{208}\mathrm{Pb}$, along with decent NS properties, featuring a maximum mass of M$_{max}=$ 2.04 M$_\odot$ and dimensionless tidal deformability of $\Lambda_{1.4}=584$ within the GW170817 constraint\cite{abbott2018gw170817}. Additionally, we introduce CPREX2 with an intermediate $b'_4=0.386$, closely aligned with Skyrme models like UNEDF1 and SV. Model parameters and other properties of CPREX are tabulated in the supplementary material. Their corresponding NS EOSs with crust calculated from the droplet model are available from the public GitHub repository \cite{Zhao_CPREX}.

\emph{Conclusion and discussion --}
In this letter, we systematically investigated the influence of $S_\mathrm{V}$, L and $b'_4$ on the $\Delta F$ of $^{48}\mathrm{Ca}$ and $^{208}\mathrm{Pb}$ based on Bayesian inference.

We identified a quasi linear relationship between the symmetry energy parameters ($S_\mathrm{V}$ and $L$) and the $\Delta F$. This relationship exhibits a weak dependency on nuclear models. Unlike $L$, which is positively related to the $\Delta F$ values of both the $^{208}$Pb and $^{48}$Ca, $S_\mathrm{V}$ impacts $^{208}$Pb and $^{48}$Ca differently. This suggests that $S_\mathrm{V}\gtrsim 40$ MeV is favored by PREX-2/CREX measurements. The large $S_\textrm{V}$ compared with empirical value (around 32 MeV) suggests the tension between the current nuclear models and joint PREX-2/CREX experiments.

Additionally, we found that the isovector parts of the spin-orbit interactions in both Skyrme and RMF models influence the neutron skin of $^{208}$Pb and $^{48}$Ca \emph{differently}. Recently, the authors of Ref. \cite{Yue:2024srj} observed similar behaviors in the extended Skyrme models. Although the correlation between $b'_4$ and $\Delta F^{\mathrm{Ca}48/\mathrm{Pb}208}$ may not be strong, as observed in \cite{Reinhard:2022inh}, $b'_4$ can be quite sensitive to a linear combination of $\Delta F^{\mathrm{Ca}48}$ and $\Delta F^{\mathrm{Pb}208}$, such as $a'' \Delta F^{\mathrm{Ca}48}+b''\Delta F^{\mathrm{Pb}208}$, and  {by tuning $b'_4$ the agreement between theoretical predictions and joint PREX-2/CREX measurements is efficiently improved.} However, We realized that in RMF models, the increase of $b'_4$ usually comes with the amplification of the (isovector) nucleon density fluctuation, altering of single particle energy spectrum as discussed in \cite{Kunjipurayil:2025xss}, and an increase of $g_\delta$ and $g_\rho$ which break the common quadratic dependence of symmetry energy on asymmetry. The Skyrme / RMF posterior dataset may help in the search of optimized models with more physical constraints in future investigations.  

Finally, among numerous RMF models that fall within the 1-$\sigma$ of the PREX-2/CREX respectively, we identify CPREX1 that maintains good agreement ($<1\%$ deviation) with the binding energy, the charge radii of $^{208}\mathrm{Pb}$, $^{90}\mathrm{Zr}$ and $^{48}\mathrm{Ca}$ as well as tidal deformability and maximum mass of NSs. Similar to DINO with large $g_\delta$ and $g_\rho$, CPREX1 deviates from the quadratic dependence of symmetry energy on isospin asymmetry, yielding $S_\mathrm{V}=32.9$ MeV around SNM, but $S_\mathrm{V}=54.3$ MeV when defined with respect to PNM.

Follow-up neutron skin experiments, like the Mainz Radius EXperiment (MREX), will continue to refine the possible uncertainty of the PREX-2 experiment \cite{MREX}. The linear relations between the isovector parameters ($S_\mathrm{V}$, $L$, and $b'_4$) and the $\Delta F$ values of $^{208}$Pb and $^{48}$Ca  may serve as a simple ``emulator'' for rapidly and reliably approximating $S_\mathrm{V}$, $L$, and $b'_4$ in the investigations of neutron skin thickness experiments.

\begin{acknowledgments}
 We acknowledge Jorge Piekarewicz and Jim Lattimer for the helpful discussions during the INT Workshop INT-22R-2A. T.Z. was supported by the Network for Neutrinos, Nuclear Astrophysics and Symmetries, through the National Science Foundation Physics Frontier Center Grant No. PHY-2020275. M.P. and T.Z acknowledges  support by the Department of Energy, Award No. DE-FG02-93ER40756. Z.L. and A.W.S. were supported by NSF PHY 21-16686. A.W.S. was also supported by the Department of Energy Office of Nuclear Physics. B.K. acknowledges partial support from the Department of Science and Technology, Government of India, with grant no. CRG/2021/000101. All the Bayesian samples and data for making plots are available from the GitHub repository\cite{Zhao_CPREX}.

TZ and ZL contributed equally to this work.
\end{acknowledgments}

\bibliographystyle{apsrev4-1}
\bibliography{aps}


\appendix
\newpage
\newpage 
\section*{Supplemental Material}

\input{EDF_resub}
\input{SO_RMF_resub}
\input{weak_form_resub}
\input{bayesian_resub}
\input{fit_SvL_FchFw_resub}
\input{density_profile}

\end{document}

%% file: EDF_resub.tex
\subsection{Relativistic mean field model and Skyrme model}
\input{tab_edf_resub}
In this section, we provide an overview of the theoretical framework utilized. For a more comprehensive understanding, interested readers can refer to the detailed discussions available in various sources such as \cite{Kumar_2017,IOPB,Yang_2020} along with their cited references. The starting point for our relativistic calculation of nuclear response is based on the covariant model presented in Ref. \cite{Chen_2014}. Additionally, we incorporate the $\delta$-meson to accommodate the CREX experiment, which prefers softer symmetry energy \cite{adhikari2022precision}. The interacting Lagrangian density can be expressed as:
\begin{widetext}
\begin{eqnarray}
 {\mathscr L}_{\rm int} &=&
\bar\psi \left[g_{\sigma}\sigma+\frac{g_{\delta}}{2}{\mbox{\boldmath $\tau$}}\cdot {\mbox{\boldmath $\delta$}}   \!-\! 
         \left(g_{\omega}\omega_\mu  \!+\!
    \frac{g_{\rho}}{2}{\mbox{\boldmath $\tau$}}\cdot{\mbox{\boldmath $\rho$}}_{\mu} 
                               \!+\!    
    \frac{e}{2}(1\!+\!\tau_{3})A_{\mu}\right)\gamma^{\mu}
         \right]\psi \nonumber \\
                   &-& 
    \frac{\kappa}{3!} (g_{\sigma}\sigma)^3 \!-\!
    \frac{\lambda}{4!}(g_{\sigma}\sigma)^4 \!+\!
    \frac{\zeta_\omega}{4!}   g_{\omega}^4(\omega_{\mu}\omega^\mu)^2 +
   \Lambda_{\omega\rho}\Big(g_{\rho}^{2}\,{\mbox{\boldmath$\rho$}}_{\mu}\cdot{\mbox{\boldmath$\rho$}}^{\mu}\Big)
                           \Big(g_{\omega}^{2}\omega_{\nu}\omega^{\nu}\Big)\;.
 \label{LDensity}
\end{eqnarray}
\end{widetext}
The Lagrangian density used in our study includes the isodoublet nucleon field $\psi$ as the basic degree of freedom. These nucleons interact via photon ($A_{\mu}$) exchange and through the exchange of four massive mesons: a scalar-isoscalar ($\sigma$), scalar-isovector ($\delta$), a vector-isoscalar ($\omega^{\mu}$), and a vector-isovector ($\rho^{\mu}$). This formulation is based on the works of \cite{Waleck_1974,Serot:1984ey,Serot_1997}. The coupling constant $\Lambda_{\omega\rho}$ for the isoscalar-isovector interaction plays a crucial role in determining the density dependence of the symmetry energy, particularly its slope at saturation density $L$ as discussed in \cite{Horowitz_2001}. It is noteworthy that the model can be calibrated using an analytic one-to-one correspondence between bulk parameters of infinite nuclear matter and the various coupling constants. Specifically, the values of the symmetry energy $S_\mathrm{V}$, its slope $L$, and curvature $K_{sym}$ at saturation density are correlated to the isovector parameters ($g_{\delta}$, $g_{\rho}$, and $\Lambda_{\omega\rho}$) as explained in \cite{Chen_2014, singh2014effects}. Table \ref{tab:special_coupling} shows the $\sigma$ meson mass and 8 coupling constants for CPREX1 and CPREX2 proposed in this letter. The masses of other massive mesons and the free nucleon are $m_\delta=980$ MeV, $m_\omega= 782.5$ MeV, $m_\rho= 763$ MeV, and $m=939$ MeV. The nuclear properties of $^{48}$Ca and $^{208}$Pb are shown in Table \ref{tab:BE_RC} along with experimental measurement and our Hartree calculations for various models for comparison. The nuclear properties of $^{48}$Ca and $^{208}$Pb are shown in Table~\ref{tab:BE_RC}, along with experimental measurements and our Hartree calculations for various models for comparison. These two nuclei, together with $^{90}$Zr (not shown), are selected for inclusion in our Bayesian likelihood to limit computational cost, as the evaluation of finite nuclei properties constitutes the main bottleneck in the sampling. While the CPREX models reproduce the binding energies and charge radii of these nuclei with high accuracy, they are not optimized for global nuclear fits. For example, the binding energy per nucleon of $^{16}$O for CPREX1 is predicted with an error of about 2\%, which, while larger than for the nuclei included in the fit, remains competitive with or better than the uncertainty range of chiral EFT (1--10\%). We emphasize that prioritizing consistency with PREX2/CREX and astrophysical constraints necessarily limits the model’s ability to fit additional nuclei, given its finite degrees of freedom. Fitting to a broader range of nuclei would risk overfitting and reduce the sensitivity to the observables we aim to constrain. Table \ref{tab:SAT_NS} shows the infinite nuclear matter properties at saturation density and the neutron star properties calculated from EOSs in $\beta$-equilibrium. The crust EOSs are constructed with the compressible liquid droplet model with fixed surface tension parameters $\sigma_s=1.2$ MeV fm$^{-2}$, $S_S=48$ MeV\cite{lattimer1991generalized}. The tabulated complete EOSs of these RMF models can be found in the GitHub repository\cite{Zhao_CPREX}.

To gain a complementary understanding of both PREX-2/CREX measurements, we also study the Skyrme interactions inferred from those neutron skin thickness measurements. The Skyrme Hamiltonian for infinite nuclear matter is written as,  
\begin{equation}
H_{\mathrm{Sk}} = \frac{k_{Fn}^5} {10 \pi^2 m_n^{*}} + \frac{k_{Fp}^5} {10 \pi^2 m_p^{*}} + H_{\mathrm{pot}}(n_n,n_p),
\end{equation}
where the first two terms are the kinetic terms for the neutrons and the protons with effective mass contribution and $H_{\mathrm{pot}}$ is the potential term  expressed as:
\begin{equation}
    \begin{split}
        H_{\mathrm{pot}}=&\frac{1}{2}n^2t_0\left[1+\frac{x_0}{2}\right) - \frac{1}{2}(n_n^2+n_p^2) t_0\left(\frac{1}{2}+x_0\right) \\
        +&\frac{1}{24}n^\gamma t_3 \left(n^2(2+x_3) - (n_n^2+n_p^2)(1+2 x_3)\right].
    \end{split}
\end{equation}
The neutron and proton effective masses are functions of densities as,
\begin{equation}
\begin{split}
    \frac{m_{n/p}}{m^*_{n/p}}=&1+\frac{m_{n/p}}{4}\Big\{n\big[t_1(2+ x_1)+t_2(2+x_2)\big]\\
    +& n_{n/p}\big[-t_1(1+2 x_1)+t_2(1+2 x_2)\big]\Big\} \, .
\end{split}
\end{equation}
To describe a finite nucleus, two additional terms are included in the Skyrme Hamiltonian \cite{Reinhard:1995zz,Chabanat:1997un}. These terms are:
\begin{equation}
\begin{split}
    H_{\mathrm{SO}}=b_4 \textbf{J}\cdot\nabla n+b'_4(\textbf{J}_n\cdot\nabla n_n+\textbf{J}_p\cdot\nabla n_p),
\end{split}
\end{equation}
and \begin{equation}
\begin{split}
    H_{\mathrm{J}}=-\frac{1}{16}(t_1x_1+t_2x_2)\textbf{J}^2+
    \frac{1}{16}(t_1-t_2)(\textbf{J}^2_n+\textbf{J}^2_p),
\end{split}
\end{equation}
where $H_{\mathrm{SO}}$ are spin-orbit interactions and $H_{\mathrm{J}}$ are the central tensor terms. The $\bf{J}_{n/p}$ is the spin-orbit density $\bf{J}_{n/p}=\psi^\dag_{n/p}\overrightarrow{\sigma}\times\overrightarrow{\triangledown}\psi_{n/p}$ and $\bf{J}=\bf{J}_n+\bf{J}_p$.

%% file: tab_edf_resub.tex
\begin{table*}[htb]
    \centering
\begin{tabular}{llccccccccccccc}
\hline\hline
&&Experiment& NL3& FSU2& IOPB-I& IUFSU& BigApple& HPUC& BSRV& DINOa& DINOb& DINOc& CPREX1& CPREX2\\
\hline
&$B/A$ [MeV]&~7.87 & 7.88 & 7.87 & 7.86 & 7.88 & 7.85 & 7.85 & 7.84 & 7.87 & 7.87 & 7.87 & 7.84 & 7.86\\
&$R_{ch}$ [fm]&~5.50 & 5.51 & 5.49 & 5.52 & 5.49 & 5.50 & 5.56 & 5.53 & 5.51 & 5.51 & 5.51 & 5.49 & 5.49\\
{\bf$^{208}$Pb~~}&$\Delta R_{np}$ [fm]&~0.159$\pm$ 0.017 & 0.2797 & 0.2862 & 0.2195 & 0.1618 & 0.1508 & 0.1196 & 0.2595 & 0.1746 & 0.1993 & 0.2235 & 0.1905 & 0.1525\\
&$F_{ch}$ []&~0.409 & 0.4067 & 0.4094 & 0.4052 & 0.4106 & 0.4080 & 0.3992 & 0.4043 & 0.4074 & 0.4075 & 0.4073 & 0.4100 & 0.4092\\
&$\Delta F$ []&~0.041$\pm$0.013 & 0.0414 & 0.0423 & 0.0319 & 0.0233 & 0.0214 & 0.0168 & 0.0378 & 0.0262 & 0.0303 & 0.0342 & 0.0282 & 0.0222\\
\hline
&$B/A$ [MeV]&~8.67 & 8.65 & 8.62 & 8.64 & 8.53 & 8.52 & 8.65 & 8.66 & 8.67 & 8.67 & 8.67 & 8.64 & 8.66\\
&$R_{ch}$ [fm]&~3.48 & 3.45 & 3.43 & 3.45 & 3.44 & 3.46 & 3.46 & 3.44 & 3.47 & 3.47 & 3.47 & 3.48 & 3.46\\
{\bf$^{48}$Ca~~}&$\Delta R_{np}$ [fm]&~0.137$\pm$0.015 & 0.2255 & 0.2318 & 0.1995 & 0.1736 & 0.1690 & 0.1479 & 0.2196 & 0.0994 & 0.1054 & 0.1141 & 0.1252 & 0.1357\\
&$F_{ch}$ []&~0.158 & 0.1604 & 0.1665 & 0.1616 & 0.1647 & 0.1582 & 0.1577 & 0.1621 & 0.1591 & 0.1589 & 0.1585 & 0.1537 & 0.1571\\
&$\Delta F$ []&~0.0277$\pm$0.0055 & 0.0551 & 0.0564 & 0.0490 & 0.0435 & 0.0413 & 0.0391 & 0.0527 & 0.0330 & 0.0345 & 0.0364 & 0.0335 & 0.0362\\
\hline\hline
\end{tabular}
    \caption{Experimental data for the binding energy per nucleon\cite{Wang:2012}, charge radii\cite{Angeli:2013}, neutron skins (excluding PREX-2/CREX)\cite{lattimer2023constraints}, charge from factor and form factor difference from PREX\cite{adhikari2021accurate} for $^{208}$Pb and CREX\cite{adhikari2022precision} for $^{48}$Ca. Also displayed are the theoretical results obtained with NL3\cite{lalazissis1997new}, FSUGold2\cite{Chen_2014}, IOPB-I\cite{IOPB}, IUFSU\cite{fattoyev2010relativistic}, BigApple\cite{fattoyev2020gw190814}, HPUC\cite{sharma2023new}, BSRV\cite{kumar2023crex}, DINOa-c\cite{reed2024density} and the two new parameterizations, CPREX1 and CPREX2.}
    \label{tab:BE_RC}
\end{table*}
\begin{table*}[htb]
    \centering
\begin{tabular}{llcccccccccccc}
\hline\hline
&& NL3& FSU2& IOPB-I& IUFSU& BigApple& HPUC& BSRV& DINOa& DINOb& DINOc& CPREX1& CPREX2\\
\hline
&n$_s$ [fm$^{-3}$] & 0.1483 & 0.1504 & 0.1495 & 0.1546 & 0.1546 & 0.1490 & 0.1480 & 0.1522 & 0.1525 & 0.1519 & 0.1516 & 0.1518\\
&M$^D$ [MeV] & 558.7 & 557.0 & 557.2 & 572.1 & 572.8 & 572.9 & 565.3 & 587.4 & 593.0 & 593.9 & 692.8 & 648.1\\
&B [MeV] & 16.24 & 16.26 & 16.10 & 16.40 & 16.34 & 15.98 & 16.10 & 16.16 & 16.21 & 16.21 & 16.29 & 16.14\\
SNM&K [MeV] & 271.6 & 237.7 & 222.6 & 231.3 & 227.0 & 220.2 & 227.2 & 210.0 & 207.0 & 206.0 & 223.8 & 223.5\\
&S$_V$ [MeV] & 37.3 & 37.6 & 33.3 & 31.3 & 31.3 & 28.4 & 36.1 & 31.4 & 33.1 & 34.6 & 32.9 & 29.8\\
&L [MeV] & 118.2 & 112.7 & 63.6 & 47.2 & 39.8 & 41.6 & 84.6 & 50.0 & 70.0 & 90.0 & -3.5 & 0.4\\
&K$_{sym}$ [MeV] & 101.0 & 25.4 & -37.0 & 28.5 & 87.5 & 81.1 & -73.2 & 506.0 & 609.1 & 714.8 & -418.4 & -239.8\\
\hline
&M$^D_n$ [MeV] & 569.2 & 566.0 & 566.7 & 580.5 & 582.8 & 581.4 & 573.3 & 352.1 & 333.0 & 320.5 & 377.4 & 465.6\\
&M$^D_p$ [MeV] & 569.2 & 566.0 & 566.7 & 580.5 & 582.8 & 581.4 & 574.8 & 908.8 & 948.2 & 969.1 & 1062.5 & 870.1\\
PNM&S$_V$ [MeV] & 38.3 & 38.6 & 34.7 & 32.9 & 33.1 & 29.9 & 37.2 & 46.5 & 50.6 & 53.4 & 54.3 & 38.4\\
&L [MeV] & 121.2 & 115.9 & 67.7 & 49.5 & 40.6 & 42.7 & 88.7 & 172.1 & 216.4 & 247.8 & 211.2 & 75.9\\
&K$_{sym}$ [MeV] & 100.3 & 27.2 & -45.5 & 23.1 & 74.3 & 89.2 & -70.6 & 726.7 & 907.2 & 1021.2 & 801.8 & 76.4\\
\hline
&M$_{max}$ [M$_\odot$] & 2.77 & 2.07 & 2.15 & 1.94 & 2.60 & 2.05 & 2.04 & 2.17 & 2.15 & 2.15 & 2.04 & 2.12\\
&R$_{1.0}$ [km] & 14.4 & 14.1 & 13.2 & 12.6 & 12.8 & 12.6 & 13.6 & 14.4 & 14.8 & 15.1 & 13.9 & 12.9\\
NS&R$_{1.4}$ [km] & 14.5 & 13.9 & 13.2 & 12.6 & 13.1 & 12.8 & 13.4 & 14.4 & 14.6 & 14.9 & 13.4 & 12.9\\
&$\Lambda_{1.0}$ [] & 7797 & 6473 & 4347 & 3384 & 3918 & 3752 & 4903 & 6623 & 7572 & 8579 & 4543 & 3544\\
&$\Lambda_{1.4}$ [] & 1275 & 876 & 687 & 500 & 719 & 593 & 689 & 1065 & 1150 & 1256 & 584 & 570\\
\hline\hline
\end{tabular}
    \caption{Saturation properties and neutron star properties of RMF models listed in Table \ref{tab:BE_RC}. Saturation properties for SNM and PNM are defined in the letter. For all models except DINOa–c and CPREX1–2, the two definitions coincide. In these cases, the discrepancy signals the breakdown of the quadratic approximation of symmetry energy. Neutron star properties are calculated with the crust EOSs constructed with the compressible liquid droptlet model repectively for various RMF models with fixed surface tension parameters $\sigma_s=1.2$ MeV fm$^{-2}$, $S_S=48$ MeV\cite{lattimer1991generalized}.}
    \label{tab:SAT_NS}
\end{table*}
\begin{table*}[bth]
\caption{$\sigma$ meson mass and 7 coupling constant of RMF models.
\label{tab:special_coupling}}
\begin{tabular}{lccccccccc}
\hline \hline
 & $m_\sigma$ (MeV)& $g_\sigma^2$& $g_\delta^2$& $g_\omega^2$& $g_\rho^2$& $\kappa$ (MeV)& $\lambda$& $\Lambda_{\omega\rho}$& $\zeta$ \\
\hline
CPREX1 & 452.852 & 60.4489 & 1348.64 & 96.8552 & 948.802 & 12.1881 & -0.0342058 & 0.00227575 & 0.00587231\\
CPREX2 & 484.272 & 77.601 & 754.613 & 120.463 & 589.842 & 7.6226 & -0.0211199 & 0.00335068 & 0.0116056\\
\hline\hline
\end{tabular}
\end{table*}

%% file: SO_RMF_resub.tex
\subsection{The approximated spin-orbit potential in RMF models}
The large component of the Dirac equation follows,
\begin{eqnarray}
    \left[\vec{\sigma}\cdot\vec{p}\frac{1}{2m+\epsilon-S-V}\vec{\sigma}\cdot\vec{p}-S+V\right]g(r)&=& \epsilon g(r)
\end{eqnarray}
where $S=g_\sigma\ \sigma_0\pm g_\delta \delta_0$, $V=g_\omega\ \omega\pm g_\rho \rho_0$, are mean-field potential from classical expectation value of meson fields. By reducing the spherically symmetric Dirac equation with scalar and vector potentials to the non-relativistic Schrödinger equation to order $(p/m)^2$\cite{horowitz1981self,reinhard1989relativistic}, the effective single nucleon classical Hamiltonian is,
\begin{eqnarray}
    H &=& \vec{p}\frac{1}{2M^*}\vec{p}-S(r)+V(r)+V_{SO}\vec{s}\cdot\vec{l}\\
    V_{SO}&=&-\frac{1}{r}\frac{d}{dr}\left[\frac{1}{M^*}\right]
\end{eqnarray}
where $\vec{l}=\vec{r}\times \vec{p}$, $\vec{s}=\vec{\sigma}/2$ and $M^*$ could be defined as the Dirac mass $M^D=m-S(r)$ assuming that the relativistic single particle energy $\epsilon+S(r)-V(r)<<2(m-S)$. However, it's possible to rearrange terms of higher order in $(p/m)^2$ to obtain an equivalent form, $M^*=m-V(r)$, as shown in the Appendix of \cite{reinhard1989relativistic}. Although these two expressions are equivalent to order $(p/m)^2$, their ratio can be of the order of 10\%, as $S(0)-V(0)\approx 50$ MeV, the empirical depth of the nucleon potential well. Therefore, we consider the average of these two choices originally introduced in \cite{koepf1991spin},
\begin{eqnarray}
    M^* &=&m-\frac{S(r)+V(r)}{2},
\end{eqnarray}
which can also be justified by taking $\epsilon<<2m-S-V$ \cite{ring1996relativistic,ebran2016spin}. Note that this definition of $M^*$ should not be confused with the Dirac or Landau effective mass. Our qualitative conclusions remain unchanged regardless of the specific definition of $M^*$ used. Since scalar and vector potentials are mainly determined by the corresponding densities of the two-form current, modified by Coulomb interaction and the self and mixing couplings of mesons, we can simplify by considering only Yukawa coupling to generate the potentials. Furthermore, vector densities are strongly correlated with scalar densities, allowing us to approximate scalar densities with vector densities:
\begin{eqnarray}
    S(r) &=& \frac{g_\sigma^2}{m^2_\sigma}n_B \pm \frac{g_\delta^2}{4m^2_\delta}(n_p-n_n)\\
    V(r) &=& \frac{g_\omega^2}{m^2_\omega}n_B \pm \frac{g_\rho^2}{4m^2_\rho}(n_p-n_n)
\end{eqnarray}
where the upper sign is for protons and the lower sign is for neutrons. The difference of isospin potential between proton and neutron is,
\begin{eqnarray}
    V^p_{SO}+V^n_{SO}&=&-\frac{d}{rdr}\left[\frac{M^*_p + M^*_n}{M^*_p M^*_n}\right]\\
    V^p_{SO}-V^n_{SO}&=&\frac{d}{rdr}\left[\frac{M^*_p - M^*_n}{M^*_p M^*_n}\right]\\
    M^*_p - M^*_n  &=& -(\frac{g_\delta^2}{m^2_\delta}+\frac{g_\rho^2}{m^2_\rho})\frac{n_p-n_n}{4}\\
     M^*_p + M^*_n &=& 2m-(\frac{g_\sigma^2}{m^2_\sigma}+\frac{g_\omega^2}{m^2_\omega})n_B
\end{eqnarray}
since $(M^*_p + M^*_n)^2>>(M^*_p - M^*_n)^2$, we can take $M^*_p M^*_n\approx (M^*_p + M^*_n)^2/4$, keeping only linear term in vector densities,
\begin{eqnarray}
    V^p_{SO}+V^n_{SO}
   &\approx& -\frac{\frac{g_\sigma^2}{m^2_\sigma}+\frac{g_\omega^2}{m^2_\omega}}{rm^2}\frac{dn_B}{dr}\label{eq:spinobit_sum_rmf}\\
    V^p_{SO}-V^n_{SO}&\approx& -\frac{\frac{g_\delta^2}{m^2_\delta}+\frac{g_\rho^2}{m^2_\rho}}{4rm^2}\frac{d(n_p-n_n)}{dr}\label{eq:spinobit_diff_rmf} .
\end{eqnarray}
A similar form without the delta meson has been used to study the isospin dependence of the spin-orbit interaction, contrasting with the spin-orbit force in non-relativistic Skyrme models \cite{sharma1995isospin,lalazissis1998reduction,pearson2001skyrme}. Isospin dependence of spin-orbit interaction is introduced to Skyrme models as:
\begin{eqnarray}
    V^{p/n}_{SO}&=&-\frac{2b_4}{r} \frac{d n_B}{dr}- \frac{b'_4}{r}\frac{d}{dr} \left[n_B\pm(n_p-n_n)\right]
\end{eqnarray}
to address the problem of isotope shifts in the Pb region\cite{reinhard1995nuclear}. Note that older models with only one spin-orbit potential parameter, $W_0=2b_4=2b'_4$ is about 100 MeV$\cdot$fm$^{5}$\cite{chabanat1997skyrme,pearson2001skyrme}.
\begin{eqnarray}
    V^p_{SO}+V^n_{SO}&=&-\frac{4b_4+2b'_4}{r}\frac{dn_B }{dr} \label{eq:spinobit_sum_skyrme}\\
    V^p_{SO}-V^n_{SO}&=&-\frac{2b'_4}{r}\frac{d}{dr} \left[n_p-n_n\right] \label{eq:spinobit_diff_skyrme}
\end{eqnarray}
By comparing Eq. (\ref{eq:spinobit_sum_skyrme}-\ref{eq:spinobit_diff_skyrme}) with Eq. (\ref{eq:spinobit_sum_rmf}-\ref{eq:spinobit_diff_rmf}), the model parameter $b_4$ and $b_4'$ can expressed as:
\begin{eqnarray}
    b_4&\approx&\frac{1}{4m^2}\left (\frac{g_\sigma^2}{m^2_\sigma}+\frac{g_\omega^2}{m^2_\omega}-\frac{g_\delta^2}{4m^2_\delta}-\frac{g_\rho^2}{4m^2_\rho}\right ),\\
    b'_4&\approx&\frac{1}{8m^2}\left (\frac{g_\delta^2}{m^2_\delta}+\frac{g_\rho^2}{m^2_\rho}\right ). \label{eq:b4'}
\end{eqnarray}
In RMF models, we obtain $b_4$ and $b_4'$ similar to those in non-relativistic models, as described in the above equations. In principle, $b_4$ and $b_4'$ are density-dependent in RMF models. We take the leading order term so that $b_4$ and $b_4'$ are essentially a linear combination of coupling Yukawa coupling strength.

%% file: weak_form_resub.tex
\subsection{Weak Charge form factors with spin-orbit contributions}
Form factors are key observable in electron scattering experiments. To calculate the weak charge form factor measured in PREX-2/CREX. We need to perform finite nuclei calculations to obtain the proton and neutron profiles then perform Fourier transform on the profile to get the corresponding form factor. In this work, we consider the impact of spin-orbit currents on the electromagnetic and weak charge form factors of $^{48}\mathrm{Ca}$ and $^{208}\mathrm{Pb}$. It has been shown the contributions from spin-orbit currents are comparable to the present CREX experimental error bars\cite{horowitz2012impact}, and therefore should be necessary to incorporate spin-orbit currents in the calculation of weak form factors as:
\begin{equation*}
    \begin{split}
        Z\mathrm{F}_\mathrm{ch}(q)&=\sum_{i=p,n}\{\mathrm{G}_\mathrm{E}^i(q^2)\mathrm{F}_\mathrm{V}^i(q)+\frac{\mathrm{G}_\mathrm{M}^i(q^2)-\mathrm{G}_\mathrm{E}^i(q^2)}{1+\tau}\\&\times [\tau\mathrm{F}_\mathrm{V}^i(q)+\frac{q}{2\mathrm{m}}\mathrm{F}_\mathrm{T}^i(q)]\}~,
    \end{split}
\end{equation*}
and 
\begin{equation*}
    \begin{split}
        Q_\mathrm{W}\mathrm{F}_\mathrm{W}(q)&=\sum_{i=p,n}\{\tilde{\mathrm{G}}_\mathrm{E}^i(q^2)\mathrm{F}_\mathrm{V}^i(q)+\frac{\tilde{\mathrm{G}}_\mathrm{M}^i(q^2)-\tilde{\mathrm{G}}_\mathrm{E}^i(q^2)}{1+\tau}\\&\times [\tau\mathrm{F}_\mathrm{V}^i(q)+\frac{q}{2\mathrm{m}}\mathrm{F}_\mathrm{T}^i(q)]\}~,
    \end{split}
\end{equation*}
where $\tau=q^2/4m^2$, the $\mathrm{G}^i_\mathrm{E,M}$ are electric and magnetic single nucleon form factors and the $\tilde{\mathrm{G}}^i_\mathrm{E,M}$ are electric and magnetic form factors for the weak-neutral current. Similar to \cite{horowitz2012impact}, we adopted a simple dipole parametrization of these single-nucleon form factors.  

The vector and tensor form factors $\mathrm{F}_\mathrm{v}(q)$ and $\mathrm{F}_\mathrm{T}(q)$ are obtained by:
\begin{equation*}
    \mathrm{F}_\mathrm{V}(q)=\sum_{n\kappa}(2j+1)\int_0^\infty[g^2_{n\kappa}(r)+f^2_{n\kappa}(r)]j_0(qr)dr ~,
\end{equation*}
and
\begin{equation*}
    \mathrm{F}_\mathrm{T}(q)=\sum_{n\kappa}2(2j+1)\int_0^\infty g_{n\kappa}(r) f_{n\kappa}(r) j_1(qr)dr ~,
\end{equation*}
where $g_{n\kappa}(r)$ and $f_{n\kappa}(r)$ are the upper and lower components of Dirac spinor in RMF models. Wave functions in the non-relativistic Skyrme model correspond to the larger upper component $g_{n\kappa}(r)$. The lower component $f_{n\kappa}(r)$ is obtained by assuming a free space relation \cite{horowitz2012impact,adhikari2022precision}.
\begin{equation*}
    f_{n\kappa}(r)=\frac{1}{2\mathrm{m}}(\frac{d}{dr}+\frac{\kappa}{r})g_{n\kappa}(r).
\end{equation*}

%% file: bayesian_resub.tex
\input{tab_prior_resub}
\subsection{Prior}
In this and the following two sections, we describe the prior, likelihood, and posterior in our Bayesian analysis. Our RMF model has 12 parameters. We fixed the $\delta$, $\omega$, and $\rho$ meson masses at $m_\delta=980$ MeV, $m_\omega=782.5$ MeV, and $m_\rho=763$ MeV, as in many FSU-type models \cite{Chen_2014}, leaving us with 9 free parameters. The $\sigma$ meson mass is a free parameter with a uniform distribution in the range $[450,550]$ MeV, corresponding to mass of $f_0(500)$\cite{tanabashi2018review}. We also take a uniform prior for the symmetric nuclear matter (SNM) properties $n_s$, 
$B$, $M^D$, and $K$, as shown in Table \ref{tab:prior}. These 4 bulk nuclear properties can be mapped to 4 isoscalar coupling constants, as in \cite{Chen_2014}, except for the $\omega$ meson self-coupling $\zeta_{\omega}$, which takes a flat distribution in the range $[0,0.03]$ , since $\zeta_{\omega}>0.03$ is suppressed by astronomical observations \cite{huang2024constraining}, particularly due to the maximum-mass constraint. The isovector sector is governed by $g_\delta$, $g_\rho$, and $\Lambda_{\omega\rho}$, which can be mapped to the symmetry energy properties $S_\mathrm{V}$ and $L$, assuming a flat distribution of $g_\delta$ in the range $[0,1500]$, covering higtest value in DINO models\cite{reed2024density}. $S_\mathrm{V}$ takes a uniform prior, while $L$ is uniformly distributed in $[L^-,L^+]$. Because $L^-$ and $L^+$ are not fixed but determined by other parameters, the flat distributions of these parameters result in a non-flat prior distribution of $L$:
\begin{eqnarray}
P(L)=\int^{\zeta_\omega^{+}}_{\zeta_\omega^{-}}\dots \int^{m_\sigma^{+}}_{m_\sigma^{-}} \int_{L^{-}(m_\sigma,\dots,\zeta_\omega)}^{L^{+}(m_\sigma,\dots,\zeta_\omega)}\delta(L'-L)dL' d m_\sigma \dots d\zeta_\omega
\end{eqnarray}
$L^-$ is a physical bound beyond which the quadratic formula for mapping $L$ to $\Lambda_{\omega\rho}$ has no solution \cite{Chen_2014}. $L^+$ can be solved by setting $\Lambda_{\omega\rho}=0$, so that $L<L^+$ corresponds to $\Lambda_{\omega\rho}>0$, ensuring that $\beta$-equilibrium matter remains neutron-rich ($n_p<n_n$) at high density. Positive $\Lambda_{\omega\rho}$ is introduced to reduce the symmetry energy, similar to the positive $\zeta_{\omega}$, which serves to reduce the SNM energy. Thus, we take $\Lambda_{\omega\rho}=0$ as our natural upper bound for the symmetry energy slope in the RMF model. 

The prior distribution of Skyrme parameters is chosen similarly to ensure consistency with the prior distribution used for the RMF models.
We use flat prior distributions for $n_s$, $B$, 
$m^*$, $K$, and $S_\mathrm{V}$, as well as for the $x_i$, $b_4$, and $b'_4$ parameters with bounds in Table \ref{tab:prior}.
The range of the $x_i$, $b_4$, and $b_4^{\prime}$ parameters in the Skyrme prior distributions are determined as follows. First, we examined 11 widely used Skyrme models (SGII\cite{SGII}, NRAPR\cite{NRAPR}, UNEDF0\cite{UNEDF0}, UNEDF2\cite{UNEDFII}, SLy4\cite{SLy4}, SV-min\cite{SV-min}, SkO\cite{Sko}, SkOp\cite{Skop}, BSk16\cite{Bsk16}, KDE0v\cite{Kde0}, and Gs\cite{Gs}).
We then identified the maximum and minimum $x_i$, $b_4$, and $b'_4$ parameters in these 11 models, denoted as $x_i^{\mathrm{max}}$, $x_i^{\mathrm{min}}$, $b_4^{\mathrm{max}}$, $b_4^{\mathrm{min}}$, $b_4^{\prime \mathrm{max}}$, and $b_4^{\prime \mathrm{min}}$.
To cover a wide range of possible Skyrme parameters, we determine the prior distribution of $x_i$, $b_4$, and $b'_4$ parameters as $[x_i^{\mathrm{min}} - \Delta x_i, x_i^{\mathrm{max}} + \Delta x_i]$, $[b_4^{\mathrm{min}} - \Delta b_4, b_4^{\mathrm{max}} + \Delta b_4]$, and $[b_4^{\prime \mathrm{min}} - \Delta b_4^{\prime}, b_4^{\prime \mathrm{max}} + \Delta b_4^{\prime}]$, where $\Delta x_i = x_i^{\mathrm{max}} - x_i^{\mathrm{min}}$, $\Delta b_4 = b_4^{\mathrm{max}} - b_4^{\mathrm{min}}$, and $\Delta b_4^{\prime} = b_4^{\prime \mathrm{max}} - b_4^{\prime \mathrm{min}}$.

Additionally, we constrain our RMF and Skyrme models to ensure that the pressure of matter in $\beta$-equilibrium is always increasing with increasing energy density and to ensure that the proton fraction in $\beta$-equilibrium never vanishes. Note that this doesn't guarantee that symmetry energy slope $L$ defined around SNM has to be positive.

\input{tab_likelihood_resub}
\subsection{Likelihood}
\begin{figure}
    \centering
    \includegraphics[width=\linewidth]{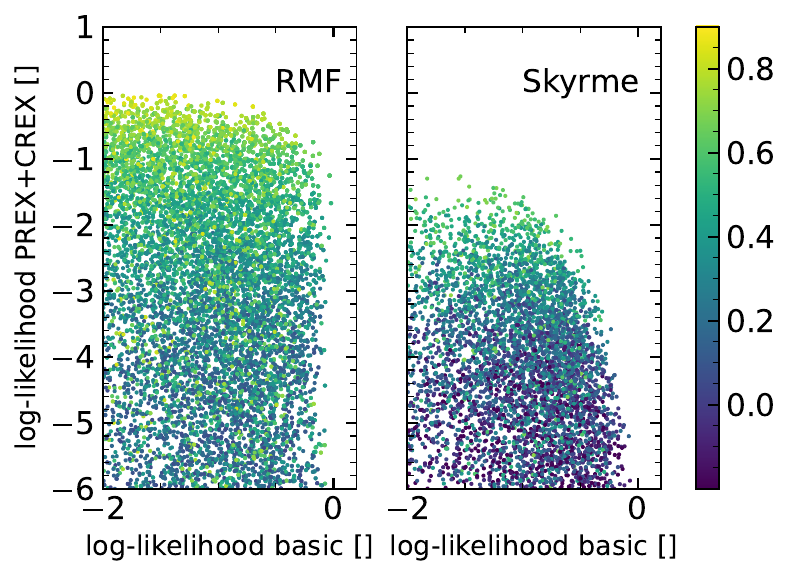}
    \caption{The likelihood posterior samples of Skyrme and RMF models colored by the parameter $b'_4$ from -0.36 (deep blue) to 0.9 (yellow) $\mathrm{fm^4}$.
    The x-axis shows the log-likelihood from basic nuclei constraints including binding energy, charge radii, and charge form factor, while the y-axis shows the log-likelihood of PREX-2/CREX experiments in Table \ref{tab:likelihood}.}
    \label{fig:likelihood}
\end{figure}
In the Bayesian analysis of RMF and Skyrme models, we use the same constraints and uncertainties. To account for theoretical uncertainty, we incorporate a 2\% uncertainty in the measured value of $R_{ch}$ and a 5\% uncertainty in the standard deviation of $B/A$. It is also necessary to constrain $F_{ch}$ accurately to ensure that the electric weak form factor difference, $F_{ch}-F_W$, can be interpreted as a measurement in PREX-2/CREX. The constraints on $B/A$, $R_{ch}$, and $F_{ch}$ together form the basic nuclear constraints in our Bayesian analysis, as listed in Table \ref{tab:likelihood}.

Sampling with basic nuclear constraints results in drastically different posteriors for $L$. In the Skyrme model, negative $L$ is suppressed by the constraint that the pressure of matter in $\beta$-equilibrium is always increasing with energy density. However, RMF models with negative $L$ around SNM can have a large symmetry energy slope around PNM due to the breakdown of the quadratic formula of symmetry energy with large $g_\delta$ and $g_\rho$. Since our goal is not to constrain the nuclear model with basic nuclear constraints, we add a likelihood factor to our RMF models to ensure the $S_\mathrm{V}$-$L$ distribution is compatible between Skyrme and RMF models with only basic nuclei constraints, as shown by the 
horizontally shaded ellipse in Fig. 1 in the letter.

In addition to the basic nuclear constraints, we impose additional PREX-2/CREX constraints separately and jointly, forming four sets of likelihoods: basic nuclei, +PREX, +CREX, and +PREX+CREX. We take the electric weak form factor difference $F_{ch}-F_W$ instead of the weak form factor $F_W$ as constraints of PREX-2/CREX experiments, as tabulated in Table \ref{tab:likelihood}. In principle, the weak form factor $F_W$ can be determined from the parity-violating asymmetry $A_{PV}$ measured in PREX-2/CREX experiments through Coulomb distortion calculations, assuming an electric charge distribution. Since the electric charge distribution has already been determined experimentally with great accuracy, both $F_W$ and $F_{ch}-F_W$ are determined experimentally from the measured $A_{PV}$ without any model dependence. However, theoretical models like RMF and Skyrme models, optimized with charge radius, do not have enough accuracy to ensure consistency with the charge distribution assumed in Coulomb distortion calculations. Therefore, we need to make sure that the value of $F_{ch}$ used in our calculations is close to that used in the analysis of PREX-2/CREX. To achieve this, we take $\sigma_{F_{ch}}=0.005$, which is much smaller than $\sigma_{F_W}$. Due to the remaining uncertainty in $F_{ch}$, we consider $F_{ch}-F_W$ in the likelihood of its better constraints compared to $F_W$.

Figure \ref{fig:likelihood} shows the likelihood of basic nuclei constraints vs the likelihood of PREX-2/CREX experiments listed in Table. \ref{tab:likelihood}. The lack of points in the upper right corner indicates the tension between basic nuclei properties and the PREX-2/CREX results. The fact that the points close to the upper right corner all have relatively large $\mathrm{b'_4}$ indicates that larger $\mathrm{b'_4}$ in mean field models may mitigate the tension between basic nuclei constraints and the PREX-2/CREX experiments.

Finally, we want to comment on additional constraints from giant monopole resonances (GMR) and dipole polarizability, which we do not consider explicitly in the likelihood. Instead, we use a tight prior on the compressibility $K$ within the empirical window $[210,250]$ MeV to account for GMR. Although we don't include dipole polarizability explicitly, it is possible to consider dipole polarizability as an additional independent constraint on $S_\mathrm{V}$ \cite{Piekarewicz2012}:
\begin{eqnarray}
S_\mathrm{V}=\frac{L+\left[146\pm1 \right]\rm{~MeV}}{\left[6.11\pm0.316\right]} 
\end{eqnarray}
This is shown in Fig. 1 and applied in Fig. 2 in the letter.
\begin{figure*}
    \centering
    \begin{overpic}[width=\linewidth]{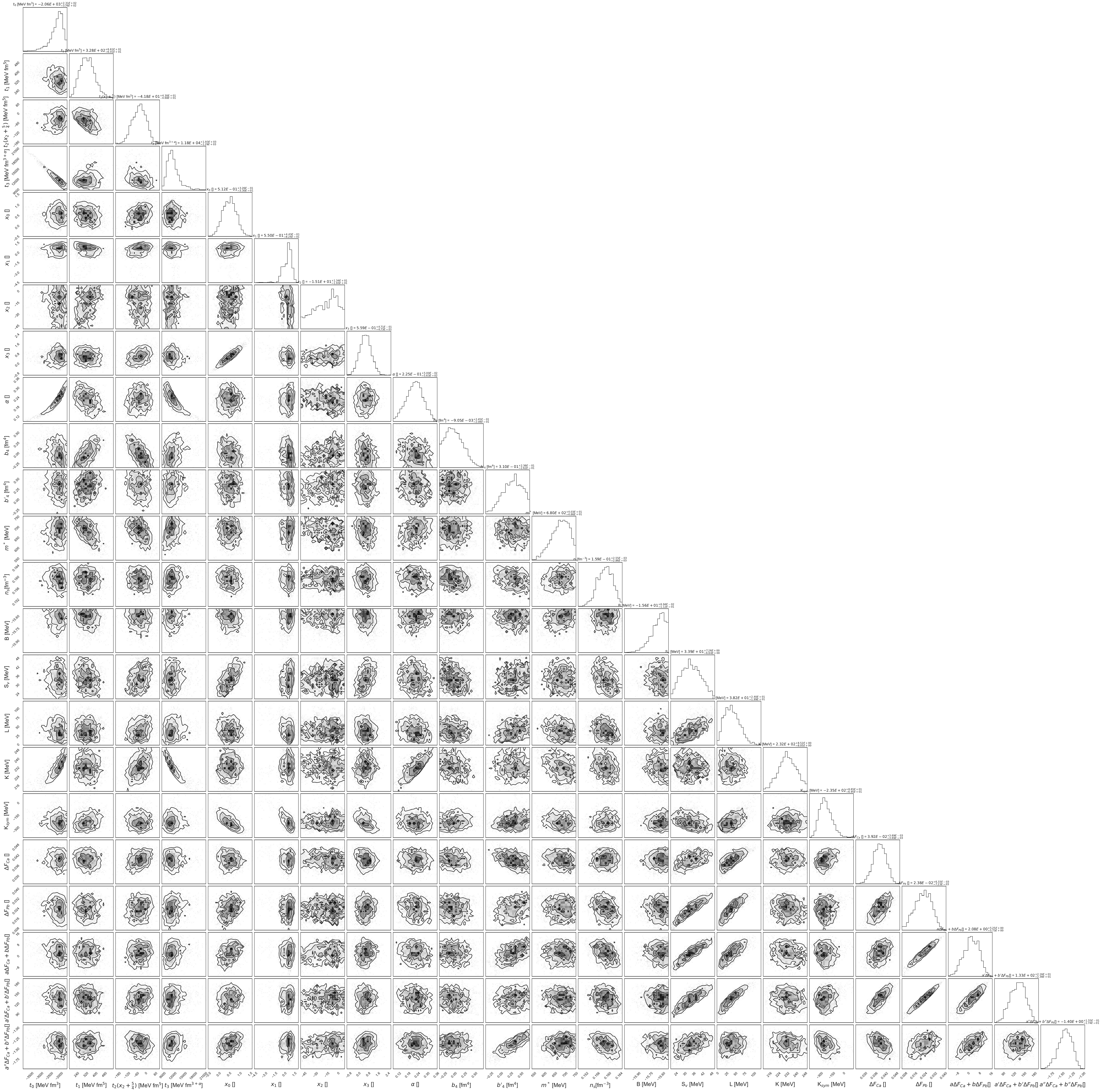}
    \put(0.53\linewidth,0.53\linewidth){\includegraphics[width=0.45\linewidth]{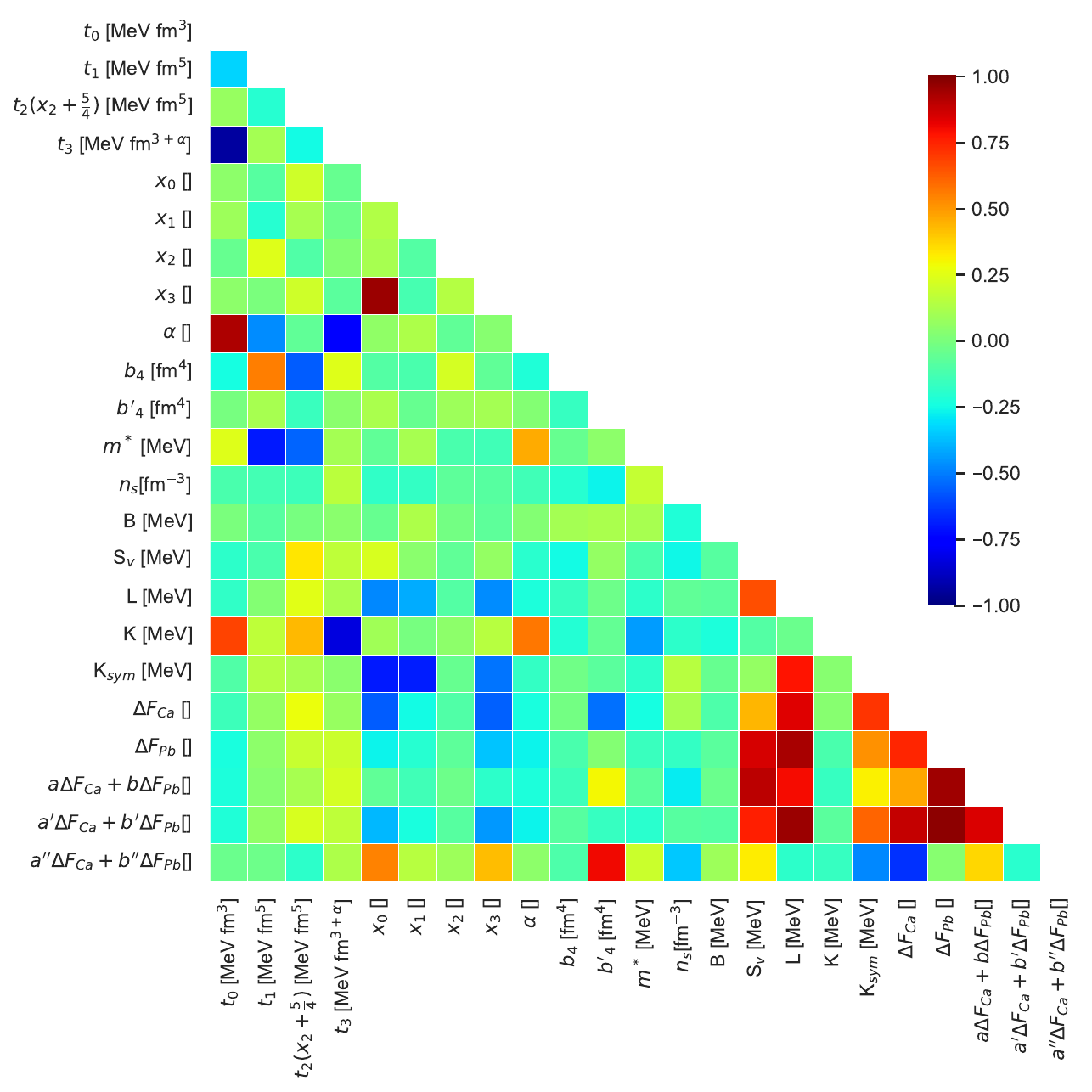}}
    \end{overpic}
    \caption{The posterior distributions of model parameters, saturation properties and $\Delta F$ measured in PREX-2/CREX for Skyrme model. The Pearson correlation coefficient between each pair of variables is displayed in the upper right corner.}
    \label{fig:corner_posterior1}
\end{figure*}
\begin{figure*}
    \centering
    \begin{overpic}[width=\linewidth]{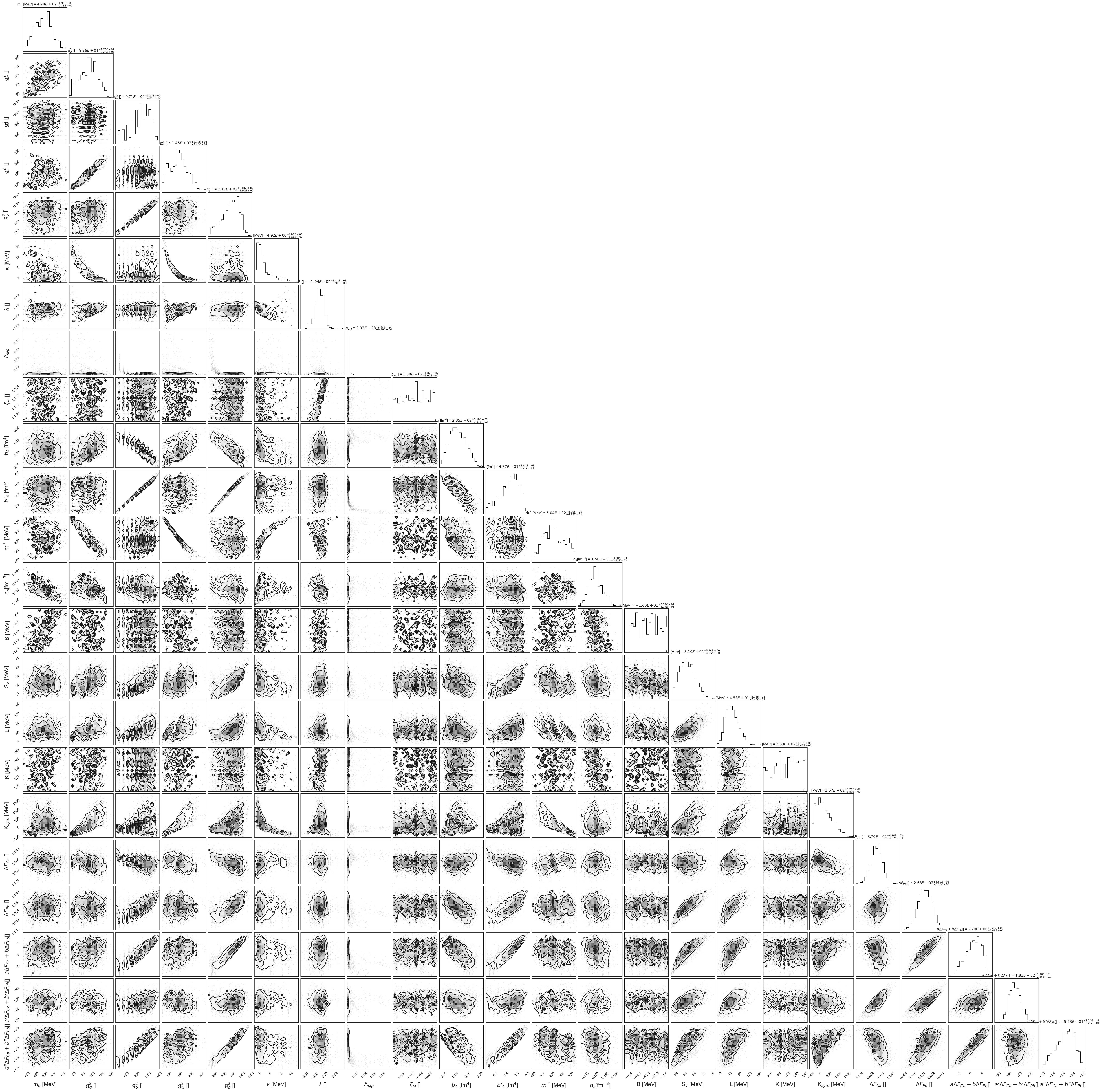}
    \put(0.53\linewidth,0.53\linewidth){\includegraphics[width=0.45\linewidth]{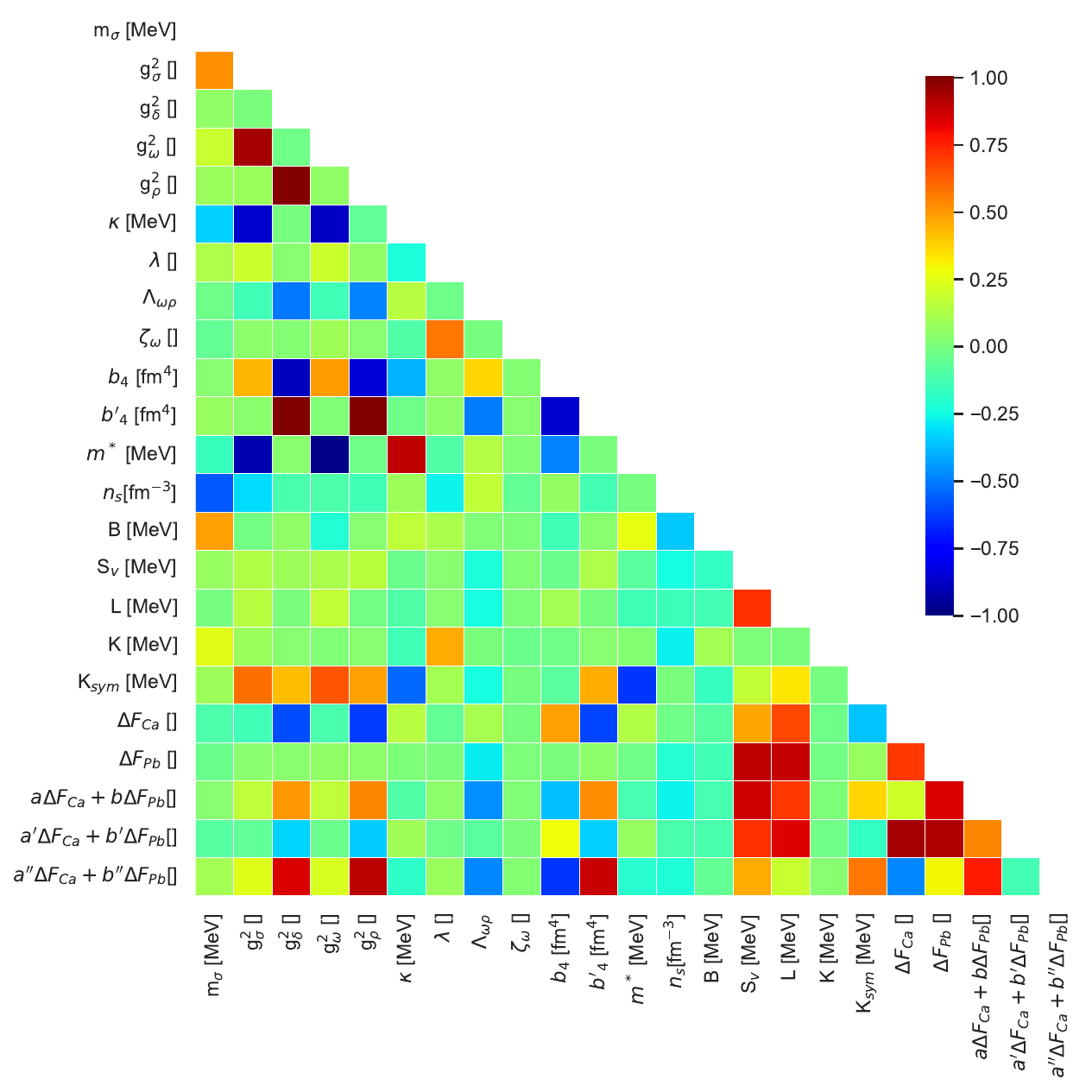}}
    \end{overpic}
    \caption{The same as Fig.\ref{fig:corner_posterior1} but for RMF model.}
    \label{fig:corner_posterior2}
\end{figure*}
\begin{figure}
    \centering
    \includegraphics[width=\linewidth]{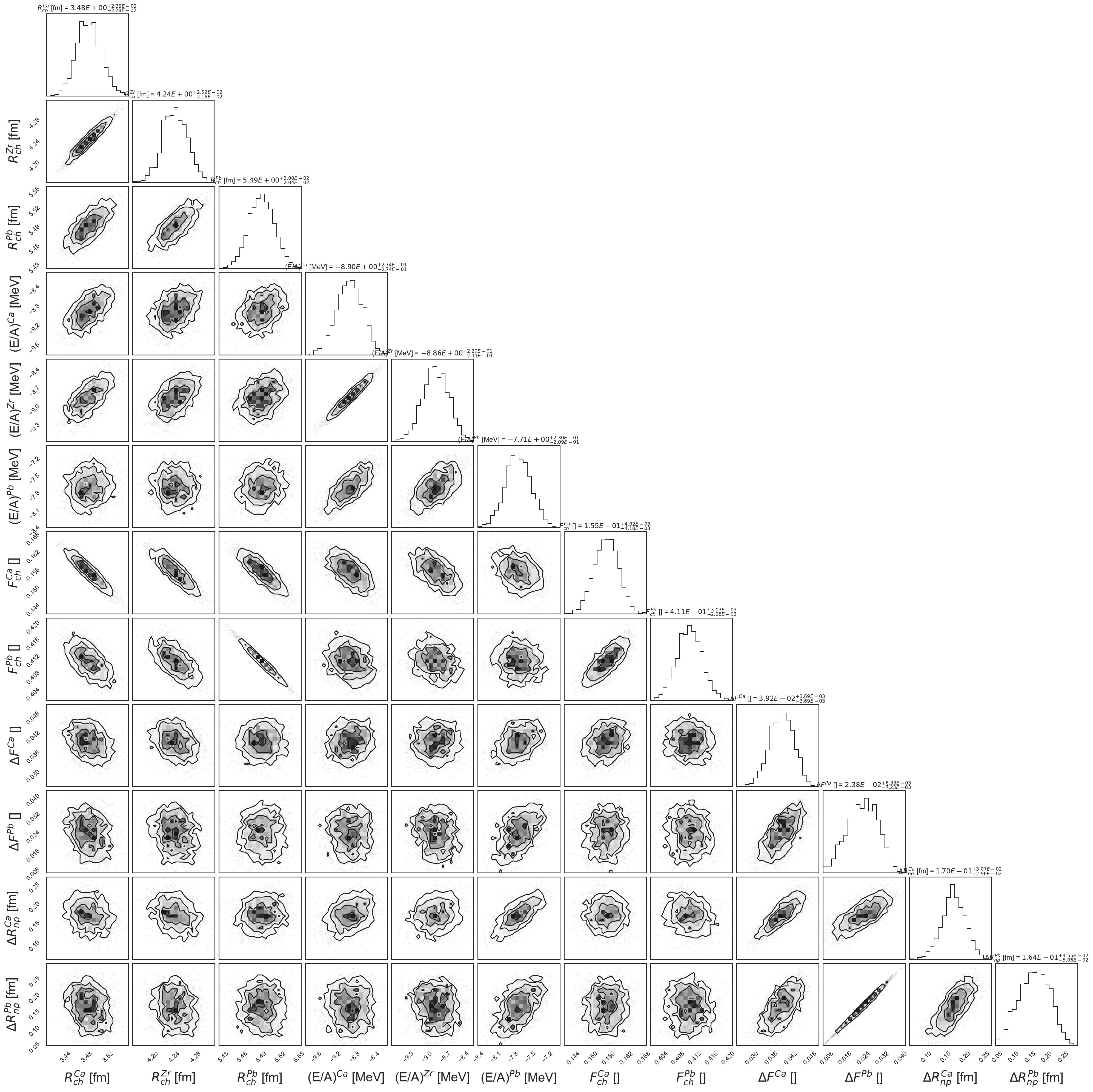}
    \includegraphics[width=\linewidth]{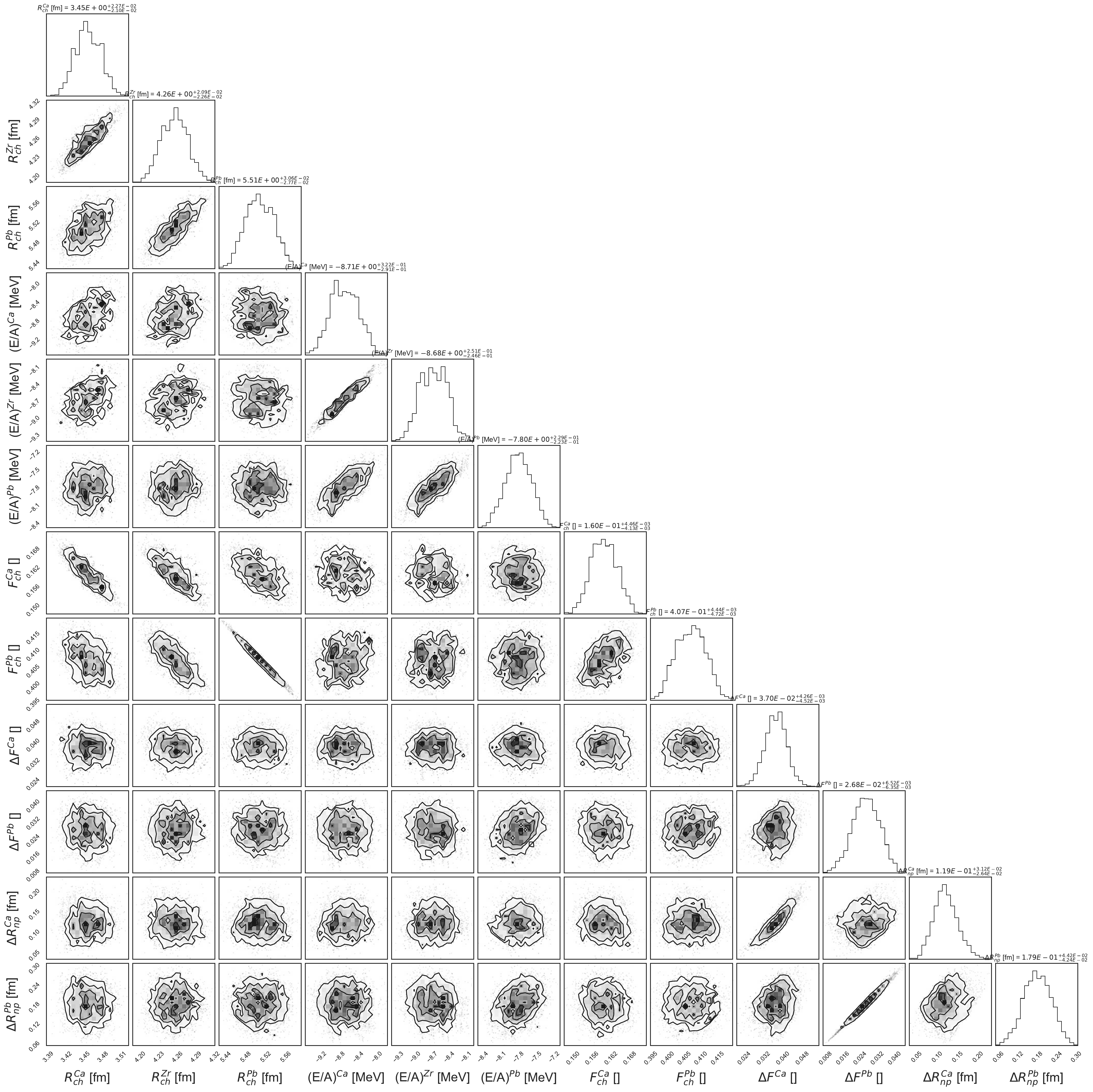}
    \caption{The posterior distributions of basic nuclear properties used in the likelihood as well as neutron skins of $^{48}$Ca and $^{208}$Pb for Skyrme model (upper) and RMF (lower) model.}
    \label{fig:corner_posterior_finite}
\end{figure}
\subsection{Posterior}
With the prior and likelihood discussed above, we obtain the posterior distributions of model parameters, saturation properties, and $\Delta F$ measured in PREX-2/CREX, as shown in Fig. \ref{fig:corner_posterior1} and Fig. \ref{fig:corner_posterior2}. To highlight the significant aspects of the posterior distribution, we also plot the Pearson correlation coefficient between any two parameters, as shown in the upper right corner of in Fig. \ref{fig:corner_posterior1} and Fig. \ref{fig:corner_posterior2}. Some model parameters, such as $t_0$, $t_3$, and $\alpha=\gamma-2$, are tightly correlated due to the mapping between saturation properties used in our prior. Isoscalar and isovector Yukawa couplings are also tightly correlated due to the priors on binding energy $B$ and effective mass $M^*$. For both Skyrme and RMF models, the symmetry energy $S_\mathrm{V}$ and its slope $L$ are very sensitive parameters related to the form factor difference measured in PREX-2/CREX. From the posterior of Bayesian analysis, we obtain the mean and covariance of $S_\mathrm{V}$ and $L$ with +PREX+CREX constraints,
\begin{eqnarray}
   (\bar S_\mathrm{V}, \bar L)&=& (34.1,~~40.3)_{Skyrme},~~~(31.4,~~49.1)_{RMF} \label{eq:SvL_posterior_cov} \\
   \sqrt{\mathrm{cov}} &=&
   \begin{pmatrix}
   6.5& 9.1\\
   9.1& 21.3
   \end{pmatrix}_{Skyrme},~~
   \begin{pmatrix}
   5.2& 9.6\\
   9.6& 27.4
   \end{pmatrix}_{RMF} \nonumber
\end{eqnarray}
compared with the posterior with basic nuclei constraints only,
\begin{eqnarray}
   (\bar S_\mathrm{V}, \bar L)&=& (32.8,~~66.8),~~~
   \sqrt{\mathrm{cov}} =
   \begin{pmatrix}
   5.9& 11.7\\
   11.7& 31.2
   \end{pmatrix} \label{eq:SvL_posterior_basic_cov}
\end{eqnarray}.

Interestingly, despite the weak correlation between $b'_4$ and $\Delta F^{\mathrm{Ca}48/\mathrm{Pb}208}$ in the Skyrme model, as observed in \cite{Reinhard:2022inh}, $b'_4$ can be quite sensitive to a linear combination of $\Delta F^{\mathrm{Ca}48}$ and $\Delta F^{\mathrm{Pb}208}$, such as $a'' \Delta F^{\mathrm{Ca}48}+b''\Delta F^{\mathrm{Pb}208}$, significantly influencing the distribution of $\Delta F^{\mathrm{Ca}48}, \Delta F^{\mathrm{Pb}208}$. The posterior for $b'_4$ increases to $b'_4=0.310^{+0.24}_{-0.26}$ with additional PREX-2/CREX constraints from $b'_4=0.244^{+0.21}_{-0.19}$ with basic nuclei constraints only. A similar trend is also observed in RMF models. The detailed impact of $b'_4$ on the nucleon distribution is discussed in a separate section below.

Figure \ref{fig:corner_posterior_finite} shows the finite nuclei properties of $^{48}$Ca, $^{90}$Zr, and $^{208}$Pb of the Bayesian posterior samples with +PREX+CREX constraints. This distribution is consistent with the likelihood imposed in Table \ref{tab:likelihood}, with the exception that the charge posterior is more constrained than the 2\% charge radius constraints we impose. This tighter constraint on charge radius is indeed due to the charge form factor being more constraining. We also noticed a much weaker $\Delta R_{np}$-$\Delta F$ correlation in $^{48}$Ca than in $^{208}$Pb, since the momentum transfer of electron elastic scattering is much higher in CREX than in PREX due to the experimental setup. Therefore, CREX should be viewed more as a measurement of form factor difference than the neutron skin of $^{48}$Ca.

We don't expect isoscalar parameters such as binding energy $B$ incompressibility $K$, self coupling of $\omega$ meson and $\zeta$ being sensitive to PREX and CREX which probes isovector properties. Therefore, their posterior is relatively flat as in the prior. $\Lambda_{\omega\rho}$ represents meson mixing between the two vector mesons. Therefore, it's extremely correlated with the Yukawa coupling of the two vector mesons. The meson mixing is small when Yukawa couplings of the two vector mesons are large.

\subsection{Confidence level}
The aim of this section is to quantify the level of consistency between theoretical predictions of $\Delta F$ for $^{48}$Ca and $^{208}$Pb, derived from Skyrme and RMF models, and the experimental measurements reported by PREX-2/CREX. This is accomplished using the well-established statistical framework of confidence levels, which quantify the degree to which new measurements are compatible with existing knowledge.

Let us assume that the state of knowledge prior to the CREX measurement suggests $\Delta F^{^{48}\mathrm{Ca}} = 0.0432 \pm 0.01$. Under this assumption, we expect that approximately 68\% of subsequent measurements will fall within the 1-$\sigma$ confidence interval: $\Delta F^{^{48}\mathrm{Ca}} \in [0.0332, 0.0532]$, assuming the knowledge is correct. Suppose an experiment yields $\Delta F^{^{48}\mathrm{Ca}} = 0.0332$, which lies exactly at the lower bound of the 1-$\sigma$ interval. In that case, the probability that the current theoretical knowledge is correct, given the measurement, is 68\%, while there is a 32\% probability that the result is a statistical fluctuation inconsistent with the model—this defines the 1-$\sigma$ confidence level. Now, consider CREX measurement with error bar yields $\Delta F^{^{48}\mathrm{Ca}} = 0.0277 \pm 0.0055$. This result lies outside the original 1-$\sigma$ bound, and its 1-$\sigma$ upper bound barely overlaps with the 1-$\sigma$ lower bound of the theoretical expectation. It is a common misconception that such a scenario is equivalent to the prior one. In deed, the right approach involves computing the difference between the experimental result and the theoretical prediction as a probability distribution: $\mathcal{N}(\mu=0.0432-0.0277,\sigma=\sqrt{0.01^2+0.0055^2})$. The null hypothesis (i.e., that the discrepancy is due to statistical fluctuation) corresponds to zero deviation. In this case, zero lies approximately 1.36 standard deviations away from the mean of the difference distribution, implying a p-value (significance) of 17.6\%. In other words, there is an 82.4\% confidence level that the measurement is inconsistent with prior knowledge, once uncertainties are fully accounted for.

The previous example is straightforward because both knowledge and measurement are modeled as Gaussian distributions, and only a single observable is involved. For more general cases—arbitrary distributions with n-dimensional observables $\vec{x}$, we need to perform an overlapping integral to determine how confident level of our knowledge $P_1(\vec{x})$ is correct given the measurement $P_2(\vec{x})$. Bayesian confidence level can be defined as,
\begin{eqnarray}
    P(\Delta\vec{ x})&=&\int P_{2}(\vec{x}) P_{1}(\vec{x}+\Delta\vec{ x}) d^n\vec{x}\\
    1-\alpha&=&\frac{\int_{P(\Delta\vec{ x})>P(0)} P(\Delta\vec{ x})d^n \Delta \vec{x}}{\int P(\Delta\vec{ x})d^n \Delta \vec{x}}
\end{eqnarray}
where $1 - \alpha$ quantifies the degree of incompatibility between knowledge and experiment. In the special case where $P_2$ is a delta function centered at $\vec{x}_0$, the overlap reduces to:
\begin{eqnarray}
P(\Delta\vec{ x})&=&\int 
\delta(\vec{x}-\vec{x_0}) P_{2}(\vec{x}+\Delta\vec{ x}) d^n\vec{x}=P_{2}(\vec{x_0}+\Delta\vec{ x}) \nonumber\\
\end{eqnarray}
then $\alpha(\vec{x}_0)$ reduce to the significance level defined as the tail probability of highest posterior density region of $P_2(\vec{x})<P_2(\vec{x}_0)$ as disscuss in the previous example with Gaussian distribution. Note that this analysis applies to probability distributions $P_1$ and $P_2$ that are localized and exhibit vanishing tails. The numerical implementation can be found in the Github\cite{Zhao_CPREX}.

Table \ref{tab:overlap} presents the resulting confidence levels $1 - \alpha$ quantifying the agreement between theoretical predictions of $\Delta F$ from Skyrme and RMF models and the PREX-2/CREX measurements. The theoretical predictions are taken from the posterior distributions discussed in the previous section. 

Overall, we observe greater tension between Skyrme predictions and experimental data compared to RMF predictions. This discrepancy primarily arises from the inclusion of the scalar-isovector $\delta$ meson in RMF models, with its large Yukawa coupling ($g_\delta^2 \in [0, 1500]$) enhancing the isovector spin-orbit interaction. This leads to an effective spin-orbit parameter $b_4' \sim 1$ fm$^4$, whereas the prior used in Skyrme modeling, $b_4' \in [-0.36, 0.72]$ fm$^4$, though broader than typical ranges, remains insufficient to reproduce the central values of PREX-2/CREX.

Furthermore, the overlaps between knowledge and experiment are notably higher when only PREX is included, and lower when CREX or both measurements are considered. This indicates that the PREX-2 data is broadly consistent with current theoretical priors and imposes minimal update on the posterior, while the CREX measurement is in strong tension with prior expectations.

\begin{table}[bth]
\begin{ruledtabular}
\caption{Confidence levels ($1 - \alpha$) quantifying the compatibility between posteriors of $\Delta F$ for $^{48}$Ca and $^{208}$Pb based on Skyrme and RMF models and the PREX-2/CREX measurements. Higher values correspond to increased statistical tension between model and experiment.}
\label{tab:overlap}
\begin{tabular}{lcc}
Posterior   &  RMF &Skyrme\\
\hline
basic nuclei& 97.0\% & 99.7\%\\
+PREX       & 97.4\% & 99.3\%\\
+CREX       & 75.8\% & 97.6\%\\
+PREX+CREX  & 69.7\% & 93.5\%\\
\end{tabular}
\end{ruledtabular}
\end{table}

%% file: tab_prior_resub.tex
\begin{table}[bth]
\caption{Prior distribution of Skyrme and RMF parameters. $n_s$, $B$, $m^*$ (or $M^D$), $K$ and $S_V$ map to $t_0$, $t_1$, $t_2$, $t_3$, $\gamma$ for Skyrme models and $g_\sigma$, $g_\omega$, $g_\rho$, $\kappa$, $\lambda$ for RMF models.
\label{tab:prior}}
\begin{ruledtabular}
\begin{tabular}{lll}
 & parameter & prior \\
\hline
&$n_s$ [MeV]  & [0.14,0.165]\cite{drischler2024bayesian}  \\
&$B$ [MeV]  & [-15.5,-16.5]\cite{drischler2024bayesian}     \\
Both&$m^*$, $M^D$ [MeV] & [0.5,0.8]$\times$ 939 \cite{li2018nucleon}  \\
&$K$ [MeV] & [210,250]  \cite{xu2021bayesian}\\
&$S_V$ ([MeV] & [20,50] \cite{sun2024compiled}\\
\hline
&$m_\sigma$ [MeV] &  [450,550] \\
&$m_\delta$ [MeV] &  980 \\
&$m_\omega$ [MeV] &  782.5 \\
RMF&$m_\rho$ [MeV] &  763 \\
&$L$ [MeV] & [$L^{-}$,$L^+$] \\
&$g_\delta^2 []$ & [0,1500] \\
&$\zeta_\omega$ []& [0,0.03]  \\
\hline
&$x_0$ [] & [-1.81,2.15] \\
&$x_1$ [] & [-7.53, 3.77] \\
Skyrme&$x_2$ [] & [-49.90, 91.93] \\
&$x_3$ [] & [-3.41, 3.73] \\
&$b_4$ [$\mathrm{fm}^4$] &[-0.36, 0.72] \\
&$b'_4$ [$\mathrm{fm}^4$] &[-0.36, 0.72] \\
\end{tabular}
\end{ruledtabular}
\end{table}

%% file: tab_likelihood_resub.tex
\begin{table}[bth]
\begin{ruledtabular}
\caption{The list of experiment and observation with adopted errors. Mean value of the binding energy per nucleon\cite{Wang:2012} and charge radii\cite{Angeli:2013}. Charge and weak form factor $F_{ch}$ and $F_{W}$ listed here correspond to momentum transfer $q=0.8733$ fm$^{-1}$ for $^{48}$Ca in CREX \cite{adhikari2022precision} or $0.3977$ fm$^{-1}$ for in PREX \cite{adhikari2021accurate}. 
\label{tab:likelihood}}
\begin{tabular}{llccc}
 & Property& Mean & Standard Deviation \\
\hline
&$R_{ch} ^{^{48}Ca}$ [fm] & 3.48 & 0.070\\
&$R_{ch} ^{^{90}Zr}$ [fm] & 4.27 & 0.085\\
&$R_{ch} ^{^{208}Pb}$ [fm] & 5.50 & 0.11\\
Basic nuclei&B/A$ ^{^{48}Ca}$ [MeV] & 8.67 & 0.433 \\
constraints &B/A$ ^{^{90}Zr}$ [MeV] & 8.71 & 0.436 \\
&B/A$ ^{^{208}Pb}$ [MeV] & 7.87 & 0.393\\
&$F_{ch} ^{^{48}Ca}$ [] & 0.1581 & 0.005\\
&$F_{ch} ^{^{208}Pb}$ [] & 0.409 & 0.005\\
\hline
CREX &$F_{W} ^{^{48}Ca}-F_{ch} ^{^{48}Ca}$ [] & 0.0277 & 0.0055\\
PREX &$F_{W} ^{^{208}Pb}-F_{ch} ^{^{208}Pb}$ [] & 0.041 & 0.013\\
\end{tabular}
\end{ruledtabular}
\end{table}

%% file: fit_SvL_FchFw_resub.tex
\begin{figure}
    \centering
    \includegraphics[width=\linewidth]{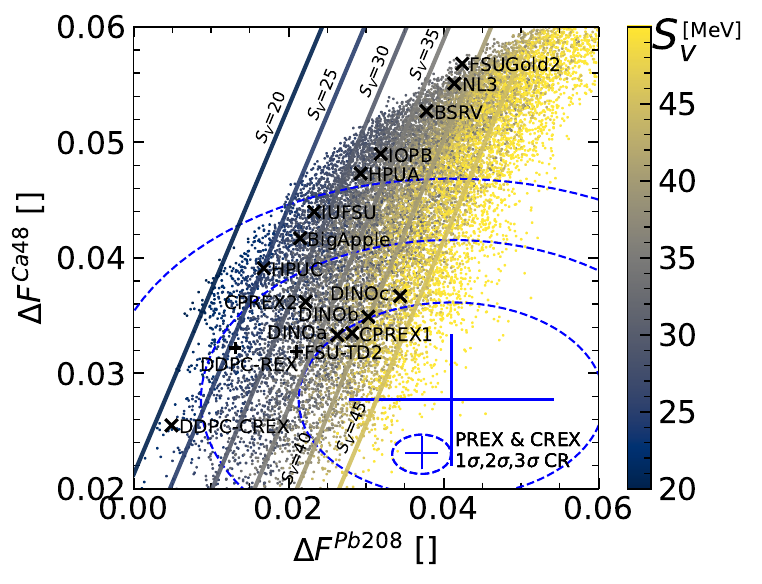}
    \includegraphics[width=\linewidth]{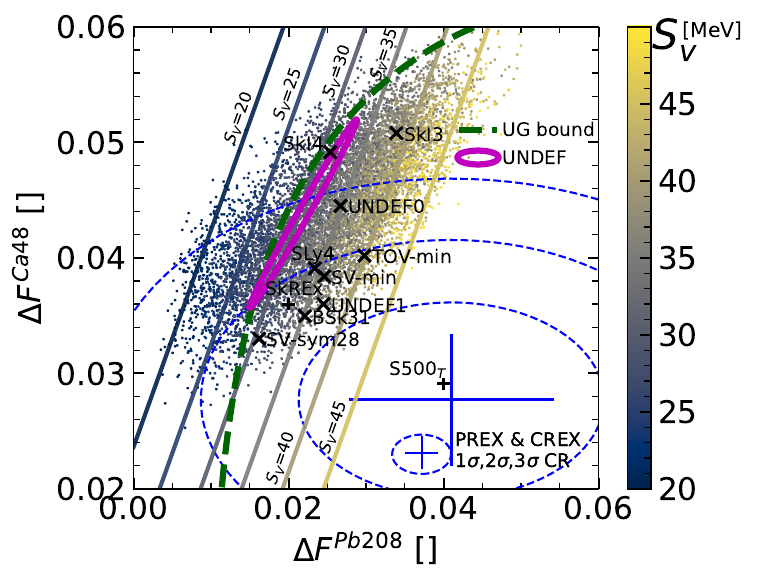}
    \caption{Similar to Fig. 3 in the letter but colored by symmetry energy parameter $S_V$. Colorful lines corresponds to the minimum $\chi^2$ fitting of Eq.(5) in the letter with $S_V=[20,25,30,35,40,45]$ MeV.}
    \label{fig:SvdeltaF}
\end{figure}
\begin{figure}
    \centering
    \includegraphics[width=\linewidth]{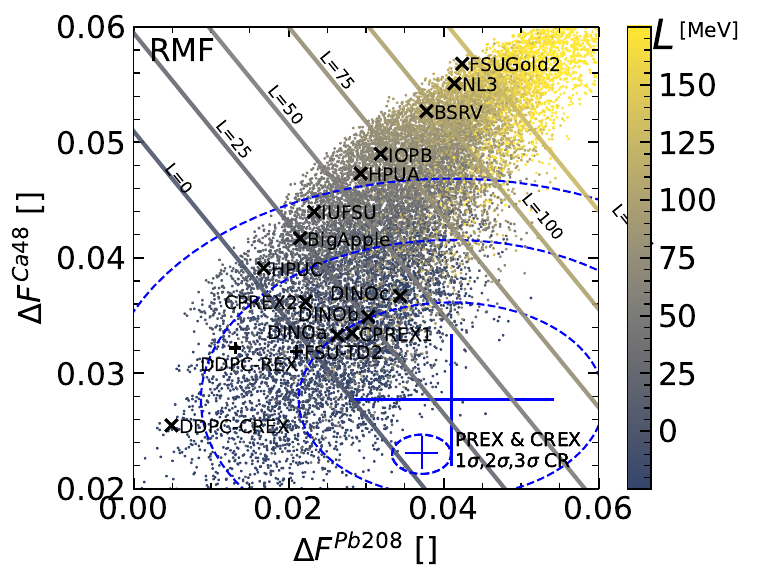}
    \includegraphics[width=\linewidth]{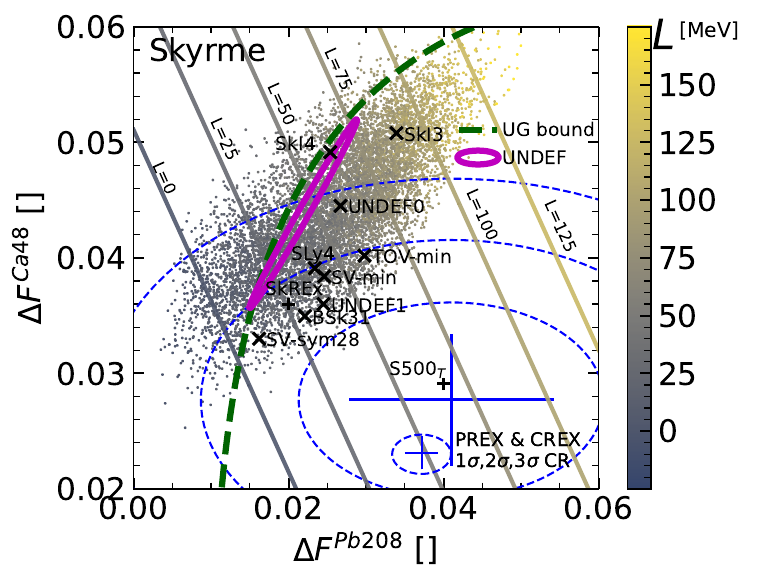}
    \caption{Similar to Fig. 3 in the letter but colored by symmetry energy parameter$L$. Colorful lines corresponds to the minimum $\chi^2$ fitting of Eq. (6) in the letter with $L=[0,25,50,75,100,125,150]$ MeV.}
    \label{fig:LdeltaF}
\end{figure}
\begin{figure}
    \centering
    \includegraphics[width=\linewidth]{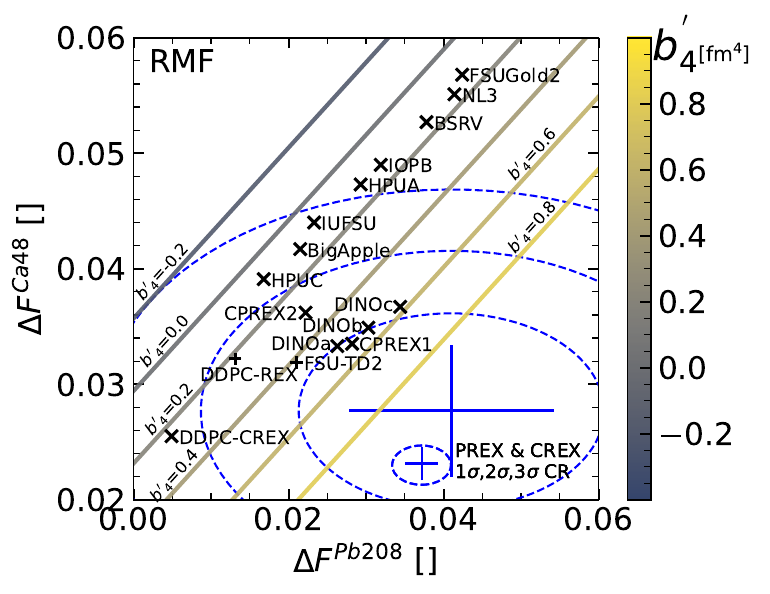}
    \includegraphics[width=\linewidth]{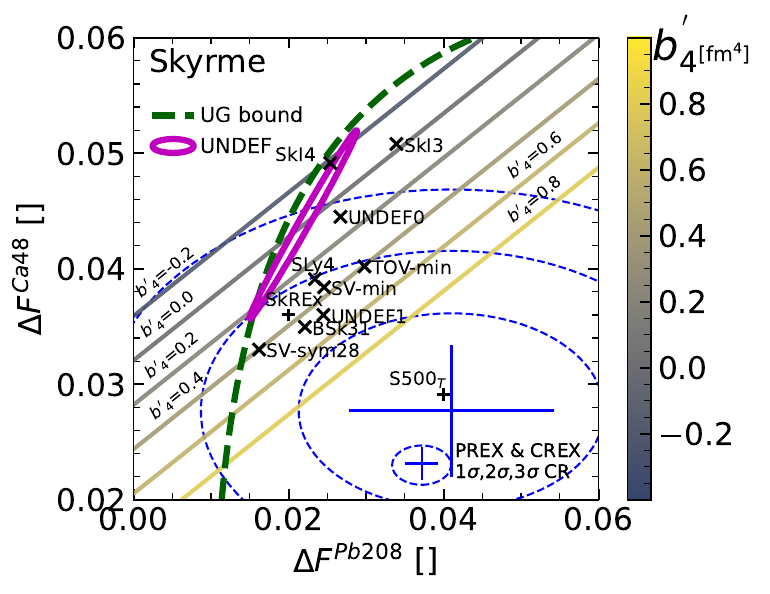}
    \caption{Similar to Fig. 3 in the letter but without colored posterior points for RMF and Skyrme models.}
    \label{fig:b4pdeltaF}
\end{figure}
\begin{figure}
    \centering
    \includegraphics[width=0.9\linewidth]{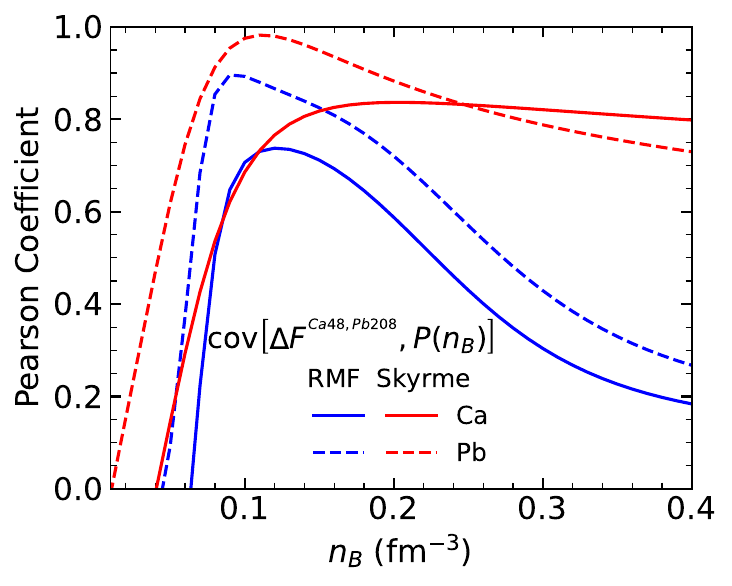}
    \includegraphics[width=0.9\linewidth]{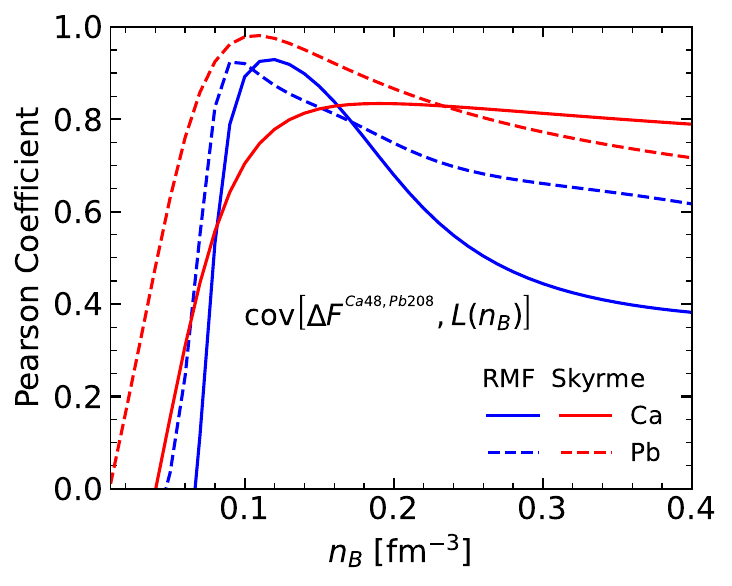}
    \caption{Pearson correlation coefficient between neutron skin (form factor difference) of $^{48}$Ca (solid) and $^{208}$Pb (dashed), and pressure (upper panel) or symmetry energy slope (lower panel) at various baryon number densities. Blue and Red correspond to RMF models and Skyrme models.
    \label{fig:pearson_FL}}
\end{figure}
\subsection{Linear Correlation Between $S_\mathrm{V}$, $L$, $b'_4$, $\Delta F^{\mathrm{Ca48}}$, and $\Delta F^{\mathrm{Pb208}}$}
Based on all the samples used in the Bayesian analysis, we found that any three of the variables $S_\mathrm{V}$, $L$, $\Delta F^{\mathrm{Ca48}}$, and $\Delta F^{\mathrm{Pb208}}$ can form semi-linear relations with each other, resulting in four relations as shown in Eqs. (5-9) in the letter. These pocket formulas could help interpret future parity-violating asymmetry experiments. In this section, we explicitly show how these formulas perform for Skyrme and RMF models.

Figures \ref{fig:SvdeltaF}-\ref{fig:LdeltaF} show the distributions of $(\Delta F^{\mathrm{Ca48}}, \Delta F^{\mathrm{Pb208}}, S_\mathrm{V})$ and $(\Delta F^{\mathrm{Ca48}}, \Delta F^{\mathrm{Pb208}}, L)$, where the linear correlations in Eqs. (5-6) in the letter are observed. The precision of the fitting formula in Eq. (5) remains within 10\% (1-$\sigma$) for both Skyrme and RMF models, while Eq. (6) shows about a 10 MeV deviation for the Skyrme model and a 25 MeV deviation in the RMF model. An interesting trend is noted in Fig. \ref{fig:SvdeltaF} for both RMF and Skyrme models: the predicted points of ${\Delta F^{\mathrm{Ca48}}, \Delta F^{\mathrm{Pb208}}}$ move \emph{collectively} from left to right (towards the mean of joint PREX-2/CREX measurements) with increasing $S_\mathrm{V}$. This collective movement results from $S_\mathrm{V}$ being sensitive to the neutron skin of $^{208}$Pb while being insensitive to the neutron skin of $^{48}$Ca in both RMF and Skyrme models. The difference in sensitivity of $\Delta F^{\mathrm{Pb208}}$ and $\Delta F^{\mathrm{Ca48}}$ to $S_\mathrm{V}$ is reflected by the slope coefficients $a$ and $b$ in Eq. (5) in the letter, where $b$ is much larger than $a$ in both Skyrme and RMF fittings.

In Fig. \ref{fig:LdeltaF}, we present the corresponding $L$ values of our sampled points of ${\Delta F^{\mathrm{Ca48}}, \Delta F^{\mathrm{Pb208}}}$. As expected, $L$ is sensitive to both the neutron skin of $^{48}$Ca and $^{208}$Pb, with the points moving collectively from lower left to upper right with increasing $L$. This movement highlights the challenge of finding parameterizations that satisfy both PREX-2/CREX measurements by only tuning parameters controlling the $L$ values. Notably, for the same $L$, Skyrme models predict slightly smaller $\Delta F$ due to the larger slope of the constant $L$ relation for $\Delta F$ compared to RMF models, as seen in Fig. \ref{fig:LdeltaF}. This can be understood from the Pearson correlation coefficient between $\Delta F$ and $L(n_B)$ in Fig. \ref{fig:pearson_FL}. 

In Fig. \ref{fig:b4pdeltaF}, we present a more comprehensive demonstration of the $b'_4$ correlation with ${\Delta F^{\mathrm{Ca48}}, \Delta F^{\mathrm{Pb208}}}$ comparing with existing Skyrme and RMF models. The widely used SLy4 model has $W_0 =2b_4=2b_4'= 123$ MeV fm$^5$\cite{Chabanat:1997un}, equivalent to $b_4 = b'_4 = 0.312$ fm$^{4}$, which lies on the 2-$\sigma$ contour of the PREX-2/CREX measurement. The first Skyrme models with $b_4 \neq b'_4$, SkI3 and SkI4 are introduced with $b'_4\leq 0<b_4$, placing them further away from the PREX-2/CREX measurement \cite{reinhard1995nuclear}. SV classes of Skyrme model prefer $0<b'_4/b_4<0.5$, see Fig. 9 in \cite{klupfel2009variations}. TOV-min takes a relatively large $b'_4=0.426$\cite{erler2013energy}, placing it closer towards the right bottom corner. More recent calibrated models UNEDF0 with $b_4 = 0.634$, $b'_4 = -0.463$ and UNEDF1 with $b_4 = 0.194$, $b'_4 = 0.362$ have $b'_4$ of opposite sign\cite{kortelainen2012nuclear}. Therefore, UNEDF1 is much closer to the PREX-2/CREX measurements compared to UNEDF0. Most recently, the extended Skyrme model with $b'_4 = 1.27$ is proposed to reach the center of PREX-2/CREX measurement\cite{Yue:2024srj}, which is consistent with our correlation which gives $b'_4 = 1.37 \pm 0.49$ fm$^4$ for Skyrme models from PREX-2/CREX measurements of $\Delta F$. In the case of RMF models, the effective parameter is $b'_4 = 1.02 \pm 0.37$, corresponding to a 90\% lower bound of $b'_4 > 0.544$ fm$^4$ or $g_\delta^2 + 1.65 g_\rho^2 > 2433$ after incorporating the masses of nucleons and mesons. Widely used RMF models without delta mesons, such as FSUGold2 \cite{Chen_2014} with $g_\rho^2= 80.4656$ and FSUGarnet \cite{chen2015searching} with $g_\rho^2= 192.9274$, typically exhibit a small effective $b'_4 < 0.1$. Other RMF models with $b'_4\lesssim 0.1$, such as NL3\cite{lalazissis1997new}, IOPB-I\cite{IOPB}, IUFSU\cite{fattoyev2010relativistic}, BigApple\cite{fattoyev2020gw190814}, HPUA-C\cite{sharma2023new}, all lie beyond the 2-$\sigma$ region, although these models span a wide range of symmetry energy slope L. New RMF models featuring $\delta$ mesons with small isovector Yukawa coupling, such as BSRV with $g_\delta^2 < 20$ and $g_\rho^2 < 265$, corresponding to $b'_4 \lesssim 0.1$, also do not show significant improvement \cite{kumar2023crex}. However, models like DINOa-c, which display large isovector Yukawa couplings ($g_\delta^2 > 1100$ and $g_\rho^2 > 800$), all surpass the 90\% lower bound of $b'_4 > 0.544$ \cite{reed2024density}. Consequently, DINOa-c lies close to the 1-$\sigma$ confidence region of PREX-2/CREX.

In Fig.\ref{fig:pearson_FL}, the Pearson correlation peaks around two-thirds of the saturation density for $^{208}$Pb, where the neutron skin correlates well with the form factor difference $\Delta F$ for $^{208}$Pb. The neutron skin is known to correlate best with the symmetry energy slope at two-thirds of the saturation density for the Skyrme model \cite{brown2000neutron}, a correlation confirmed by RMF models as well \cite{typel2001neutron}. Note that for both Skyrme and RMF models, the Pearson correlation between P and $\Delta F^{Pb208}$ is higher than that between P and $\Delta F^{Ca208}$ below and around the saturation density. It clearly shows the neutron skin of $^{208}\mathrm{Pb}$ is more sensitive to the pressures in neutron stars than $^{208}\mathrm{Ca}$ does.
Additionally, our Pearson correlation study for $\Delta F^{\mathrm{Ca48}}$ suggests that $\Delta F^{\mathrm{Ca48}}$ measured in CREX is most sensitive to the symmetry energy slope $L$ at $n_B=0.12$ fm$^{-3}$ for the RMF model and $n_B=0.16$ fm$^{-3}$ for the Skyrme model, around saturation density. The Pearson correlation with L are about the same between the two nuclei in the RMF models, while the Pearson correlation with L and for $^{208}$Pb is much higher than that for $^{48}$Ca in Skyrme models, explaining the larger slope of the constant $L$ relation for $\Delta F$.
%

%% file: density_profile.tex
\begin{figure*}
    \centering
    \includegraphics[width=0.3\linewidth]{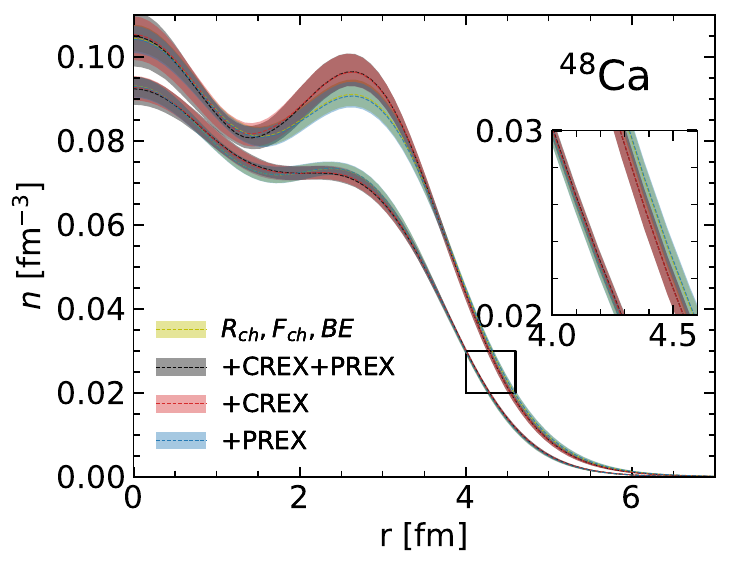}
    \includegraphics[width=0.3\linewidth]{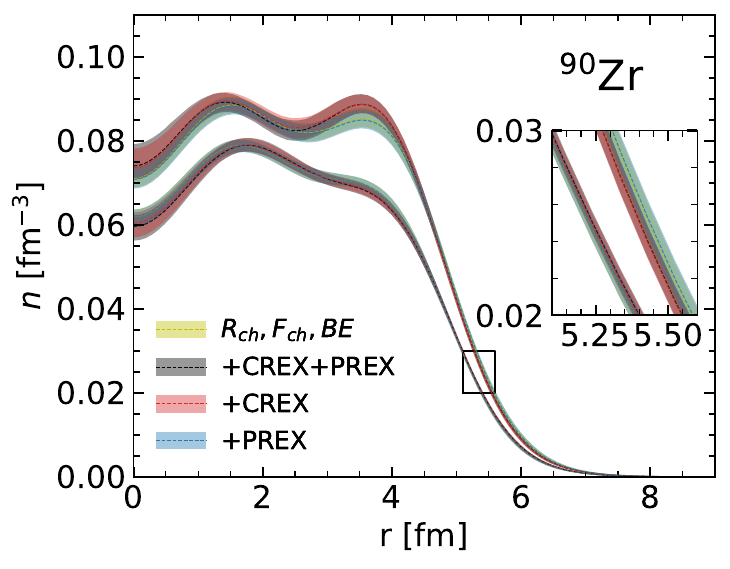}
    \includegraphics[width=0.3\linewidth]{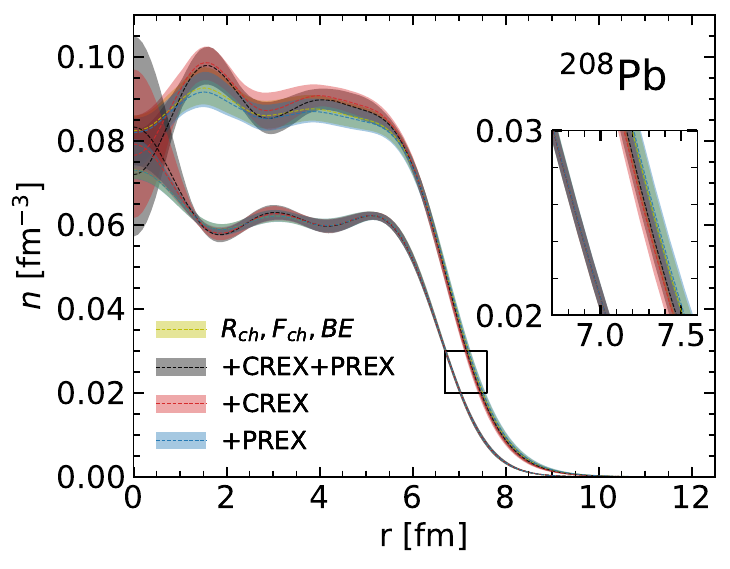}
    \caption{Vector densities of protons (lower) and neutrons (upper) in $^{48}$Ca, $^{90}$Zr, and $^{208}$Pb. The different colored bands and central lines represent the 18\%, 50\%, and 82\% percentiles under various constraints specified in the legend.}
    \label{fig:profile_likelihood}
\end{figure*}
\begin{figure*}
    \centering
    \includegraphics[width=0.3\linewidth]{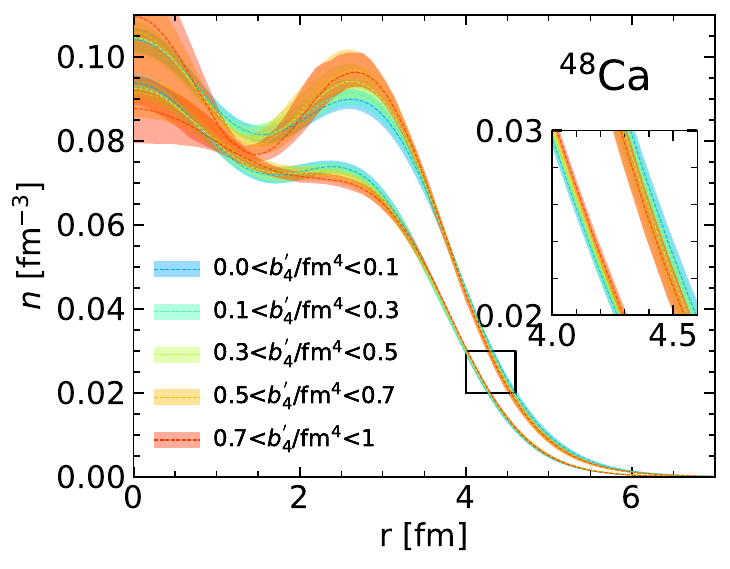}
    \includegraphics[width=0.3\linewidth]{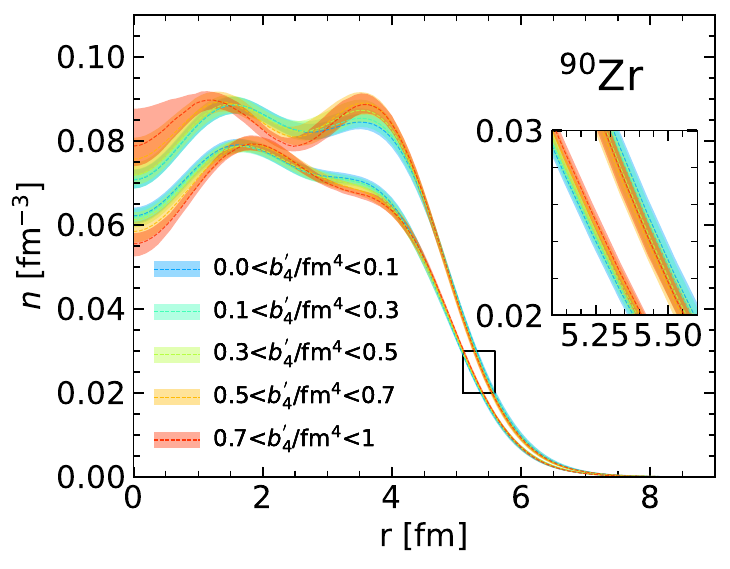}
    \includegraphics[width=0.3\linewidth]{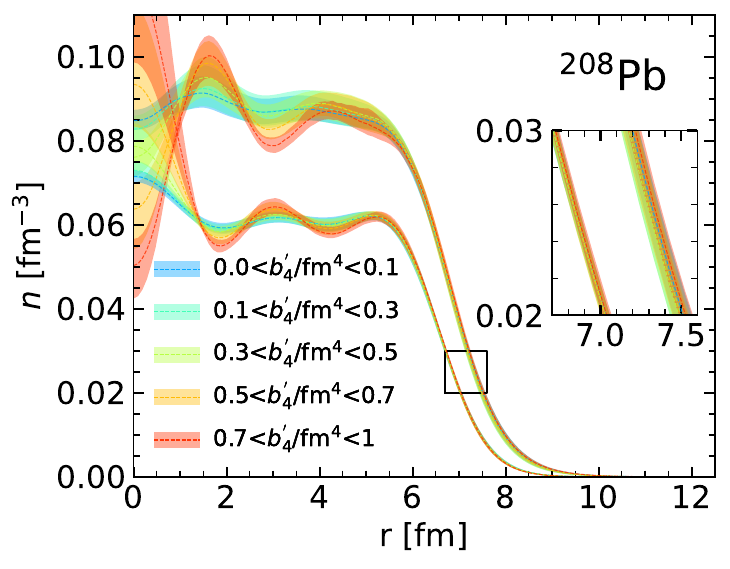}
    \caption{Similar to Fig. \ref{fig:profile_likelihood}. The different colored bands and central lines represent  RMF models with various $b'_4$ defined in Eq.(\ref{eq:b4'}). }
    \label{fig:profile_b4'}
\end{figure*}
\subsection{Nucleon density distribution of $^{48}$Ca and $^{208}$Pb in RMF model}
In this section, we delve into the proton and neutron density distributions of $^{48}$Ca and $^{208}$Pb within RMF models, utilizing samples with four different likelihoods. We computed the baryon number density, denoted as $n(r)$, as a function of radius $r$ and assessed the 18\%, 50\%, and 82\% percentiles of $n(r)$ across all radii.

Figure \ref{fig:profile_likelihood} illustrates the 18\%, 50\%, and 82\% percentiles of proton and neutron densities in $^{48}$Ca and $^{208}$Pb under various likelihood scenarios. The yellow band represents the basic nuclear constraints outlined in Table \ref{tab:likelihood}. Interestingly, this band significantly overlaps with the blue band, which incorporates additional constraints from PREX. This overlap is expected due to the abundance of RMF models in our prior with large $S_\mathrm{V}$ and $L$, coupled with the relatively larger statistical error of PREX compared to CREX. Similar overlaps are observed between the black and red bands. However, the imposition of CREX constraints notably alters the proton and neutron distributions, as evidenced by the discernible differences between the yellow and red bands (or black and blue bands). In $^{48}$Ca, the neutron bulge around 2.5 fm increases by approximately 0.005 fm$^{-3}$ with the imposition of CREX constraints, while the distributions for $r<1$ fm remain unchanged. Consequently, the neutron distribution around the surface contracts inward by about 0.05 fm, resulting in smaller values for $\Delta R_{np}$ and $\Delta F$. Conversely, in $^{208}$Pb, the imposition of CREX constraints enhances neutron bulges at 1.5 fm and 4 fm, while also increasing neutron deficits and proton bulges at the center. Since the central region contributes minimally to the neutron skin, the overall neutron distribution shifts inward, albeit not as significantly as in $^{48}$Ca.

As highlighted in our letter, the degeneracy (correlation) between $\Delta R_{np}$ (or $\Delta F$) of $^{48}$Ca and $^{208}$Pb may be mitigated by breaking $S_\mathbf{V}$-$L$ relations or adjusting $b'_4$. We categorized the sample into five bins based on $b'_4$ and assessed the 18\%, 50\%, and 82\% percentiles of $n(r)$ accordingly. Figure \ref{fig:profile_b4'} elucidates how varying $b'_4$ modulates the density distribution of protons and neutrons. For $^{48}$Ca, increasing $b'_4$ amplifies the neutron bulge and suppresses the proton bulge around 2.5 fm. Given that the bulge around 2.5 fm exerts the greatest influence on the overall nucleon distribution due to the $r^2$ factor in the volume weighting, the heightened isovector density in the bulge precipitates a notable decrease in $\Delta R_{np}$ and $\Delta F$ for $^{48}$Ca. The behavior of the intermediate nucleus $^{90}$Zr appears to mirror that of $^{48}$Ca, albeit with slightly smaller variations in magnitude. However, for $^{208}$Pb, increasing $b'_4$ deepens the neutron deficit and enhances the proton bulge at the center, while also diminishing the isovector density at 3 fm. Although neutron excesses increase around 1.5 fm and 4 fm, the overall neutron excess remains relatively unchanged. Notably, $\Delta R_{np}$ and $\Delta F$ exhibit insensitivity or slight positive correlation with $b'_4$. This can also be seen from Pearson correlation of $\Delta F$ in Fig. \ref{fig:corner_posterior1} and Fig. \ref{fig:corner_posterior2}.

Across all nuclei, a larger $b'_4$ accentuates the isovector density oscillations throughout the nuclei. However, in $^{48}$Ca, these oscillations manifest as a neutron excess in the bulk interior, while maintaining a relatively consistent average neutron excess in the interior of $^{208}$Pb. Since a substantial neutron excess in the nuclear interior correlates with a smaller neutron skin and a diminished form factor difference, adjusting $b'_4$ may break the correlation between $\Delta R_{np}$ (or $\Delta F$) of $^{48}$Ca and $^{208}$Pb.

In summary, the CREX experiment necessitates a small $\Delta R_{np}$ and $\Delta F$ for $^{48}$Ca, while the PREX experiment demands a large $\Delta R_{np}$ and $\Delta F$ for $^{208}$Pb. Given the larger error bars in $\Delta F$ for the PREX experiment (three times standard deviation) compared to the CREX experiment, the joint PREX-2/CREX results are predominantly influenced by the CREX experiment, thereby shifting $\Delta R_{np}$ and $\Delta F$ for both $^{48}$Ca and $^{208}$Pb towards smaller values, consistent with other Bayesian analyses\cite{Zhang2022,salinas2023bayesian}. If future experiments, such as MREX, confirm the large $\Delta F^{Pb208}$ by narrowing down the statistical error, one potential solution involves introducing a significant isospin-dependent term of the spin-orbit interaction to amplify the variation of isovector density. Such variation impacts the average interior neutron excess differently between $^{48}$Ca and $^{208}$Pb. In this study, we achieve this by adjusting $b'_4$, although other interactions such as the isovector tensor interaction, may serve a similar purpose \cite{salinas2023impact,Yue:2024srj}. Accounting for multiple interactions that excite isovector density oscillations may offer insights into resolving the tension between PREX-2/CREX.

%% file: arxiv.bbl
\begin{thebibliography}{99}%
\makeatletter
\providecommand \@ifxundefined [1]{%
 \@ifx{#1\undefined}
}%
\providecommand \@ifnum [1]{%
 \ifnum #1\expandafter \@firstoftwo
 \else \expandafter \@secondoftwo
 \fi
}%
\providecommand \@ifx [1]{%
 \ifx #1\expandafter \@firstoftwo
 \else \expandafter \@secondoftwo
 \fi
}%
\providecommand \natexlab [1]{#1}%
\providecommand \enquote  [1]{``#1''}%
\providecommand \bibnamefont  [1]{#1}%
\providecommand \bibfnamefont [1]{#1}%
\providecommand \citenamefont [1]{#1}%
\providecommand \href@noop [0]{\@secondoftwo}%
\providecommand \href [0]{\begingroup \@sanitize@url \@href}%
\providecommand \@href[1]{\@@startlink{#1}\@@href}%
\providecommand \@@href[1]{\endgroup#1\@@endlink}%
\providecommand \@sanitize@url [0]{\catcode `\\12\catcode `\$12\catcode `\&12\catcode `\#12\catcode `\^12\catcode `\_12\catcode `\%12\relax}%
\providecommand \@@startlink[1]{}%
\providecommand \@@endlink[0]{}%
\providecommand \url  [0]{\begingroup\@sanitize@url \@url }%
\providecommand \@url [1]{\endgroup\@href {#1}{\urlprefix }}%
\providecommand \urlprefix  [0]{URL }%
\providecommand \Eprint [0]{\href }%
\providecommand \doibase [0]{http://dx.doi.org/}%
\providecommand \selectlanguage [0]{\@gobble}%
\providecommand \bibinfo  [0]{\@secondoftwo}%
\providecommand \bibfield  [0]{\@secondoftwo}%
\providecommand \translation [1]{[#1]}%
\providecommand \BibitemOpen [0]{}%
\providecommand \bibitemStop [0]{}%
\providecommand \bibitemNoStop [0]{.\EOS\space}%
\providecommand \EOS [0]{\spacefactor3000\relax}%
\providecommand \BibitemShut  [1]{\csname bibitem#1\endcsname}%
\let\auto@bib@innerbib\@empty
\bibitem [{\citenamefont {Lalazissis}\ \emph {et~al.}(1997{\natexlab{a}})\citenamefont {Lalazissis}, \citenamefont {Konig},\ and\ \citenamefont {Ring}}]{Lalazissis:1996rd}%
  \BibitemOpen
  \bibfield  {author} {\bibinfo {author} {\bibfnamefont {G.~A.}\ \bibnamefont {Lalazissis}}, \bibinfo {author} {\bibfnamefont {J.}~\bibnamefont {Konig}}, \ and\ \bibinfo {author} {\bibfnamefont {P.}~\bibnamefont {Ring}},\ }\href {\doibase 10.1103/PhysRevC.55.540} {\bibfield  {journal} {\bibinfo  {journal} {Phys. Rev. C}\ }\textbf {\bibinfo {volume} {55}},\ \bibinfo {pages} {540} (\bibinfo {year} {1997}{\natexlab{a}})}\BibitemShut {NoStop}%
\bibitem [{\citenamefont {K\"onig}\ \emph {et~al.}(2023)\citenamefont {K\"onig} \emph {et~al.}}]{Konig:2023bzp}%
  \BibitemOpen
  \bibfield  {author} {\bibinfo {author} {\bibfnamefont {K.}~\bibnamefont {K\"onig}} \emph {et~al.},\ }\href {\doibase 10.1103/PhysRevLett.131.102501} {\bibfield  {journal} {\bibinfo  {journal} {Phys. Rev. Lett.}\ }\textbf {\bibinfo {volume} {131}},\ \bibinfo {pages} {102501} (\bibinfo {year} {2023})}\BibitemShut {NoStop}%
\bibitem [{\citenamefont {Sommer}\ \emph {et~al.}(2022)\citenamefont {Sommer} \emph {et~al.}}]{Sommer:2022sok}%
  \BibitemOpen
  \bibfield  {author} {\bibinfo {author} {\bibfnamefont {F.}~\bibnamefont {Sommer}} \emph {et~al.},\ }\href {\doibase 10.1103/PhysRevLett.129.132501} {\bibfield  {journal} {\bibinfo  {journal} {Phys. Rev. Lett.}\ }\textbf {\bibinfo {volume} {129}},\ \bibinfo {pages} {132501} (\bibinfo {year} {2022})},\ \Eprint {http://arxiv.org/abs/2210.01924} {2210.01924} \BibitemShut {NoStop}%
\bibitem [{\citenamefont {Malbrunot-Ettenauer}\ \emph {et~al.}(2022)\citenamefont {Malbrunot-Ettenauer} \emph {et~al.}}]{Malbrunot-Ettenauer:2021fnr}%
  \BibitemOpen
  \bibfield  {author} {\bibinfo {author} {\bibfnamefont {S.}~\bibnamefont {Malbrunot-Ettenauer}} \emph {et~al.},\ }\href {\doibase 10.1103/PhysRevLett.128.022502} {\bibfield  {journal} {\bibinfo  {journal} {Phys. Rev. Lett.}\ }\textbf {\bibinfo {volume} {128}},\ \bibinfo {pages} {022502} (\bibinfo {year} {2022})}\BibitemShut {NoStop}%
\bibitem [{\citenamefont {Reinhard}\ and\ \citenamefont {Nazarewicz}(2022)}]{Reinhard:2022jby}%
  \BibitemOpen
  \bibfield  {author} {\bibinfo {author} {\bibfnamefont {P.-G.}\ \bibnamefont {Reinhard}}\ and\ \bibinfo {author} {\bibfnamefont {W.}~\bibnamefont {Nazarewicz}},\ }\href {\doibase 10.1103/PhysRevC.105.L021301} {\bibfield  {journal} {\bibinfo  {journal} {Phys. Rev. C}\ }\textbf {\bibinfo {volume} {105}},\ \bibinfo {pages} {L021301} (\bibinfo {year} {2022})}\BibitemShut {NoStop}%
\bibitem [{\citenamefont {Koepf}\ and\ \citenamefont {Ring}(1991)}]{koepf1991spin}%
  \BibitemOpen
  \bibfield  {author} {\bibinfo {author} {\bibfnamefont {W.}~\bibnamefont {Koepf}}\ and\ \bibinfo {author} {\bibfnamefont {P.}~\bibnamefont {Ring}},\ }\href@noop {} {\bibfield  {journal} {\bibinfo  {journal} {Zeitschrift f{\"u}r Physik A Hadrons and Nuclei}\ }\textbf {\bibinfo {volume} {339}},\ \bibinfo {pages} {81} (\bibinfo {year} {1991})}\BibitemShut {NoStop}%
\bibitem [{\citenamefont {Horowitz}\ and\ \citenamefont {Piekarewicz}(2001)}]{Horowitz_2001}%
  \BibitemOpen
  \bibfield  {author} {\bibinfo {author} {\bibfnamefont {C.~J.}\ \bibnamefont {Horowitz}}\ and\ \bibinfo {author} {\bibfnamefont {J.}~\bibnamefont {Piekarewicz}},\ }\href {\doibase 10.1103/PhysRevLett.86.5647} {\bibfield  {journal} {\bibinfo  {journal} {Phys. Rev. Lett.}\ }\textbf {\bibinfo {volume} {86}},\ \bibinfo {pages} {5647} (\bibinfo {year} {2001})}\BibitemShut {NoStop}%
\bibitem [{\citenamefont {Adhikari}\ \emph {et~al.}(2021{\natexlab{a}})\citenamefont {Adhikari} \emph {et~al.}}]{PREX:2021umo}%
  \BibitemOpen
  \bibfield  {author} {\bibinfo {author} {\bibfnamefont {D.}~\bibnamefont {Adhikari}} \emph {et~al.} (\bibinfo {collaboration} {PREX}),\ }\href {\doibase 10.1103/PhysRevLett.126.172502} {\bibfield  {journal} {\bibinfo  {journal} {Phys. Rev. Lett.}\ }\textbf {\bibinfo {volume} {126}},\ \bibinfo {pages} {172502} (\bibinfo {year} {2021}{\natexlab{a}})}\BibitemShut {NoStop}%
\bibitem [{\citenamefont {Adhikari}\ \emph {et~al.}(2022{\natexlab{a}})\citenamefont {Adhikari} \emph {et~al.}}]{CREX:2022kgg}%
  \BibitemOpen
  \bibfield  {author} {\bibinfo {author} {\bibfnamefont {D.}~\bibnamefont {Adhikari}} \emph {et~al.} (\bibinfo {collaboration} {CREX}),\ }\href {\doibase 10.1103/PhysRevLett.129.042501} {\bibfield  {journal} {\bibinfo  {journal} {Phys. Rev. Lett.}\ }\textbf {\bibinfo {volume} {129}},\ \bibinfo {pages} {042501} (\bibinfo {year} {2022}{\natexlab{a}})}\BibitemShut {NoStop}%
\bibitem [{\citenamefont {Reed}\ \emph {et~al.}(2024)\citenamefont {Reed}, \citenamefont {Fattoyev}, \citenamefont {Horowitz},\ and\ \citenamefont {Piekarewicz}}]{reed2024density}%
  \BibitemOpen
  \bibfield  {author} {\bibinfo {author} {\bibfnamefont {B.~T.}\ \bibnamefont {Reed}}, \bibinfo {author} {\bibfnamefont {F.}~\bibnamefont {Fattoyev}}, \bibinfo {author} {\bibfnamefont {C.}~\bibnamefont {Horowitz}}, \ and\ \bibinfo {author} {\bibfnamefont {J.}~\bibnamefont {Piekarewicz}},\ }\href@noop {} {\bibfield  {journal} {\bibinfo  {journal} {Physical Review C}\ }\textbf {\bibinfo {volume} {109}},\ \bibinfo {pages} {035803} (\bibinfo {year} {2024})}\BibitemShut {NoStop}%
\bibitem [{\citenamefont {Reed}\ \emph {et~al.}(2021)\citenamefont {Reed}, \citenamefont {Fattoyev}, \citenamefont {Horowitz},\ and\ \citenamefont {Piekarewicz}}]{Reed2021}%
  \BibitemOpen
  \bibfield  {author} {\bibinfo {author} {\bibfnamefont {B.~T.}\ \bibnamefont {Reed}}, \bibinfo {author} {\bibfnamefont {F.~J.}\ \bibnamefont {Fattoyev}}, \bibinfo {author} {\bibfnamefont {C.~J.}\ \bibnamefont {Horowitz}}, \ and\ \bibinfo {author} {\bibfnamefont {J.}~\bibnamefont {Piekarewicz}},\ }\href {\doibase 10.1103/PhysRevLett.126.172503} {\bibfield  {journal} {\bibinfo  {journal} {Physical Review Letters}\ }\textbf {\bibinfo {volume} {126}} (\bibinfo {year} {2021}),\ 10.1103/PhysRevLett.126.172503}\BibitemShut {NoStop}%
\bibitem [{\citenamefont {Reinhard}\ \emph {et~al.}(2022{\natexlab{a}})\citenamefont {Reinhard}, \citenamefont {Roca-Maza},\ and\ \citenamefont {Nazarewicz}}]{reinhard2022combined}%
  \BibitemOpen
  \bibfield  {author} {\bibinfo {author} {\bibfnamefont {P.-G.}\ \bibnamefont {Reinhard}}, \bibinfo {author} {\bibfnamefont {X.}~\bibnamefont {Roca-Maza}}, \ and\ \bibinfo {author} {\bibfnamefont {W.}~\bibnamefont {Nazarewicz}},\ }\href@noop {} {\bibfield  {journal} {\bibinfo  {journal} {Physical Review Letters}\ }\textbf {\bibinfo {volume} {129}},\ \bibinfo {pages} {232501} (\bibinfo {year} {2022}{\natexlab{a}})}\BibitemShut {NoStop}%
\bibitem [{\citenamefont {Hagen}\ \emph {et~al.}(2015)\citenamefont {Hagen} \emph {et~al.}}]{Hagen:2015yea}%
  \BibitemOpen
  \bibfield  {author} {\bibinfo {author} {\bibfnamefont {G.}~\bibnamefont {Hagen}} \emph {et~al.},\ }\href {\doibase 10.1038/nphys3529} {\bibfield  {journal} {\bibinfo  {journal} {Nature Phys.}\ }\textbf {\bibinfo {volume} {12}},\ \bibinfo {pages} {186} (\bibinfo {year} {2015})},\ \Eprint {http://arxiv.org/abs/1509.07169} {arXiv:1509.07169 [nucl-th]} \BibitemShut {NoStop}%
\bibitem [{\citenamefont {Hu}\ \emph {et~al.}(2022)\citenamefont {Hu}, \citenamefont {Jiang}, \citenamefont {Miyagi}, \citenamefont {Sun}, \citenamefont {Ekström}, \citenamefont {Forssén}, \citenamefont {Hagen}, \citenamefont {Holt}, \citenamefont {Papenbrock}, \citenamefont {Stroberg},\ and\ \citenamefont {Vernon}}]{Hu2022}%
  \BibitemOpen
  \bibfield  {author} {\bibinfo {author} {\bibfnamefont {B.}~\bibnamefont {Hu}}, \bibinfo {author} {\bibfnamefont {W.}~\bibnamefont {Jiang}}, \bibinfo {author} {\bibfnamefont {T.}~\bibnamefont {Miyagi}}, \bibinfo {author} {\bibfnamefont {Z.}~\bibnamefont {Sun}}, \bibinfo {author} {\bibfnamefont {A.}~\bibnamefont {Ekström}}, \bibinfo {author} {\bibfnamefont {C.}~\bibnamefont {Forssén}}, \bibinfo {author} {\bibfnamefont {G.}~\bibnamefont {Hagen}}, \bibinfo {author} {\bibfnamefont {J.~D.}\ \bibnamefont {Holt}}, \bibinfo {author} {\bibfnamefont {T.}~\bibnamefont {Papenbrock}}, \bibinfo {author} {\bibfnamefont {S.~R.}\ \bibnamefont {Stroberg}}, \ and\ \bibinfo {author} {\bibfnamefont {I.}~\bibnamefont {Vernon}},\ }\href {\doibase 10.1038/s41567-022-01715-8} {\bibfield  {journal} {\bibinfo  {journal} {Nature Physics}\ } (\bibinfo {year} {2022}),\ 10.1038/s41567-022-01715-8}\BibitemShut {NoStop}%
\bibitem [{\citenamefont {Mahzoon}\ \emph {et~al.}(2017)\citenamefont {Mahzoon}, \citenamefont {Atkinson}, \citenamefont {Charity},\ and\ \citenamefont {Dickhoff}}]{mahzoon2017neutron}%
  \BibitemOpen
  \bibfield  {author} {\bibinfo {author} {\bibfnamefont {M.}~\bibnamefont {Mahzoon}}, \bibinfo {author} {\bibfnamefont {M.}~\bibnamefont {Atkinson}}, \bibinfo {author} {\bibfnamefont {R.}~\bibnamefont {Charity}}, \ and\ \bibinfo {author} {\bibfnamefont {W.}~\bibnamefont {Dickhoff}},\ }\href@noop {} {\bibfield  {journal} {\bibinfo  {journal} {Physical review letters}\ }\textbf {\bibinfo {volume} {119}},\ \bibinfo {pages} {222503} (\bibinfo {year} {2017})}\BibitemShut {NoStop}%
\bibitem [{\citenamefont {Atkinson}\ \emph {et~al.}(2020)\citenamefont {Atkinson}, \citenamefont {Mahzoon}, \citenamefont {Keim}, \citenamefont {Bordelon}, \citenamefont {Pruitt}, \citenamefont {Charity},\ and\ \citenamefont {Dickhoff}}]{atkinson2020dispersive}%
  \BibitemOpen
  \bibfield  {author} {\bibinfo {author} {\bibfnamefont {M.}~\bibnamefont {Atkinson}}, \bibinfo {author} {\bibfnamefont {M.}~\bibnamefont {Mahzoon}}, \bibinfo {author} {\bibfnamefont {M.}~\bibnamefont {Keim}}, \bibinfo {author} {\bibfnamefont {B.}~\bibnamefont {Bordelon}}, \bibinfo {author} {\bibfnamefont {C.}~\bibnamefont {Pruitt}}, \bibinfo {author} {\bibfnamefont {R.}~\bibnamefont {Charity}}, \ and\ \bibinfo {author} {\bibfnamefont {W.}~\bibnamefont {Dickhoff}},\ }\href@noop {} {\bibfield  {journal} {\bibinfo  {journal} {Physical Review C}\ }\textbf {\bibinfo {volume} {101}},\ \bibinfo {pages} {044303} (\bibinfo {year} {2020})}\BibitemShut {NoStop}%
\bibitem [{\citenamefont {Zhao}()}]{Zhao_CPREX}%
  \BibitemOpen
  \bibfield  {author} {\bibinfo {author} {\bibfnamefont {T.}~\bibnamefont {Zhao}},\ }\href {https://github.com/sotzee/CPREX} {\enquote {\bibinfo {title} {{GitHub repository: CPREX}},}\ }\BibitemShut {NoStop}%
\bibitem [{\citenamefont {Zhang}\ and\ \citenamefont {Chen}(2023)}]{Zhang2022}%
  \BibitemOpen
  \bibfield  {author} {\bibinfo {author} {\bibfnamefont {Z.}~\bibnamefont {Zhang}}\ and\ \bibinfo {author} {\bibfnamefont {L.-W.}\ \bibnamefont {Chen}},\ }\href {\doibase 10.1103/PhysRevC.108.024317} {\bibfield  {journal} {\bibinfo  {journal} {Phys. Rev. C}\ }\textbf {\bibinfo {volume} {108}},\ \bibinfo {pages} {024317} (\bibinfo {year} {2023})}\BibitemShut {NoStop}%
\bibitem [{\citenamefont {Salinas}\ and\ \citenamefont {Piekarewicz}(2023)}]{salinas2023bayesian}%
  \BibitemOpen
  \bibfield  {author} {\bibinfo {author} {\bibfnamefont {M.}~\bibnamefont {Salinas}}\ and\ \bibinfo {author} {\bibfnamefont {J.}~\bibnamefont {Piekarewicz}},\ }\href@noop {} {\bibfield  {journal} {\bibinfo  {journal} {Physical Review C}\ }\textbf {\bibinfo {volume} {107}},\ \bibinfo {pages} {045802} (\bibinfo {year} {2023})}\BibitemShut {NoStop}%
\bibitem [{\citenamefont {Gandolfi}\ \emph {et~al.}(2014)\citenamefont {Gandolfi}, \citenamefont {Carlson}, \citenamefont {Reddy}, \citenamefont {Steiner},\ and\ \citenamefont {Wiringa}}]{gandolfi2014equation}%
  \BibitemOpen
  \bibfield  {author} {\bibinfo {author} {\bibfnamefont {S.}~\bibnamefont {Gandolfi}}, \bibinfo {author} {\bibfnamefont {J.}~\bibnamefont {Carlson}}, \bibinfo {author} {\bibfnamefont {S.}~\bibnamefont {Reddy}}, \bibinfo {author} {\bibfnamefont {A.}~\bibnamefont {Steiner}}, \ and\ \bibinfo {author} {\bibfnamefont {R.}~\bibnamefont {Wiringa}},\ }\href@noop {} {\bibfield  {journal} {\bibinfo  {journal} {The European Physical Journal A}\ }\textbf {\bibinfo {volume} {50}},\ \bibinfo {pages} {1} (\bibinfo {year} {2014})}\BibitemShut {NoStop}%
\bibitem [{\citenamefont {Lee}\ \emph {et~al.}(1998)\citenamefont {Lee}, \citenamefont {Kuo}, \citenamefont {Li},\ and\ \citenamefont {Brown}}]{lee1998nuclear}%
  \BibitemOpen
  \bibfield  {author} {\bibinfo {author} {\bibfnamefont {C.-H.}\ \bibnamefont {Lee}}, \bibinfo {author} {\bibfnamefont {T.}~\bibnamefont {Kuo}}, \bibinfo {author} {\bibfnamefont {G.}~\bibnamefont {Li}}, \ and\ \bibinfo {author} {\bibfnamefont {G.}~\bibnamefont {Brown}},\ }\href@noop {} {\bibfield  {journal} {\bibinfo  {journal} {Physical Review C}\ }\textbf {\bibinfo {volume} {57}},\ \bibinfo {pages} {3488} (\bibinfo {year} {1998})}\BibitemShut {NoStop}%
\bibitem [{\citenamefont {Zuo}\ \emph {et~al.}(1999)\citenamefont {Zuo}, \citenamefont {Bombaci},\ and\ \citenamefont {Lombardo}}]{zuo1999asymmetric}%
  \BibitemOpen
  \bibfield  {author} {\bibinfo {author} {\bibfnamefont {W.}~\bibnamefont {Zuo}}, \bibinfo {author} {\bibfnamefont {I.}~\bibnamefont {Bombaci}}, \ and\ \bibinfo {author} {\bibfnamefont {U.}~\bibnamefont {Lombardo}},\ }\href@noop {} {\bibfield  {journal} {\bibinfo  {journal} {Physical Review C}\ }\textbf {\bibinfo {volume} {60}},\ \bibinfo {pages} {024605} (\bibinfo {year} {1999})}\BibitemShut {NoStop}%
\bibitem [{\citenamefont {Steiner}(2006)}]{Steiner06hs}%
  \BibitemOpen
  \bibfield  {author} {\bibinfo {author} {\bibfnamefont {A.~W.}\ \bibnamefont {Steiner}},\ }\href {\doibase 10.1103/PhysRevC.74.045808} {\bibfield  {journal} {\bibinfo  {journal} {Phys. Rev. C}\ }\textbf {\bibinfo {volume} {74}},\ \bibinfo {pages} {045808} (\bibinfo {year} {2006})},\ \Eprint {http://arxiv.org/abs/nucl-th/0607040} {nucl-th/0607040} \BibitemShut {NoStop}%
\bibitem [{\citenamefont {Drischler}\ \emph {et~al.}(2014)\citenamefont {Drischler}, \citenamefont {Soma},\ and\ \citenamefont {Schwenk}}]{drischler2014microscopic}%
  \BibitemOpen
  \bibfield  {author} {\bibinfo {author} {\bibfnamefont {C.}~\bibnamefont {Drischler}}, \bibinfo {author} {\bibfnamefont {V.}~\bibnamefont {Soma}}, \ and\ \bibinfo {author} {\bibfnamefont {A.}~\bibnamefont {Schwenk}},\ }\href@noop {} {\bibfield  {journal} {\bibinfo  {journal} {Physical Review C}\ }\textbf {\bibinfo {volume} {89}},\ \bibinfo {pages} {025806} (\bibinfo {year} {2014})}\BibitemShut {NoStop}%
\bibitem [{\citenamefont {Cai}\ and\ \citenamefont {Chen}(2019)}]{cai2019relativistic}%
  \BibitemOpen
  \bibfield  {author} {\bibinfo {author} {\bibfnamefont {B.-J.}\ \bibnamefont {Cai}}\ and\ \bibinfo {author} {\bibfnamefont {L.-W.}\ \bibnamefont {Chen}},\ }\href@noop {} {\bibfield  {journal} {\bibinfo  {journal} {Physical Review C}\ }\textbf {\bibinfo {volume} {100}},\ \bibinfo {pages} {024303} (\bibinfo {year} {2019})}\BibitemShut {NoStop}%
\bibitem [{\citenamefont {Salinas}\ and\ \citenamefont {Piekarewicz}(2024)}]{salinas2023impact}%
  \BibitemOpen
  \bibfield  {author} {\bibinfo {author} {\bibfnamefont {M.}~\bibnamefont {Salinas}}\ and\ \bibinfo {author} {\bibfnamefont {J.}~\bibnamefont {Piekarewicz}},\ }\href {\doibase 10.1103/PhysRevC.109.045807} {\bibfield  {journal} {\bibinfo  {journal} {Phys. Rev. C}\ }\textbf {\bibinfo {volume} {109}},\ \bibinfo {pages} {045807} (\bibinfo {year} {2024})}\BibitemShut {NoStop}%
\bibitem [{\citenamefont {Steiner}\ \emph {et~al.}(2005{\natexlab{a}})\citenamefont {Steiner}, \citenamefont {Prakash}, \citenamefont {Lattimer},\ and\ \citenamefont {Ellis}}]{Steiner05ia}%
  \BibitemOpen
  \bibfield  {author} {\bibinfo {author} {\bibfnamefont {A.~W.}\ \bibnamefont {Steiner}}, \bibinfo {author} {\bibfnamefont {M.}~\bibnamefont {Prakash}}, \bibinfo {author} {\bibfnamefont {J.~M.}\ \bibnamefont {Lattimer}}, \ and\ \bibinfo {author} {\bibfnamefont {P.~J.}\ \bibnamefont {Ellis}},\ }\href {\doibase 10.1016/j.physrep.2005.02.004} {\bibfield  {journal} {\bibinfo  {journal} {Phys. Rep.}\ }\textbf {\bibinfo {volume} {411}},\ \bibinfo {pages} {325} (\bibinfo {year} {2005}{\natexlab{a}})},\ \Eprint {http://arxiv.org/abs/nucl-th/0410066} {nucl-th/0410066} \BibitemShut {NoStop}%
\bibitem [{\citenamefont {Horowitz}\ \emph {et~al.}(2001)\citenamefont {Horowitz}, \citenamefont {Pollock}, \citenamefont {Souder},\ and\ \citenamefont {Michaels}}]{horowitz2001parity}%
  \BibitemOpen
  \bibfield  {author} {\bibinfo {author} {\bibfnamefont {C.}~\bibnamefont {Horowitz}}, \bibinfo {author} {\bibfnamefont {S.~J.}\ \bibnamefont {Pollock}}, \bibinfo {author} {\bibfnamefont {P.~A.}\ \bibnamefont {Souder}}, \ and\ \bibinfo {author} {\bibfnamefont {R.}~\bibnamefont {Michaels}},\ }\href@noop {} {\bibfield  {journal} {\bibinfo  {journal} {Physical Review C}\ }\textbf {\bibinfo {volume} {63}},\ \bibinfo {pages} {025501} (\bibinfo {year} {2001})}\BibitemShut {NoStop}%
\bibitem [{\citenamefont {Myers}\ and\ \citenamefont {Swiatecki}(1980)}]{Myers:1980iht}%
  \BibitemOpen
  \bibfield  {author} {\bibinfo {author} {\bibfnamefont {W.~D.}\ \bibnamefont {Myers}}\ and\ \bibinfo {author} {\bibfnamefont {W.~J.}\ \bibnamefont {Swiatecki}},\ }\href {\doibase 10.1016/0375-9474(80)90623-5} {\bibfield  {journal} {\bibinfo  {journal} {Nucl. Phys. A}\ }\textbf {\bibinfo {volume} {336}},\ \bibinfo {pages} {267} (\bibinfo {year} {1980})}\BibitemShut {NoStop}%
\bibitem [{\citenamefont {Centelles}\ \emph {et~al.}(2009)\citenamefont {Centelles}, \citenamefont {Roca-Maza}, \citenamefont {Vinas},\ and\ \citenamefont {Warda}}]{Centelles:2008vu}%
  \BibitemOpen
  \bibfield  {author} {\bibinfo {author} {\bibfnamefont {M.}~\bibnamefont {Centelles}}, \bibinfo {author} {\bibfnamefont {X.}~\bibnamefont {Roca-Maza}}, \bibinfo {author} {\bibfnamefont {X.}~\bibnamefont {Vinas}}, \ and\ \bibinfo {author} {\bibfnamefont {M.}~\bibnamefont {Warda}},\ }\href {\doibase 10.1103/PhysRevLett.102.122502} {\bibfield  {journal} {\bibinfo  {journal} {Phys. Rev. Lett.}\ }\textbf {\bibinfo {volume} {102}},\ \bibinfo {pages} {122502} (\bibinfo {year} {2009})}\BibitemShut {NoStop}%
\bibitem [{\citenamefont {Warda}\ \emph {et~al.}(2009)\citenamefont {Warda}, \citenamefont {Viñas}, \citenamefont {Roca-Maza},\ and\ \citenamefont {Centelles}}]{Warda2009}%
  \BibitemOpen
  \bibfield  {author} {\bibinfo {author} {\bibfnamefont {M.}~\bibnamefont {Warda}}, \bibinfo {author} {\bibfnamefont {X.}~\bibnamefont {Viñas}}, \bibinfo {author} {\bibfnamefont {X.}~\bibnamefont {Roca-Maza}}, \ and\ \bibinfo {author} {\bibfnamefont {M.}~\bibnamefont {Centelles}},\ }\href {\doibase 10.1103/PhysRevC.80.024316} {\bibfield  {journal} {\bibinfo  {journal} {Physical Review C - Nuclear Physics}\ }\textbf {\bibinfo {volume} {80}} (\bibinfo {year} {2009}),\ 10.1103/PhysRevC.80.024316}\BibitemShut {NoStop}%
\bibitem [{\citenamefont {Tews}\ \emph {et~al.}(2017)\citenamefont {Tews}, \citenamefont {Lattimer}, \citenamefont {Ohnishi},\ and\ \citenamefont {Kolomeitsev}}]{tews2017symmetry}%
  \BibitemOpen
  \bibfield  {author} {\bibinfo {author} {\bibfnamefont {I.}~\bibnamefont {Tews}}, \bibinfo {author} {\bibfnamefont {J.~M.}\ \bibnamefont {Lattimer}}, \bibinfo {author} {\bibfnamefont {A.}~\bibnamefont {Ohnishi}}, \ and\ \bibinfo {author} {\bibfnamefont {E.~E.}\ \bibnamefont {Kolomeitsev}},\ }\href@noop {} {\bibfield  {journal} {\bibinfo  {journal} {The Astrophysical Journal}\ }\textbf {\bibinfo {volume} {848}},\ \bibinfo {pages} {105} (\bibinfo {year} {2017})}\BibitemShut {NoStop}%
\bibitem [{\citenamefont {Piekarewicz}\ \emph {et~al.}(2012)\citenamefont {Piekarewicz}, \citenamefont {Agrawal}, \citenamefont {Colò}, \citenamefont {Nazarewicz}, \citenamefont {Paar}, \citenamefont {Reinhard}, \citenamefont {Roca-Maza},\ and\ \citenamefont {Vretenar}}]{Piekarewicz2012}%
  \BibitemOpen
  \bibfield  {author} {\bibinfo {author} {\bibfnamefont {J.}~\bibnamefont {Piekarewicz}}, \bibinfo {author} {\bibfnamefont {B.~K.}\ \bibnamefont {Agrawal}}, \bibinfo {author} {\bibfnamefont {G.}~\bibnamefont {Colò}}, \bibinfo {author} {\bibfnamefont {W.}~\bibnamefont {Nazarewicz}}, \bibinfo {author} {\bibfnamefont {N.}~\bibnamefont {Paar}}, \bibinfo {author} {\bibfnamefont {P.~G.}\ \bibnamefont {Reinhard}}, \bibinfo {author} {\bibfnamefont {X.}~\bibnamefont {Roca-Maza}}, \ and\ \bibinfo {author} {\bibfnamefont {D.}~\bibnamefont {Vretenar}},\ }\href {\doibase 10.1103/PhysRevC.85.041302} {\bibfield  {journal} {\bibinfo  {journal} {Physical Review C - Nuclear Physics}\ }\textbf {\bibinfo {volume} {85}} (\bibinfo {year} {2012}),\ 10.1103/PhysRevC.85.041302}\BibitemShut {NoStop}%
\bibitem [{\citenamefont {Kortelainen}\ \emph {et~al.}(2010{\natexlab{a}})\citenamefont {Kortelainen}, \citenamefont {Lesinski}, \citenamefont {Mor{\'e}}, \citenamefont {Nazarewicz}, \citenamefont {Sarich}, \citenamefont {Schunck}, \citenamefont {Stoitsov},\ and\ \citenamefont {Wild}}]{kortelainen2010nuclear}%
  \BibitemOpen
  \bibfield  {author} {\bibinfo {author} {\bibfnamefont {M.}~\bibnamefont {Kortelainen}}, \bibinfo {author} {\bibfnamefont {T.}~\bibnamefont {Lesinski}}, \bibinfo {author} {\bibfnamefont {J.}~\bibnamefont {Mor{\'e}}}, \bibinfo {author} {\bibfnamefont {W.}~\bibnamefont {Nazarewicz}}, \bibinfo {author} {\bibfnamefont {J.}~\bibnamefont {Sarich}}, \bibinfo {author} {\bibfnamefont {N.}~\bibnamefont {Schunck}}, \bibinfo {author} {\bibfnamefont {M.}~\bibnamefont {Stoitsov}}, \ and\ \bibinfo {author} {\bibfnamefont {S.}~\bibnamefont {Wild}},\ }\href@noop {} {\bibfield  {journal} {\bibinfo  {journal} {Physical Review C}\ }\textbf {\bibinfo {volume} {82}},\ \bibinfo {pages} {024313} (\bibinfo {year} {2010}{\natexlab{a}})}\BibitemShut {NoStop}%
\bibitem [{\citenamefont {Drischler}\ \emph {et~al.}(2024)\citenamefont {Drischler}, \citenamefont {Giuliani}, \citenamefont {Bezoui}, \citenamefont {Piekarewicz},\ and\ \citenamefont {Viens}}]{drischler2024bayesian}%
  \BibitemOpen
  \bibfield  {author} {\bibinfo {author} {\bibfnamefont {C.}~\bibnamefont {Drischler}}, \bibinfo {author} {\bibfnamefont {P.}~\bibnamefont {Giuliani}}, \bibinfo {author} {\bibfnamefont {S.}~\bibnamefont {Bezoui}}, \bibinfo {author} {\bibfnamefont {J.}~\bibnamefont {Piekarewicz}}, \ and\ \bibinfo {author} {\bibfnamefont {F.}~\bibnamefont {Viens}},\ }\href@noop {} {\bibfield  {journal} {\bibinfo  {journal} {arXiv preprint arXiv:2405.02748}\ } (\bibinfo {year} {2024})}\BibitemShut {NoStop}%
\bibitem [{\citenamefont {Reinhard}\ \emph {et~al.}(2021)\citenamefont {Reinhard}, \citenamefont {Roca-Maza},\ and\ \citenamefont {Nazarewicz}}]{Reinhard2021}%
  \BibitemOpen
  \bibfield  {author} {\bibinfo {author} {\bibfnamefont {P.~G.}\ \bibnamefont {Reinhard}}, \bibinfo {author} {\bibfnamefont {X.}~\bibnamefont {Roca-Maza}}, \ and\ \bibinfo {author} {\bibfnamefont {W.}~\bibnamefont {Nazarewicz}},\ }\href {\doibase 10.1103/PhysRevLett.127.232501} {\bibfield  {journal} {\bibinfo  {journal} {Physical Review Letters}\ }\textbf {\bibinfo {volume} {127}} (\bibinfo {year} {2021}),\ 10.1103/PhysRevLett.127.232501}\BibitemShut {NoStop}%
\bibitem [{\citenamefont {Y{\"u}ksel}\ and\ \citenamefont {Paar}(2023)}]{yuksel2023implications}%
  \BibitemOpen
  \bibfield  {author} {\bibinfo {author} {\bibfnamefont {E.}~\bibnamefont {Y{\"u}ksel}}\ and\ \bibinfo {author} {\bibfnamefont {N.}~\bibnamefont {Paar}},\ }\href@noop {} {\bibfield  {journal} {\bibinfo  {journal} {Physics Letters B}\ }\textbf {\bibinfo {volume} {836}},\ \bibinfo {pages} {137622} (\bibinfo {year} {2023})}\BibitemShut {NoStop}%
\bibitem [{\citenamefont {Reinhard}\ and\ \citenamefont {Flocard}(1995{\natexlab{a}})}]{Reinhard:1995zz}%
  \BibitemOpen
  \bibfield  {author} {\bibinfo {author} {\bibfnamefont {P.~G.}\ \bibnamefont {Reinhard}}\ and\ \bibinfo {author} {\bibfnamefont {H.}~\bibnamefont {Flocard}},\ }\href {\doibase 10.1016/0375-9474(94)00770-N} {\bibfield  {journal} {\bibinfo  {journal} {Nucl. Phys. A}\ }\textbf {\bibinfo {volume} {584}},\ \bibinfo {pages} {467} (\bibinfo {year} {1995}{\natexlab{a}})}\BibitemShut {NoStop}%
\bibitem [{\citenamefont {Horowitz}\ and\ \citenamefont {Piekarewicz}(2012{\natexlab{a}})}]{Horowitz:2012we}%
  \BibitemOpen
  \bibfield  {author} {\bibinfo {author} {\bibfnamefont {C.~J.}\ \bibnamefont {Horowitz}}\ and\ \bibinfo {author} {\bibfnamefont {J.}~\bibnamefont {Piekarewicz}},\ }\href {\doibase 10.1103/PhysRevC.86.045503} {\bibfield  {journal} {\bibinfo  {journal} {Phys. Rev. C}\ }\textbf {\bibinfo {volume} {86}},\ \bibinfo {pages} {045503} (\bibinfo {year} {2012}{\natexlab{a}})}\BibitemShut {NoStop}%
\bibitem [{\citenamefont {Roca-Maza}\ \emph {et~al.}(2018)\citenamefont {Roca-Maza}, \citenamefont {Col{\`o}},\ and\ \citenamefont {Sagawa}}]{roca2018nuclear}%
  \BibitemOpen
  \bibfield  {author} {\bibinfo {author} {\bibfnamefont {X.}~\bibnamefont {Roca-Maza}}, \bibinfo {author} {\bibfnamefont {G.}~\bibnamefont {Col{\`o}}}, \ and\ \bibinfo {author} {\bibfnamefont {H.}~\bibnamefont {Sagawa}},\ }\href@noop {} {\bibfield  {journal} {\bibinfo  {journal} {Physical review letters}\ }\textbf {\bibinfo {volume} {120}},\ \bibinfo {pages} {202501} (\bibinfo {year} {2018})}\BibitemShut {NoStop}%
\bibitem [{\citenamefont {Sagawa}\ \emph {et~al.}(2024)\citenamefont {Sagawa}, \citenamefont {Naito}, \citenamefont {Roca-Maza}, \citenamefont {Hatsuda} \emph {et~al.}}]{sagawa2024qcd}%
  \BibitemOpen
  \bibfield  {author} {\bibinfo {author} {\bibfnamefont {H.}~\bibnamefont {Sagawa}}, \bibinfo {author} {\bibfnamefont {T.}~\bibnamefont {Naito}}, \bibinfo {author} {\bibfnamefont {X.}~\bibnamefont {Roca-Maza}}, \bibinfo {author} {\bibfnamefont {T.}~\bibnamefont {Hatsuda}},  \emph {et~al.},\ }\href@noop {} {\bibfield  {journal} {\bibinfo  {journal} {Physical Review C}\ }\textbf {\bibinfo {volume} {109}},\ \bibinfo {pages} {L011302} (\bibinfo {year} {2024})}\BibitemShut {NoStop}%
\bibitem [{\citenamefont {Yue}\ \emph {et~al.}(2024)\citenamefont {Yue}, \citenamefont {Zhang},\ and\ \citenamefont {Chen}}]{Yue:2024srj}%
  \BibitemOpen
  \bibfield  {author} {\bibinfo {author} {\bibfnamefont {T.-G.}\ \bibnamefont {Yue}}, \bibinfo {author} {\bibfnamefont {Z.}~\bibnamefont {Zhang}}, \ and\ \bibinfo {author} {\bibfnamefont {L.-W.}\ \bibnamefont {Chen}},\ }\href@noop {} {\  (\bibinfo {year} {2024})},\ \Eprint {http://arxiv.org/abs/2406.03844} {arXiv:2406.03844 [nucl-th]} \BibitemShut {NoStop}%
\bibitem [{\citenamefont {Chabanat}\ \emph {et~al.}(1998{\natexlab{a}})\citenamefont {Chabanat}, \citenamefont {Bonche}, \citenamefont {Haensel}, \citenamefont {Meyer},\ and\ \citenamefont {Schaeffer}}]{Chabanat:1997un}%
  \BibitemOpen
  \bibfield  {author} {\bibinfo {author} {\bibfnamefont {E.}~\bibnamefont {Chabanat}}, \bibinfo {author} {\bibfnamefont {P.}~\bibnamefont {Bonche}}, \bibinfo {author} {\bibfnamefont {P.}~\bibnamefont {Haensel}}, \bibinfo {author} {\bibfnamefont {J.}~\bibnamefont {Meyer}}, \ and\ \bibinfo {author} {\bibfnamefont {R.}~\bibnamefont {Schaeffer}},\ }\href {\doibase 10.1016/S0375-9474(98)00180-8} {\bibfield  {journal} {\bibinfo  {journal} {Nucl. Phys. A}\ }\textbf {\bibinfo {volume} {635}},\ \bibinfo {pages} {231} (\bibinfo {year} {1998}{\natexlab{a}})},\ \bibinfo {note} {[Erratum: Nucl.Phys.A 643, 441--441 (1998)]}\BibitemShut {NoStop}%
\bibitem [{\citenamefont {Reinhard}\ and\ \citenamefont {Flocard}(1995{\natexlab{b}})}]{reinhard1995nuclear}%
  \BibitemOpen
  \bibfield  {author} {\bibinfo {author} {\bibfnamefont {P.-G.}\ \bibnamefont {Reinhard}}\ and\ \bibinfo {author} {\bibfnamefont {H.}~\bibnamefont {Flocard}},\ }\href@noop {} {\bibfield  {journal} {\bibinfo  {journal} {Nuclear Physics A}\ }\textbf {\bibinfo {volume} {584}},\ \bibinfo {pages} {467} (\bibinfo {year} {1995}{\natexlab{b}})}\BibitemShut {NoStop}%
\bibitem [{\citenamefont {Chabanat}\ \emph {et~al.}(1997)\citenamefont {Chabanat}, \citenamefont {Bonche}, \citenamefont {Haensel}, \citenamefont {Meyer},\ and\ \citenamefont {Schaeffer}}]{chabanat1997skyrme}%
  \BibitemOpen
  \bibfield  {author} {\bibinfo {author} {\bibfnamefont {E.}~\bibnamefont {Chabanat}}, \bibinfo {author} {\bibfnamefont {P.}~\bibnamefont {Bonche}}, \bibinfo {author} {\bibfnamefont {P.}~\bibnamefont {Haensel}}, \bibinfo {author} {\bibfnamefont {J.}~\bibnamefont {Meyer}}, \ and\ \bibinfo {author} {\bibfnamefont {R.}~\bibnamefont {Schaeffer}},\ }\href@noop {} {\bibfield  {journal} {\bibinfo  {journal} {Nuclear Physics A}\ }\textbf {\bibinfo {volume} {627}},\ \bibinfo {pages} {710} (\bibinfo {year} {1997})}\BibitemShut {NoStop}%
\bibitem [{\citenamefont {Erler}\ \emph {et~al.}(2013)\citenamefont {Erler}, \citenamefont {Horowitz}, \citenamefont {Nazarewicz}, \citenamefont {Rafalski},\ and\ \citenamefont {Reinhard}}]{erler2013energy}%
  \BibitemOpen
  \bibfield  {author} {\bibinfo {author} {\bibfnamefont {J.}~\bibnamefont {Erler}}, \bibinfo {author} {\bibfnamefont {C.}~\bibnamefont {Horowitz}}, \bibinfo {author} {\bibfnamefont {W.}~\bibnamefont {Nazarewicz}}, \bibinfo {author} {\bibfnamefont {M.}~\bibnamefont {Rafalski}}, \ and\ \bibinfo {author} {\bibfnamefont {P.-G.}\ \bibnamefont {Reinhard}},\ }\href@noop {} {\bibfield  {journal} {\bibinfo  {journal} {Physical Review C}\ }\textbf {\bibinfo {volume} {87}},\ \bibinfo {pages} {044320} (\bibinfo {year} {2013})}\BibitemShut {NoStop}%
\bibitem [{\citenamefont {Kl{\"u}pfel}\ \emph {et~al.}(2009)\citenamefont {Kl{\"u}pfel}, \citenamefont {Reinhard}, \citenamefont {B{\"u}rvenich},\ and\ \citenamefont {Maruhn}}]{klupfel2009variations}%
  \BibitemOpen
  \bibfield  {author} {\bibinfo {author} {\bibfnamefont {P.}~\bibnamefont {Kl{\"u}pfel}}, \bibinfo {author} {\bibfnamefont {P.-G.}\ \bibnamefont {Reinhard}}, \bibinfo {author} {\bibfnamefont {T.}~\bibnamefont {B{\"u}rvenich}}, \ and\ \bibinfo {author} {\bibfnamefont {J.}~\bibnamefont {Maruhn}},\ }\href@noop {} {\bibfield  {journal} {\bibinfo  {journal} {Physical Review C}\ }\textbf {\bibinfo {volume} {79}},\ \bibinfo {pages} {034310} (\bibinfo {year} {2009})}\BibitemShut {NoStop}%
\bibitem [{\citenamefont {Kortelainen}\ \emph {et~al.}(2012)\citenamefont {Kortelainen}, \citenamefont {McDonnell}, \citenamefont {Nazarewicz}, \citenamefont {Reinhard}, \citenamefont {Sarich}, \citenamefont {Schunck}, \citenamefont {Stoitsov},\ and\ \citenamefont {Wild}}]{kortelainen2012nuclear}%
  \BibitemOpen
  \bibfield  {author} {\bibinfo {author} {\bibfnamefont {M.}~\bibnamefont {Kortelainen}}, \bibinfo {author} {\bibfnamefont {J.}~\bibnamefont {McDonnell}}, \bibinfo {author} {\bibfnamefont {W.}~\bibnamefont {Nazarewicz}}, \bibinfo {author} {\bibfnamefont {P.-G.}\ \bibnamefont {Reinhard}}, \bibinfo {author} {\bibfnamefont {J.}~\bibnamefont {Sarich}}, \bibinfo {author} {\bibfnamefont {N.}~\bibnamefont {Schunck}}, \bibinfo {author} {\bibfnamefont {M.}~\bibnamefont {Stoitsov}}, \ and\ \bibinfo {author} {\bibfnamefont {S.}~\bibnamefont {Wild}},\ }\href@noop {} {\bibfield  {journal} {\bibinfo  {journal} {Physical Review C}\ }\textbf {\bibinfo {volume} {85}},\ \bibinfo {pages} {024304} (\bibinfo {year} {2012})}\BibitemShut {NoStop}%
\bibitem [{\citenamefont {Goriely}\ \emph {et~al.}(2016)\citenamefont {Goriely}, \citenamefont {Chamel},\ and\ \citenamefont {Pearson}}]{goriely2016further}%
  \BibitemOpen
  \bibfield  {author} {\bibinfo {author} {\bibfnamefont {S.}~\bibnamefont {Goriely}}, \bibinfo {author} {\bibfnamefont {N.}~\bibnamefont {Chamel}}, \ and\ \bibinfo {author} {\bibfnamefont {J.}~\bibnamefont {Pearson}},\ }\href@noop {} {\bibfield  {journal} {\bibinfo  {journal} {Physical Review C}\ }\textbf {\bibinfo {volume} {93}},\ \bibinfo {pages} {034337} (\bibinfo {year} {2016})}\BibitemShut {NoStop}%
\bibitem [{\citenamefont {Chen}\ and\ \citenamefont {Piekarewicz}(2014)}]{Chen_2014}%
  \BibitemOpen
  \bibfield  {author} {\bibinfo {author} {\bibfnamefont {W.-C.}\ \bibnamefont {Chen}}\ and\ \bibinfo {author} {\bibfnamefont {J.}~\bibnamefont {Piekarewicz}},\ }\href {\doibase 10.1103/PhysRevC.90.044305} {\bibfield  {journal} {\bibinfo  {journal} {Phys. Rev. C}\ }\textbf {\bibinfo {volume} {90}},\ \bibinfo {pages} {044305} (\bibinfo {year} {2014})}\BibitemShut {NoStop}%
\bibitem [{\citenamefont {Fattoyev}\ \emph {et~al.}(2010)\citenamefont {Fattoyev}, \citenamefont {Horowitz}, \citenamefont {Piekarewicz},\ and\ \citenamefont {Shen}}]{fattoyev2010relativistic}%
  \BibitemOpen
  \bibfield  {author} {\bibinfo {author} {\bibfnamefont {F.~J.}\ \bibnamefont {Fattoyev}}, \bibinfo {author} {\bibfnamefont {C.~J.}\ \bibnamefont {Horowitz}}, \bibinfo {author} {\bibfnamefont {J.}~\bibnamefont {Piekarewicz}}, \ and\ \bibinfo {author} {\bibfnamefont {G.}~\bibnamefont {Shen}},\ }\href@noop {} {\bibfield  {journal} {\bibinfo  {journal} {Physical Review C}\ }\textbf {\bibinfo {volume} {82}},\ \bibinfo {pages} {055803} (\bibinfo {year} {2010})}\BibitemShut {NoStop}%
\bibitem [{\citenamefont {Sharma}\ \emph {et~al.}(2023)\citenamefont {Sharma}, \citenamefont {Kumar}, \citenamefont {Kumar}, \citenamefont {Thakur}, \citenamefont {Kumar},\ and\ \citenamefont {Dhiman}}]{sharma2023new}%
  \BibitemOpen
  \bibfield  {author} {\bibinfo {author} {\bibfnamefont {A.}~\bibnamefont {Sharma}}, \bibinfo {author} {\bibfnamefont {M.}~\bibnamefont {Kumar}}, \bibinfo {author} {\bibfnamefont {S.}~\bibnamefont {Kumar}}, \bibinfo {author} {\bibfnamefont {V.}~\bibnamefont {Thakur}}, \bibinfo {author} {\bibfnamefont {R.}~\bibnamefont {Kumar}}, \ and\ \bibinfo {author} {\bibfnamefont {S.~K.}\ \bibnamefont {Dhiman}},\ }\href@noop {} {\bibfield  {journal} {\bibinfo  {journal} {Nuclear Physics A}\ }\textbf {\bibinfo {volume} {1040}},\ \bibinfo {pages} {122762} (\bibinfo {year} {2023})}\BibitemShut {NoStop}%
\bibitem [{\citenamefont {Kumar}\ \emph {et~al.}(2023)\citenamefont {Kumar}, \citenamefont {Kumar}, \citenamefont {Thakur}, \citenamefont {Kumar}, \citenamefont {Agrawal},\ and\ \citenamefont {Dhiman}}]{kumar2023crex}%
  \BibitemOpen
  \bibfield  {author} {\bibinfo {author} {\bibfnamefont {M.}~\bibnamefont {Kumar}}, \bibinfo {author} {\bibfnamefont {S.}~\bibnamefont {Kumar}}, \bibinfo {author} {\bibfnamefont {V.}~\bibnamefont {Thakur}}, \bibinfo {author} {\bibfnamefont {R.}~\bibnamefont {Kumar}}, \bibinfo {author} {\bibfnamefont {B.}~\bibnamefont {Agrawal}}, \ and\ \bibinfo {author} {\bibfnamefont {S.~K.}\ \bibnamefont {Dhiman}},\ }\href@noop {} {\bibfield  {journal} {\bibinfo  {journal} {Physical Review C}\ }\textbf {\bibinfo {volume} {107}},\ \bibinfo {pages} {055801} (\bibinfo {year} {2023})}\BibitemShut {NoStop}%
\bibitem [{\citenamefont {Abbott}\ \emph {et~al.}(2018)\citenamefont {Abbott}, \citenamefont {Abbott}, \citenamefont {Abbott}, \citenamefont {Acernese}, \citenamefont {Ackley}, \citenamefont {Adams}, \citenamefont {Adams}, \citenamefont {Addesso}, \citenamefont {Adhikari}, \citenamefont {Adya} \emph {et~al.}}]{abbott2018gw170817}%
  \BibitemOpen
  \bibfield  {author} {\bibinfo {author} {\bibfnamefont {B.~P.}\ \bibnamefont {Abbott}}, \bibinfo {author} {\bibfnamefont {R.}~\bibnamefont {Abbott}}, \bibinfo {author} {\bibfnamefont {T.}~\bibnamefont {Abbott}}, \bibinfo {author} {\bibfnamefont {F.}~\bibnamefont {Acernese}}, \bibinfo {author} {\bibfnamefont {K.}~\bibnamefont {Ackley}}, \bibinfo {author} {\bibfnamefont {C.}~\bibnamefont {Adams}}, \bibinfo {author} {\bibfnamefont {T.}~\bibnamefont {Adams}}, \bibinfo {author} {\bibfnamefont {P.}~\bibnamefont {Addesso}}, \bibinfo {author} {\bibfnamefont {R.~X.}\ \bibnamefont {Adhikari}}, \bibinfo {author} {\bibfnamefont {V.~B.}\ \bibnamefont {Adya}},  \emph {et~al.},\ }\href@noop {} {\bibfield  {journal} {\bibinfo  {journal} {Physical review letters}\ }\textbf {\bibinfo {volume} {121}},\ \bibinfo {pages} {161101} (\bibinfo {year} {2018})}\BibitemShut {NoStop}%
\bibitem [{\citenamefont {Reinhard}\ \emph {et~al.}(2022{\natexlab{b}})\citenamefont {Reinhard}, \citenamefont {Roca-Maza},\ and\ \citenamefont {Nazarewicz}}]{Reinhard:2022inh}%
  \BibitemOpen
  \bibfield  {author} {\bibinfo {author} {\bibfnamefont {P.-G.}\ \bibnamefont {Reinhard}}, \bibinfo {author} {\bibfnamefont {X.}~\bibnamefont {Roca-Maza}}, \ and\ \bibinfo {author} {\bibfnamefont {W.}~\bibnamefont {Nazarewicz}},\ }\href {\doibase 10.1103/PhysRevLett.129.232501} {\bibfield  {journal} {\bibinfo  {journal} {Phys. Rev. Lett.}\ }\textbf {\bibinfo {volume} {129}},\ \bibinfo {pages} {232501} (\bibinfo {year} {2022}{\natexlab{b}})}\BibitemShut {NoStop}%
\bibitem [{\citenamefont {Kunjipurayil}\ \emph {et~al.}(2025)\citenamefont {Kunjipurayil}, \citenamefont {Salinas},\ and\ \citenamefont {Piekarewicz}}]{Kunjipurayil:2025xss}%
  \BibitemOpen
  \bibfield  {author} {\bibinfo {author} {\bibfnamefont {A.}~\bibnamefont {Kunjipurayil}}, \bibinfo {author} {\bibfnamefont {M.}~\bibnamefont {Salinas}}, \ and\ \bibinfo {author} {\bibfnamefont {J.}~\bibnamefont {Piekarewicz}},\ }\href@noop {} {\  (\bibinfo {year} {2025})},\ \Eprint {http://arxiv.org/abs/2503.07405} {arXiv:2503.07405 [nucl-th]} \BibitemShut {NoStop}%
\bibitem [{\citenamefont {Sfienti}()}]{MREX}%
  \BibitemOpen
  \bibfield  {author} {\bibinfo {author} {\bibfnamefont {C.}~\bibnamefont {Sfienti}},\ }\href {https://gepris.dfg.de/gepris/projekt/454637981} {\bibinfo  {journal} {Research Grants}\ }\BibitemShut {NoStop}%
\bibitem [{\citenamefont {Wang}\ \emph {et~al.}(2012)\citenamefont {Wang}, \citenamefont {Audi}, \citenamefont {Wapstra}, \citenamefont {Kondev}, \citenamefont {MacCormick}, \citenamefont {Xu},\ and\ \citenamefont {Pfeiffer}}]{Wang:2012}%
  \BibitemOpen
\bibfield  {journal} {  }\bibfield  {author} {\bibinfo {author} {\bibfnamefont {M.}~\bibnamefont {Wang}}, \bibinfo {author} {\bibfnamefont {G.}~\bibnamefont {Audi}}, \bibinfo {author} {\bibfnamefont {A.~H.}\ \bibnamefont {Wapstra}}, \bibinfo {author} {\bibfnamefont {F.~G.}\ \bibnamefont {Kondev}}, \bibinfo {author} {\bibfnamefont {M.}~\bibnamefont {MacCormick}}, \bibinfo {author} {\bibfnamefont {X.}~\bibnamefont {Xu}}, \ and\ \bibinfo {author} {\bibfnamefont {B.}~\bibnamefont {Pfeiffer}},\ }\href@noop {} {\bibfield  {journal} {\bibinfo  {journal} {Chinese Phys. C}\ }\textbf {\bibinfo {volume} {36}},\ \bibinfo {pages} {1603} (\bibinfo {year} {2012})}\BibitemShut {NoStop}%
\bibitem [{\citenamefont {Angeli}\ and\ \citenamefont {Marinova}(2013)}]{Angeli:2013}%
  \BibitemOpen
  \bibfield  {author} {\bibinfo {author} {\bibfnamefont {I.}~\bibnamefont {Angeli}}\ and\ \bibinfo {author} {\bibfnamefont {K.}~\bibnamefont {Marinova}},\ }\href@noop {} {\bibfield  {journal} {\bibinfo  {journal} {At. Data Nucl. Data Tables}\ }\textbf {\bibinfo {volume} {99}},\ \bibinfo {pages} {69 } (\bibinfo {year} {2013})}\BibitemShut {NoStop}%
\bibitem [{\citenamefont {Lattimer}(2023)}]{lattimer2023constraints}%
  \BibitemOpen
  \bibfield  {author} {\bibinfo {author} {\bibfnamefont {J.~M.}\ \bibnamefont {Lattimer}},\ }\href@noop {} {\bibfield  {journal} {\bibinfo  {journal} {Particles}\ }\textbf {\bibinfo {volume} {6}},\ \bibinfo {pages} {30} (\bibinfo {year} {2023})}\BibitemShut {NoStop}%
\bibitem [{\citenamefont {Adhikari}\ \emph {et~al.}(2021{\natexlab{b}})\citenamefont {Adhikari}, \citenamefont {Albataineh}, \citenamefont {Androic}, \citenamefont {Aniol}, \citenamefont {Armstrong}, \citenamefont {Averett}, \citenamefont {Gayoso}, \citenamefont {Barcus}, \citenamefont {Bellini}, \citenamefont {Beminiwattha} \emph {et~al.}}]{adhikari2021accurate}%
  \BibitemOpen
  \bibfield  {author} {\bibinfo {author} {\bibfnamefont {D.}~\bibnamefont {Adhikari}}, \bibinfo {author} {\bibfnamefont {H.}~\bibnamefont {Albataineh}}, \bibinfo {author} {\bibfnamefont {D.}~\bibnamefont {Androic}}, \bibinfo {author} {\bibfnamefont {K.}~\bibnamefont {Aniol}}, \bibinfo {author} {\bibfnamefont {D.}~\bibnamefont {Armstrong}}, \bibinfo {author} {\bibfnamefont {T.}~\bibnamefont {Averett}}, \bibinfo {author} {\bibfnamefont {C.~A.}\ \bibnamefont {Gayoso}}, \bibinfo {author} {\bibfnamefont {S.}~\bibnamefont {Barcus}}, \bibinfo {author} {\bibfnamefont {V.}~\bibnamefont {Bellini}}, \bibinfo {author} {\bibfnamefont {R.}~\bibnamefont {Beminiwattha}},  \emph {et~al.},\ }\href@noop {} {\bibfield  {journal} {\bibinfo  {journal} {Physical review letters}\ }\textbf {\bibinfo {volume} {126}},\ \bibinfo {pages} {172502} (\bibinfo {year} {2021}{\natexlab{b}})}\BibitemShut {NoStop}%
\bibitem [{\citenamefont {Adhikari}\ \emph {et~al.}(2022{\natexlab{b}})\citenamefont {Adhikari}, \citenamefont {Albataineh}, \citenamefont {Androic}, \citenamefont {Aniol}, \citenamefont {Armstrong}, \citenamefont {Averett}, \citenamefont {Gayoso}, \citenamefont {Barcus}, \citenamefont {Bellini}, \citenamefont {Beminiwattha} \emph {et~al.}}]{adhikari2022precision}%
  \BibitemOpen
  \bibfield  {author} {\bibinfo {author} {\bibfnamefont {D.}~\bibnamefont {Adhikari}}, \bibinfo {author} {\bibfnamefont {H.}~\bibnamefont {Albataineh}}, \bibinfo {author} {\bibfnamefont {D.}~\bibnamefont {Androic}}, \bibinfo {author} {\bibfnamefont {K.}~\bibnamefont {Aniol}}, \bibinfo {author} {\bibfnamefont {D.}~\bibnamefont {Armstrong}}, \bibinfo {author} {\bibfnamefont {T.}~\bibnamefont {Averett}}, \bibinfo {author} {\bibfnamefont {C.~A.}\ \bibnamefont {Gayoso}}, \bibinfo {author} {\bibfnamefont {S.}~\bibnamefont {Barcus}}, \bibinfo {author} {\bibfnamefont {V.}~\bibnamefont {Bellini}}, \bibinfo {author} {\bibfnamefont {R.}~\bibnamefont {Beminiwattha}},  \emph {et~al.},\ }\href {\doibase 10.1103/PhysRevLett.129.042501} {\bibfield  {journal} {\bibinfo  {journal} {Physical Review Letters}\ }\textbf {\bibinfo {volume} {129}},\ \bibinfo {pages} {042501} (\bibinfo {year} {2022}{\natexlab{b}})}\BibitemShut {NoStop}%
\bibitem [{\citenamefont {Lalazissis}\ \emph {et~al.}(1997{\natexlab{b}})\citenamefont {Lalazissis}, \citenamefont {K{\"o}nig},\ and\ \citenamefont {Ring}}]{lalazissis1997new}%
  \BibitemOpen
  \bibfield  {author} {\bibinfo {author} {\bibfnamefont {G.}~\bibnamefont {Lalazissis}}, \bibinfo {author} {\bibfnamefont {J.}~\bibnamefont {K{\"o}nig}}, \ and\ \bibinfo {author} {\bibfnamefont {P.}~\bibnamefont {Ring}},\ }\href@noop {} {\bibfield  {journal} {\bibinfo  {journal} {Physical Review C}\ }\textbf {\bibinfo {volume} {55}},\ \bibinfo {pages} {540} (\bibinfo {year} {1997}{\natexlab{b}})}\BibitemShut {NoStop}%
\bibitem [{\citenamefont {Kumar}\ \emph {et~al.}(2018)\citenamefont {Kumar}, \citenamefont {Patra},\ and\ \citenamefont {Agrawal}}]{IOPB}%
  \BibitemOpen
  \bibfield  {author} {\bibinfo {author} {\bibfnamefont {B.}~\bibnamefont {Kumar}}, \bibinfo {author} {\bibfnamefont {S.~K.}\ \bibnamefont {Patra}}, \ and\ \bibinfo {author} {\bibfnamefont {B.~K.}\ \bibnamefont {Agrawal}},\ }\href {\doibase 10.1103/PhysRevC.97.045806} {\bibfield  {journal} {\bibinfo  {journal} {Phys. Rev. C}\ }\textbf {\bibinfo {volume} {97}},\ \bibinfo {pages} {045806} (\bibinfo {year} {2018})}\BibitemShut {NoStop}%
\bibitem [{\citenamefont {Fattoyev}\ \emph {et~al.}(2020)\citenamefont {Fattoyev}, \citenamefont {Horowitz}, \citenamefont {Piekarewicz},\ and\ \citenamefont {Reed}}]{fattoyev2020gw190814}%
  \BibitemOpen
  \bibfield  {author} {\bibinfo {author} {\bibfnamefont {F.}~\bibnamefont {Fattoyev}}, \bibinfo {author} {\bibfnamefont {C.}~\bibnamefont {Horowitz}}, \bibinfo {author} {\bibfnamefont {J.}~\bibnamefont {Piekarewicz}}, \ and\ \bibinfo {author} {\bibfnamefont {B.}~\bibnamefont {Reed}},\ }\href@noop {} {\bibfield  {journal} {\bibinfo  {journal} {Physical Review C}\ }\textbf {\bibinfo {volume} {102}},\ \bibinfo {pages} {065805} (\bibinfo {year} {2020})}\BibitemShut {NoStop}%
\bibitem [{\citenamefont {Lattimer}\ and\ \citenamefont {Swesty}(1991)}]{lattimer1991generalized}%
  \BibitemOpen
  \bibfield  {author} {\bibinfo {author} {\bibfnamefont {J.~M.}\ \bibnamefont {Lattimer}}\ and\ \bibinfo {author} {\bibfnamefont {F.~D.}\ \bibnamefont {Swesty}},\ }\href@noop {} {\bibfield  {journal} {\bibinfo  {journal} {Nuclear Physics A}\ }\textbf {\bibinfo {volume} {535}},\ \bibinfo {pages} {331} (\bibinfo {year} {1991})}\BibitemShut {NoStop}%
\bibitem [{\citenamefont {Kumar}\ \emph {et~al.}(2017)\citenamefont {Kumar}, \citenamefont {Singh}, \citenamefont {Agrawal},\ and\ \citenamefont {Patra}}]{Kumar_2017}%
  \BibitemOpen
  \bibfield  {author} {\bibinfo {author} {\bibfnamefont {B.}~\bibnamefont {Kumar}}, \bibinfo {author} {\bibfnamefont {S.}~\bibnamefont {Singh}}, \bibinfo {author} {\bibfnamefont {B.}~\bibnamefont {Agrawal}}, \ and\ \bibinfo {author} {\bibfnamefont {S.}~\bibnamefont {Patra}},\ }\href {\doibase 10.1016/j.nuclphysa.2017.07.001} {\bibfield  {journal} {\bibinfo  {journal} {Nuclear Physics A}\ }\textbf {\bibinfo {volume} {966}},\ \bibinfo {pages} {197} (\bibinfo {year} {2017})}\BibitemShut {NoStop}%
\bibitem [{\citenamefont {Yang}\ and\ \citenamefont {Piekarewicz}(2020)}]{Yang_2020}%
  \BibitemOpen
  \bibfield  {author} {\bibinfo {author} {\bibfnamefont {J.}~\bibnamefont {Yang}}\ and\ \bibinfo {author} {\bibfnamefont {J.}~\bibnamefont {Piekarewicz}},\ }\href {\doibase 10.1146/annurev-nucl-101918-023608} {\bibfield  {journal} {\bibinfo  {journal} {Annual Review of Nuclear and Particle Science}\ }\textbf {\bibinfo {volume} {70}},\ \bibinfo {pages} {21} (\bibinfo {year} {2020})}\BibitemShut {NoStop}%
\bibitem [{\citenamefont {Walecka}(1974)}]{Waleck_1974}%
  \BibitemOpen
  \bibfield  {author} {\bibinfo {author} {\bibfnamefont {J.}~\bibnamefont {Walecka}},\ }\href {\doibase https://doi.org/10.1016/0003-4916(74)90208-5} {\bibfield  {journal} {\bibinfo  {journal} {Annals of Physics}\ }\textbf {\bibinfo {volume} {83}},\ \bibinfo {pages} {491} (\bibinfo {year} {1974})}\BibitemShut {NoStop}%
\bibitem [{\citenamefont {Serot}\ and\ \citenamefont {Walecka}(1986)}]{Serot:1984ey}%
  \BibitemOpen
  \bibfield  {author} {\bibinfo {author} {\bibfnamefont {B.~D.}\ \bibnamefont {Serot}}\ and\ \bibinfo {author} {\bibfnamefont {J.~D.}\ \bibnamefont {Walecka}},\ }\href@noop {} {\bibfield  {journal} {\bibinfo  {journal} {Adv. Nucl. Phys.}\ }\textbf {\bibinfo {volume} {16}},\ \bibinfo {pages} {1} (\bibinfo {year} {1986})}\BibitemShut {NoStop}%
\bibitem [{\citenamefont {Serot}\ and\ \citenamefont {Walecka}(1997)}]{Serot_1997}%
  \BibitemOpen
  \bibfield  {author} {\bibinfo {author} {\bibfnamefont {B.~D.}\ \bibnamefont {Serot}}\ and\ \bibinfo {author} {\bibfnamefont {J.~D.}\ \bibnamefont {Walecka}},\ }\href {\doibase 10.1142/s0218301397000299} {\bibfield  {journal} {\bibinfo  {journal} {International Journal of Modern Physics E}\ }\textbf {\bibinfo {volume} {06}},\ \bibinfo {pages} {515} (\bibinfo {year} {1997})}\BibitemShut {NoStop}%
\bibitem [{\citenamefont {Singh}\ \emph {et~al.}(2014)\citenamefont {Singh}, \citenamefont {Biswal}, \citenamefont {Bhuyan},\ and\ \citenamefont {Patra}}]{singh2014effects}%
  \BibitemOpen
  \bibfield  {author} {\bibinfo {author} {\bibfnamefont {S.~K.}\ \bibnamefont {Singh}}, \bibinfo {author} {\bibfnamefont {S.}~\bibnamefont {Biswal}}, \bibinfo {author} {\bibfnamefont {M.}~\bibnamefont {Bhuyan}}, \ and\ \bibinfo {author} {\bibfnamefont {S.}~\bibnamefont {Patra}},\ }\href@noop {} {\bibfield  {journal} {\bibinfo  {journal} {Physical Review C}\ }\textbf {\bibinfo {volume} {89}},\ \bibinfo {pages} {044001} (\bibinfo {year} {2014})}\BibitemShut {NoStop}%
\bibitem [{\citenamefont {Horowitz}\ and\ \citenamefont {Serot}(1981)}]{horowitz1981self}%
  \BibitemOpen
  \bibfield  {author} {\bibinfo {author} {\bibfnamefont {C.}~\bibnamefont {Horowitz}}\ and\ \bibinfo {author} {\bibfnamefont {B.~D.}\ \bibnamefont {Serot}},\ }\href@noop {} {\bibfield  {journal} {\bibinfo  {journal} {Nuclear Physics A}\ }\textbf {\bibinfo {volume} {368}},\ \bibinfo {pages} {503} (\bibinfo {year} {1981})}\BibitemShut {NoStop}%
\bibitem [{\citenamefont {Reinhard}(1989)}]{reinhard1989relativistic}%
  \BibitemOpen
  \bibfield  {author} {\bibinfo {author} {\bibfnamefont {P.-G.}\ \bibnamefont {Reinhard}},\ }\href@noop {} {\bibfield  {journal} {\bibinfo  {journal} {Reports on Progress in Physics}\ }\textbf {\bibinfo {volume} {52}},\ \bibinfo {pages} {439} (\bibinfo {year} {1989})}\BibitemShut {NoStop}%
\bibitem [{\citenamefont {Ring}(1996)}]{ring1996relativistic}%
  \BibitemOpen
  \bibfield  {author} {\bibinfo {author} {\bibfnamefont {P.}~\bibnamefont {Ring}},\ }\href@noop {} {\bibfield  {journal} {\bibinfo  {journal} {Progress in Particle and Nuclear Physics}\ }\textbf {\bibinfo {volume} {37}},\ \bibinfo {pages} {193} (\bibinfo {year} {1996})}\BibitemShut {NoStop}%
\bibitem [{\citenamefont {Ebran}\ \emph {et~al.}(2016)\citenamefont {Ebran}, \citenamefont {Mutschler}, \citenamefont {Khan},\ and\ \citenamefont {Vretenar}}]{ebran2016spin}%
  \BibitemOpen
  \bibfield  {author} {\bibinfo {author} {\bibfnamefont {J.-P.}\ \bibnamefont {Ebran}}, \bibinfo {author} {\bibfnamefont {A.}~\bibnamefont {Mutschler}}, \bibinfo {author} {\bibfnamefont {E.}~\bibnamefont {Khan}}, \ and\ \bibinfo {author} {\bibfnamefont {D.}~\bibnamefont {Vretenar}},\ }\href@noop {} {\bibfield  {journal} {\bibinfo  {journal} {Physical Review C}\ }\textbf {\bibinfo {volume} {94}},\ \bibinfo {pages} {024304} (\bibinfo {year} {2016})}\BibitemShut {NoStop}%
\bibitem [{\citenamefont {Sharma}\ \emph {et~al.}(1995)\citenamefont {Sharma}, \citenamefont {Lalazissis}, \citenamefont {K{\"o}nig},\ and\ \citenamefont {Ring}}]{sharma1995isospin}%
  \BibitemOpen
  \bibfield  {author} {\bibinfo {author} {\bibfnamefont {M.}~\bibnamefont {Sharma}}, \bibinfo {author} {\bibfnamefont {G.}~\bibnamefont {Lalazissis}}, \bibinfo {author} {\bibfnamefont {J.}~\bibnamefont {K{\"o}nig}}, \ and\ \bibinfo {author} {\bibfnamefont {P.}~\bibnamefont {Ring}},\ }\href@noop {} {\bibfield  {journal} {\bibinfo  {journal} {Physical review letters}\ }\textbf {\bibinfo {volume} {74}},\ \bibinfo {pages} {3744} (\bibinfo {year} {1995})}\BibitemShut {NoStop}%
\bibitem [{\citenamefont {Lalazissis}\ \emph {et~al.}(1998)\citenamefont {Lalazissis}, \citenamefont {Vretenar}, \citenamefont {P{\"o}schl},\ and\ \citenamefont {Ring}}]{lalazissis1998reduction}%
  \BibitemOpen
  \bibfield  {author} {\bibinfo {author} {\bibfnamefont {G.}~\bibnamefont {Lalazissis}}, \bibinfo {author} {\bibfnamefont {D.}~\bibnamefont {Vretenar}}, \bibinfo {author} {\bibfnamefont {W.}~\bibnamefont {P{\"o}schl}}, \ and\ \bibinfo {author} {\bibfnamefont {P.}~\bibnamefont {Ring}},\ }\href@noop {} {\bibfield  {journal} {\bibinfo  {journal} {Physics Letters B}\ }\textbf {\bibinfo {volume} {418}},\ \bibinfo {pages} {7} (\bibinfo {year} {1998})}\BibitemShut {NoStop}%
\bibitem [{\citenamefont {Pearson}(2001)}]{pearson2001skyrme}%
  \BibitemOpen
  \bibfield  {author} {\bibinfo {author} {\bibfnamefont {J.}~\bibnamefont {Pearson}},\ }\href@noop {} {\bibfield  {journal} {\bibinfo  {journal} {Physics Letters B}\ }\textbf {\bibinfo {volume} {513}},\ \bibinfo {pages} {319} (\bibinfo {year} {2001})}\BibitemShut {NoStop}%
\bibitem [{\citenamefont {Horowitz}\ and\ \citenamefont {Piekarewicz}(2012{\natexlab{b}})}]{horowitz2012impact}%
  \BibitemOpen
  \bibfield  {author} {\bibinfo {author} {\bibfnamefont {C.}~\bibnamefont {Horowitz}}\ and\ \bibinfo {author} {\bibfnamefont {J.}~\bibnamefont {Piekarewicz}},\ }\href@noop {} {\bibfield  {journal} {\bibinfo  {journal} {Physical Review C}\ }\textbf {\bibinfo {volume} {86}},\ \bibinfo {pages} {045503} (\bibinfo {year} {2012}{\natexlab{b}})}\BibitemShut {NoStop}%
\bibitem [{\citenamefont {Li}\ \emph {et~al.}(2018)\citenamefont {Li}, \citenamefont {Cai}, \citenamefont {Chen},\ and\ \citenamefont {Xu}}]{li2018nucleon}%
  \BibitemOpen
  \bibfield  {author} {\bibinfo {author} {\bibfnamefont {B.-A.}\ \bibnamefont {Li}}, \bibinfo {author} {\bibfnamefont {B.-J.}\ \bibnamefont {Cai}}, \bibinfo {author} {\bibfnamefont {L.-W.}\ \bibnamefont {Chen}}, \ and\ \bibinfo {author} {\bibfnamefont {J.}~\bibnamefont {Xu}},\ }\href@noop {} {\bibfield  {journal} {\bibinfo  {journal} {Progress in Particle and Nuclear Physics}\ }\textbf {\bibinfo {volume} {99}},\ \bibinfo {pages} {29} (\bibinfo {year} {2018})}\BibitemShut {NoStop}%
\bibitem [{\citenamefont {Xu}\ \emph {et~al.}(2021)\citenamefont {Xu}, \citenamefont {Zhang},\ and\ \citenamefont {Li}}]{xu2021bayesian}%
  \BibitemOpen
  \bibfield  {author} {\bibinfo {author} {\bibfnamefont {J.}~\bibnamefont {Xu}}, \bibinfo {author} {\bibfnamefont {Z.}~\bibnamefont {Zhang}}, \ and\ \bibinfo {author} {\bibfnamefont {B.-A.}\ \bibnamefont {Li}},\ }\href@noop {} {\bibfield  {journal} {\bibinfo  {journal} {Physical Review C}\ }\textbf {\bibinfo {volume} {104}},\ \bibinfo {pages} {054324} (\bibinfo {year} {2021})}\BibitemShut {NoStop}%
\bibitem [{\citenamefont {Sun}\ \emph {et~al.}(2024)\citenamefont {Sun}, \citenamefont {Bhattiprolu},\ and\ \citenamefont {Lattimer}}]{sun2024compiled}%
  \BibitemOpen
  \bibfield  {author} {\bibinfo {author} {\bibfnamefont {B.}~\bibnamefont {Sun}}, \bibinfo {author} {\bibfnamefont {S.}~\bibnamefont {Bhattiprolu}}, \ and\ \bibinfo {author} {\bibfnamefont {J.~M.}\ \bibnamefont {Lattimer}},\ }\href@noop {} {\bibfield  {journal} {\bibinfo  {journal} {Physical Review C}\ }\textbf {\bibinfo {volume} {109}},\ \bibinfo {pages} {055801} (\bibinfo {year} {2024})}\BibitemShut {NoStop}%
\bibitem [{\citenamefont {Tanabashi}\ \emph {et~al.}(2018)\citenamefont {Tanabashi}, \citenamefont {Hagiwara}, \citenamefont {Hikasa}, \citenamefont {Nakamura}, \citenamefont {Sumino}, \citenamefont {Takahashi}, \citenamefont {Tanaka}, \citenamefont {Agashe}, \citenamefont {Aielli}, \citenamefont {Amsler} \emph {et~al.}}]{tanabashi2018review}%
  \BibitemOpen
  \bibfield  {author} {\bibinfo {author} {\bibfnamefont {M.}~\bibnamefont {Tanabashi}}, \bibinfo {author} {\bibfnamefont {K.}~\bibnamefont {Hagiwara}}, \bibinfo {author} {\bibfnamefont {K.}~\bibnamefont {Hikasa}}, \bibinfo {author} {\bibfnamefont {K.}~\bibnamefont {Nakamura}}, \bibinfo {author} {\bibfnamefont {Y.}~\bibnamefont {Sumino}}, \bibinfo {author} {\bibfnamefont {F.}~\bibnamefont {Takahashi}}, \bibinfo {author} {\bibfnamefont {J.}~\bibnamefont {Tanaka}}, \bibinfo {author} {\bibfnamefont {K.}~\bibnamefont {Agashe}}, \bibinfo {author} {\bibfnamefont {G.}~\bibnamefont {Aielli}}, \bibinfo {author} {\bibfnamefont {C.}~\bibnamefont {Amsler}},  \emph {et~al.},\ }\href@noop {} {\bibfield  {journal} {\bibinfo  {journal} {Physical Review D}\ }\textbf {\bibinfo {volume} {98}},\ \bibinfo {pages} {1} (\bibinfo {year} {2018})}\BibitemShut {NoStop}%
\bibitem [{\citenamefont {Huang}\ \emph {et~al.}(2024)\citenamefont {Huang}, \citenamefont {Raaijmakers}, \citenamefont {Watts}, \citenamefont {Tolos},\ and\ \citenamefont {Provid{\^e}ncia}}]{huang2024constraining}%
  \BibitemOpen
  \bibfield  {author} {\bibinfo {author} {\bibfnamefont {C.}~\bibnamefont {Huang}}, \bibinfo {author} {\bibfnamefont {G.}~\bibnamefont {Raaijmakers}}, \bibinfo {author} {\bibfnamefont {A.~L.}\ \bibnamefont {Watts}}, \bibinfo {author} {\bibfnamefont {L.}~\bibnamefont {Tolos}}, \ and\ \bibinfo {author} {\bibfnamefont {C.}~\bibnamefont {Provid{\^e}ncia}},\ }\href@noop {} {\bibfield  {journal} {\bibinfo  {journal} {Monthly Notices of the Royal Astronomical Society}\ }\textbf {\bibinfo {volume} {529}},\ \bibinfo {pages} {4650} (\bibinfo {year} {2024})}\BibitemShut {NoStop}%
\bibitem [{\citenamefont {van Giai}\ and\ \citenamefont {Sagawa}(1981)}]{SGII}%
  \BibitemOpen
  \bibfield  {author} {\bibinfo {author} {\bibfnamefont {N.}~\bibnamefont {van Giai}}\ and\ \bibinfo {author} {\bibfnamefont {H.}~\bibnamefont {Sagawa}},\ }\href {\doibase 10.1016/0370-2693(81)90646-8} {\bibfield  {journal} {\bibinfo  {journal} {Phys. Lett. B}\ }\textbf {\bibinfo {volume} {106}},\ \bibinfo {pages} {379} (\bibinfo {year} {1981})}\BibitemShut {NoStop}%
\bibitem [{\citenamefont {Steiner}\ \emph {et~al.}(2005{\natexlab{b}})\citenamefont {Steiner}, \citenamefont {Prakash}, \citenamefont {Lattimer},\ and\ \citenamefont {Ellis}}]{NRAPR}%
  \BibitemOpen
  \bibfield  {author} {\bibinfo {author} {\bibfnamefont {A.~W.}\ \bibnamefont {Steiner}}, \bibinfo {author} {\bibfnamefont {M.}~\bibnamefont {Prakash}}, \bibinfo {author} {\bibfnamefont {J.~M.}\ \bibnamefont {Lattimer}}, \ and\ \bibinfo {author} {\bibfnamefont {P.~J.}\ \bibnamefont {Ellis}},\ }\href {\doibase 10.1016/j.physrep.2005.02.004} {\bibfield  {journal} {\bibinfo  {journal} {Phys. Rept.}\ }\textbf {\bibinfo {volume} {411}},\ \bibinfo {pages} {325} (\bibinfo {year} {2005}{\natexlab{b}})},\ \Eprint {http://arxiv.org/abs/nucl-th/0410066} {arXiv:nucl-th/0410066} \BibitemShut {NoStop}%
\bibitem [{\citenamefont {Kortelainen}\ \emph {et~al.}(2014)\citenamefont {Kortelainen}, \citenamefont {McDonnell}, \citenamefont {Nazarewicz}, \citenamefont {Olsen}, \citenamefont {Reinhard}, \citenamefont {Sarich}, \citenamefont {Schunck}, \citenamefont {Wild}, \citenamefont {Davesne}, \citenamefont {Erler},\ and\ \citenamefont {Pastore}}]{UNEDF0}%
  \BibitemOpen
  \bibfield  {author} {\bibinfo {author} {\bibfnamefont {M.}~\bibnamefont {Kortelainen}}, \bibinfo {author} {\bibfnamefont {J.}~\bibnamefont {McDonnell}}, \bibinfo {author} {\bibfnamefont {W.}~\bibnamefont {Nazarewicz}}, \bibinfo {author} {\bibfnamefont {E.}~\bibnamefont {Olsen}}, \bibinfo {author} {\bibfnamefont {P.-G.}\ \bibnamefont {Reinhard}}, \bibinfo {author} {\bibfnamefont {J.}~\bibnamefont {Sarich}}, \bibinfo {author} {\bibfnamefont {N.}~\bibnamefont {Schunck}}, \bibinfo {author} {\bibfnamefont {S.~M.}\ \bibnamefont {Wild}}, \bibinfo {author} {\bibfnamefont {D.}~\bibnamefont {Davesne}}, \bibinfo {author} {\bibfnamefont {J.}~\bibnamefont {Erler}}, \ and\ \bibinfo {author} {\bibfnamefont {A.}~\bibnamefont {Pastore}},\ }\href {\doibase 10.1103/PhysRevC.89.054314} {\bibfield  {journal} {\bibinfo  {journal} {Phys. Rev. C}\ }\textbf {\bibinfo {volume} {89}},\ \bibinfo {pages} {054314} (\bibinfo {year} {2014})}\BibitemShut {NoStop}%
\bibitem [{\citenamefont {Kortelainen}\ \emph {et~al.}(2010{\natexlab{b}})\citenamefont {Kortelainen}, \citenamefont {Lesinski}, \citenamefont {Mor\'e}, \citenamefont {Nazarewicz}, \citenamefont {Sarich}, \citenamefont {Schunck}, \citenamefont {Stoitsov},\ and\ \citenamefont {Wild}}]{UNEDFII}%
  \BibitemOpen
  \bibfield  {author} {\bibinfo {author} {\bibfnamefont {M.}~\bibnamefont {Kortelainen}}, \bibinfo {author} {\bibfnamefont {T.}~\bibnamefont {Lesinski}}, \bibinfo {author} {\bibfnamefont {J.}~\bibnamefont {Mor\'e}}, \bibinfo {author} {\bibfnamefont {W.}~\bibnamefont {Nazarewicz}}, \bibinfo {author} {\bibfnamefont {J.}~\bibnamefont {Sarich}}, \bibinfo {author} {\bibfnamefont {N.}~\bibnamefont {Schunck}}, \bibinfo {author} {\bibfnamefont {M.~V.}\ \bibnamefont {Stoitsov}}, \ and\ \bibinfo {author} {\bibfnamefont {S.}~\bibnamefont {Wild}},\ }\href {\doibase 10.1103/PhysRevC.82.024313} {\bibfield  {journal} {\bibinfo  {journal} {Phys. Rev. C}\ }\textbf {\bibinfo {volume} {82}},\ \bibinfo {pages} {024313} (\bibinfo {year} {2010}{\natexlab{b}})}\BibitemShut {NoStop}%
\bibitem [{\citenamefont {Chabanat}\ \emph {et~al.}(1998{\natexlab{b}})\citenamefont {Chabanat}, \citenamefont {Bonche}, \citenamefont {Haensel}, \citenamefont {Meyer},\ and\ \citenamefont {Schaeffer}}]{SLy4}%
  \BibitemOpen
  \bibfield  {author} {\bibinfo {author} {\bibfnamefont {E.}~\bibnamefont {Chabanat}}, \bibinfo {author} {\bibfnamefont {P.}~\bibnamefont {Bonche}}, \bibinfo {author} {\bibfnamefont {P.}~\bibnamefont {Haensel}}, \bibinfo {author} {\bibfnamefont {J.}~\bibnamefont {Meyer}}, \ and\ \bibinfo {author} {\bibfnamefont {R.}~\bibnamefont {Schaeffer}},\ }\href {\doibase 10.1016/S0375-9474(98)00180-8} {\bibfield  {journal} {\bibinfo  {journal} {Nucl. Phys. A}\ }\textbf {\bibinfo {volume} {635}},\ \bibinfo {pages} {231} (\bibinfo {year} {1998}{\natexlab{b}})},\ \bibinfo {note} {[Erratum: Nucl.Phys.A 643, 441--441 (1998)]}\BibitemShut {NoStop}%
\bibitem [{\citenamefont {Kl\"upfel}\ \emph {et~al.}(2009)\citenamefont {Kl\"upfel}, \citenamefont {Reinhard}, \citenamefont {B\"urvenich},\ and\ \citenamefont {Maruhn}}]{SV-min}%
  \BibitemOpen
  \bibfield  {author} {\bibinfo {author} {\bibfnamefont {P.}~\bibnamefont {Kl\"upfel}}, \bibinfo {author} {\bibfnamefont {P.-G.}\ \bibnamefont {Reinhard}}, \bibinfo {author} {\bibfnamefont {T.~J.}\ \bibnamefont {B\"urvenich}}, \ and\ \bibinfo {author} {\bibfnamefont {J.~A.}\ \bibnamefont {Maruhn}},\ }\href {\doibase 10.1103/PhysRevC.79.034310} {\bibfield  {journal} {\bibinfo  {journal} {Phys. Rev. C}\ }\textbf {\bibinfo {volume} {79}},\ \bibinfo {pages} {034310} (\bibinfo {year} {2009})}\BibitemShut {NoStop}%
\bibitem [{\citenamefont {Reinhard}\ \emph {et~al.}(1999)\citenamefont {Reinhard}, \citenamefont {Dean}, \citenamefont {Nazarewicz}, \citenamefont {Dobaczewski}, \citenamefont {Maruhn},\ and\ \citenamefont {Strayer}}]{Sko}%
  \BibitemOpen
  \bibfield  {author} {\bibinfo {author} {\bibfnamefont {P.-G.}\ \bibnamefont {Reinhard}}, \bibinfo {author} {\bibfnamefont {D.~J.}\ \bibnamefont {Dean}}, \bibinfo {author} {\bibfnamefont {W.}~\bibnamefont {Nazarewicz}}, \bibinfo {author} {\bibfnamefont {J.}~\bibnamefont {Dobaczewski}}, \bibinfo {author} {\bibfnamefont {J.~A.}\ \bibnamefont {Maruhn}}, \ and\ \bibinfo {author} {\bibfnamefont {M.~R.}\ \bibnamefont {Strayer}},\ }\href {\doibase 10.1103/PhysRevC.60.014316} {\bibfield  {journal} {\bibinfo  {journal} {Phys. Rev. C}\ }\textbf {\bibinfo {volume} {60}},\ \bibinfo {pages} {014316} (\bibinfo {year} {1999})}\BibitemShut {NoStop}%
\bibitem [{\citenamefont {Danielewicz}\ and\ \citenamefont {Lee}(2009)}]{Skop}%
  \BibitemOpen
  \bibfield  {author} {\bibinfo {author} {\bibfnamefont {P.}~\bibnamefont {Danielewicz}}\ and\ \bibinfo {author} {\bibfnamefont {J.}~\bibnamefont {Lee}},\ }\href {\doibase 10.1016/j.nuclphysa.2008.11.007} {\bibfield  {journal} {\bibinfo  {journal} {Nucl. Phys. A}\ }\textbf {\bibinfo {volume} {818}},\ \bibinfo {pages} {36} (\bibinfo {year} {2009})},\ \Eprint {http://arxiv.org/abs/0807.3743} {arXiv:0807.3743 [nucl-th]} \BibitemShut {NoStop}%
\bibitem [{\citenamefont {Chamel}\ \emph {et~al.}(2008)\citenamefont {Chamel}, \citenamefont {Goriely},\ and\ \citenamefont {Pearson}}]{Bsk16}%
  \BibitemOpen
  \bibfield  {author} {\bibinfo {author} {\bibfnamefont {N.}~\bibnamefont {Chamel}}, \bibinfo {author} {\bibfnamefont {S.}~\bibnamefont {Goriely}}, \ and\ \bibinfo {author} {\bibfnamefont {J.~M.}\ \bibnamefont {Pearson}},\ }\href {\doibase 10.1016/j.nuclphysa.2008.08.015} {\bibfield  {journal} {\bibinfo  {journal} {Nucl. Phys. A}\ }\textbf {\bibinfo {volume} {812}},\ \bibinfo {pages} {72} (\bibinfo {year} {2008})},\ \Eprint {http://arxiv.org/abs/0809.0447} {arXiv:0809.0447 [nucl-th]} \BibitemShut {NoStop}%
\bibitem [{\citenamefont {Agrawal}\ \emph {et~al.}(2005)\citenamefont {Agrawal}, \citenamefont {Shlomo},\ and\ \citenamefont {Au}}]{Kde0}%
  \BibitemOpen
  \bibfield  {author} {\bibinfo {author} {\bibfnamefont {B.~K.}\ \bibnamefont {Agrawal}}, \bibinfo {author} {\bibfnamefont {S.}~\bibnamefont {Shlomo}}, \ and\ \bibinfo {author} {\bibfnamefont {V.~K.}\ \bibnamefont {Au}},\ }\href {\doibase 10.1103/PhysRevC.72.014310} {\bibfield  {journal} {\bibinfo  {journal} {Phys. Rev. C}\ }\textbf {\bibinfo {volume} {72}},\ \bibinfo {pages} {014310} (\bibinfo {year} {2005})}\BibitemShut {NoStop}%
\bibitem [{\citenamefont {Friedrich}\ and\ \citenamefont {Reinhard}(1986)}]{Gs}%
  \BibitemOpen
  \bibfield  {author} {\bibinfo {author} {\bibfnamefont {J.}~\bibnamefont {Friedrich}}\ and\ \bibinfo {author} {\bibfnamefont {P.-G.}\ \bibnamefont {Reinhard}},\ }\href {\doibase 10.1103/PhysRevC.33.335} {\bibfield  {journal} {\bibinfo  {journal} {Phys. Rev. C}\ }\textbf {\bibinfo {volume} {33}},\ \bibinfo {pages} {335} (\bibinfo {year} {1986})}\BibitemShut {NoStop}%
\bibitem [{\citenamefont {Chen}\ and\ \citenamefont {Piekarewicz}(2015)}]{chen2015searching}%
  \BibitemOpen
  \bibfield  {author} {\bibinfo {author} {\bibfnamefont {W.-C.}\ \bibnamefont {Chen}}\ and\ \bibinfo {author} {\bibfnamefont {J.}~\bibnamefont {Piekarewicz}},\ }\href@noop {} {\bibfield  {journal} {\bibinfo  {journal} {Physics Letters B}\ }\textbf {\bibinfo {volume} {748}},\ \bibinfo {pages} {284} (\bibinfo {year} {2015})}\BibitemShut {NoStop}%
\bibitem [{\citenamefont {Brown}(2000)}]{brown2000neutron}%
  \BibitemOpen
  \bibfield  {author} {\bibinfo {author} {\bibfnamefont {B.~A.}\ \bibnamefont {Brown}},\ }\href@noop {} {\bibfield  {journal} {\bibinfo  {journal} {Physical review letters}\ }\textbf {\bibinfo {volume} {85}},\ \bibinfo {pages} {5296} (\bibinfo {year} {2000})}\BibitemShut {NoStop}%
\bibitem [{\citenamefont {Typel}\ and\ \citenamefont {Brown}(2001)}]{typel2001neutron}%
  \BibitemOpen
  \bibfield  {author} {\bibinfo {author} {\bibfnamefont {S.}~\bibnamefont {Typel}}\ and\ \bibinfo {author} {\bibfnamefont {B.~A.}\ \bibnamefont {Brown}},\ }\href@noop {} {\bibfield  {journal} {\bibinfo  {journal} {Physical Review C}\ }\textbf {\bibinfo {volume} {64}},\ \bibinfo {pages} {027302} (\bibinfo {year} {2001})}\BibitemShut {NoStop}%
\end{thebibliography}%
